\documentclass[twocolumn]{aastex63}
\pdfoutput=1
\usepackage{amsmath}
\usepackage{amsfonts}
\usepackage{amssymb}
\usepackage{url}
\usepackage{xspace}
\usepackage{xcolor}
\usepackage{tikz}
\usepackage{enumitem}

\definecolor{mygrn}{rgb}{0.133, 0.545, 0.133}

\newcommand{\fedd}{\ensuremath{f_{\mathrm{Edd}}}\xspace}

\newcommand{\A}{\ensuremath{\mathrm{\AA}}\xspace}

\newcommand{\Hb}{\ensuremath{\rm{H}\beta}\xspace}
\newcommand{\Ha}{\ensuremath{\rm{H}\alpha}\xspace}

\newcommand{\OIII}{[\ion{O}{iii}]\ensuremath{\,\lambda5007}\xspace}

\newcommand{\Lop}{\ensuremath{L_{5100}}} 
\newcommand{\fcl}{\ensuremath{f_{\rm CL}}}

\newcommand{\todo}{\ifmmode \text{\color{red}\Huge{\(\bullet\)}} \else {\color{red}{\Huge$\bullet$}}\fi}
\newcommand{\tido}{\ifmmode {{\color{red}\bullet}} \else {\color{red}$\bullet$}\fi}

\newcommand{\E        }[1]{\ifmmode 10^{#1} \else $10^{#1}$\fi}
\newcommand{\tE        }[1]{\ifmmode \times10^{#1} \else $\times10^{#1}$\fi}
\newcommand{\til}{\ifmmode \sim \else $\sim$\fi}
\renewcommand{\~} {\ifmmode \sim \else $\sim$\fi}

\newcommand{\pc}	{\ifmmode {\rm pc} \else pc\fi}
\newcommand{\kpc}	{\ifmmode {\rm kpc} \else kpc\fi}
\newcommand{\ld}	{\ifmmode {\rm l.d.} \else l.d.\fi}
\newcommand{\kms}	{\ifmmode {\rm km\,s}^{-1} \else km\,s$^{-1}$\fi}
\newcommand{\cc}	{\ifmmode {\rm cm}^{-3}    \else cm$^{-3}$\fi}
\newcommand{\cmii}	{\ifmmode {\rm cm}^{-2}    \else cm$^{-2}$\fi}
\newcommand{\ergs}	{\ifmmode {\rm erg\,s}^{-1} \else erg s$^{-1}$\fi}
\newcommand{\ergcms}	{\ifmmode {\rm erg\,cm}^{-2}\,{\rm s}^{-1} \else erg\,cm$^{-2}$\,s$^{-1}$\fi}
\newcommand{\ergcmsA}	{\ifmmode {\rm erg\,cm}^{-2}\,{\rm s}^{-1}\,{\rm\AA}^{-1}
\else erg\,cm$^{-2}$\,s$^{-1}$\,\AA$^{-1}$\fi}
\newcommand{  \ergcmsHz  }{\ifmmode{\rm erg\,cm}^{-2}\,{\rm s}^{-1}\,{\rm Hz}^{-1}
                       \else ergs\,cm$^{-2}$\,s$^{-1}$\,Hz$^{-1}$\fi}
\newcommand{\kev}	{\ifmmode {\rm keV} \else keV\fi}

\newcommand{\mic}	{\ifmmode {\rm \mu m} \else $\mu$m\fi}
\newcommand{\vFWHM}	{\ifmmode v_{\mbox{\tiny FWHM}} \else $v_{\mbox{\tiny FWHM}}$\fi}
\newcommand{\vBLR}	{\ifmmode v_{\mbox{\tiny BLR}} \else $v_{\mbox{\tiny BLR}}$\fi}
\newcommand{\sigBLR}	{\ifmmode \sigma_{\mbox{\tiny BLR}} \else $\sigma_{\mbox{\tiny BLR}}$\fi}
\newcommand{\vNLR}	{\ifmmode v_{\mbox{\tiny NLR}} \else $v_{\mbox{\tiny NLR}}$\fi}
\newcommand{\tauBLR}	{\ifmmode \tau_{\mbox{\tiny BLR}} \else $\tau_{\mbox{\tiny BLR}}$\fi}

\newcommand{\Hubble}	{\ifmmode {\rm km\,s}^{-1}\,{\rm Mpc}^{-1} \else km\,s$^{-1}$\,Mpc$^{-1}$\fi}
\newcommand{\NDunit}	{\ifmmode {\rm Mpc}^{-3} \else Mpc$^{-3}$\fi}
\newcommand{\LFunit}	{\ifmmode {\rm Mpc}^{-3}\,{\rm mag}^{-1} \else Mpc$^{-3}$\,mag$^{-1}$\fi}
\newcommand{\MFunit}	{\ifmmode {\rm Mpc}^{-3}\,{\rm dex}^{-1} \else Mpc$^{-3}$\,dex$^{-1}$\fi}

\newcommand{\Msun}{\ifmmode M_{\odot} \else $M_{\odot}$\fi}
\newcommand{\Lsun}{\ifmmode L_{\odot} \else $L_{\odot}$\fi}
\newcommand{\Zsun}{\ifmmode Z_{\odot} \else $Z_{\odot}$\fi}
\newcommand{\mpyr}{\ifmmode \Msun\,{\rm yr}^{-1} \else $\Msun\,{\rm yr}^{-1}$\fi}

\newcommand{\qnote}{\ifmmode q_{0} \else $q_{0}$\fi}
\newcommand{\Hnote}{\ifmmode H_{0} \else $H_{0}$\fi}
\newcommand{\hnote}{\ifmmode h_{0} \else $h_{0}$\fi}
\newcommand{\anote}{\ifmmode a_{0} \else $a_{0}$\fi}
\newcommand{\tnote}{\ifmmode t_{0} \else $t_{0}$\fi}

\def\gsim{\;\rlap{\lower 2.5pt \hbox{$\sim$}}\raise 1.5pt\hbox{$>$}\;}
\def\lsim{\;\rlap{\lower 2.5pt \hbox{$\sim$}}\raise 1.5pt\hbox{$<$}\;}

\newcommand{  \Halpha   }{\ifmmode {\rm H}\alpha \else H$\alpha$\fi}

\newcommand{  \ha       }{\Halpha}
\newcommand{  \Hbeta    }{\ifmmode {\rm H}\beta \else H$\beta$\fi}

\newcommand{  \hb       }{\Hbeta}
\newcommand{  \Hgamma   }{\ifmmode {\rm H}\gamma \else H$\gamma$\fi}
\newcommand{  \Hdelta   }{\ifmmode {\rm H}\delta \else H$\delta$\fi}
\newcommand{  \Lya      }{\ifmmode {\rm Ly}\alpha \else Ly$\alpha$\fi}
\newcommand{  \Lyb      }{\ifmmode {\rm Ly}\beta \else Ly$\beta$\fi}
\newcommand{  \Pa       }{\ifmmode {\rm P}\alpha \else P$\alpha$\fi}
\newcommand{  \Pb       }{\ifmmode {\rm P}\beta \else P$\beta$\fi}
\newcommand{  \Bra      }{\ifmmode {\rm Br}\alpha \else Br$\alpha$\fi}
\newcommand{  \Brg      }{\ifmmode {\rm Br}\gamma \else Br$\gamma$\fi}
\newcommand{  \hii      }{\ifmmode {\rm H}\,\textsc{ii} \else H\,\textsc{ii}\fi}
\newcommand{  \hei      }{\ifmmode {\rm He}\,\textsc{i} \else He\,\textsc{i}\fi}
\newcommand{  \heii     }{\ifmmode {\rm He}\,\textsc{ii} \else He\,\textsc{ii}\fi}
\newcommand{  \HeIIuv   }{\ifmmode {\rm He}\,\textsc{ii}\,\lambda1640 \else He\,\textsc{ii}\,$\lambda1640$\fi}
\newcommand{  \HeIIop   }{\ifmmode {\rm He}\,\textsc{ii}\,\lambda4686 \else He\,\textsc{ii}\,$\lambda4686$\fi}
\newcommand{  \CII	}{\ifmmode \left[{\rm C}\,\textsc{ii}\right]\,\lambda157.74\,\mu{\rm m} \else [C\,{\sc ii}]\ $\lambda157.74\,\mu{\rm m}$\fi}
\newcommand{  \cii	}{\ifmmode \left[{\rm C}\,\textsc{ii}\right] \else [C\,{\sc ii}]\fi}

\newcommand{  \ciii     }{\ifmmode \left.{\rm C}\,\textsc{iii}\right] \else C\,\textsc{iii}]\fi}
\newcommand{  \CIII     }{\ifmmode \left.{\rm C}\,\textsc{iii}\right]\,\lambda1909 \else C\,\textsc{iii}]\,$\lambda1909$\fi}
\newcommand{  \civ      }{\ifmmode {\rm C}\,\textsc{iv}  \else C\,\textsc{iv}\fi}
\newcommand{  \CIV      }{\ifmmode {\rm C}\,\textsc{iv}\,\lambda1549 \else C\,\textsc{iv}\,$\lambda1549$\fi}
\newcommand{  \NIIopt   }{\ifmmode \left[{\rm N}\,\textsc{ii}\right]\,\lambda6584 \else [N\,\textsc{ii}]\,$\lambda6584$\fi}
\newcommand{  \nii      }{\ifmmode \left[{\rm N}\,\textsc{ii}\right]  \else [N\,\textsc{ii}]\fi}
\newcommand{  \niii     }{\ifmmode {\rm N}\,\textsc{iii} \else N\,\textsc{iii}\fi}
\newcommand{  \NIII     }{\ifmmode {\rm N}\,\textsc{iii}\,\lambda4640 \else N\,\textsc{iii}\,$\lambda4640$\fi}
\newcommand{  \niv      }{\ifmmode {\rm N}\,\textsc{iv}  \else N\,\textsc{iv}\fi}
\newcommand{  \NIVuv    }{\ifmmode {\rm N}\,\textsc{iv}\,\lambda1486 \else N\,\textsc{iv}\,$\lambda1486$\fi}
\newcommand{  \nv       }{\ifmmode {\rm N}\,\textsc{v}   \else N\,\textsc{v}\fi}
\newcommand{\oi}{\ifmmode \left[{\rm O}\,\textsc{i}\right] \else [O\,{\sc i}]\fi}
\newcommand{\OI}{\ifmmode \left[{\rm O}\,\textsc{i}\right]\,\lambda6300 \else [O\,{\sc i}]$\,\lambda6300$\fi}
\newcommand{\oii}{\ifmmode \left[{\rm O}\,\textsc{ii}\right] \else [O\,{\sc ii}]\fi}
\newcommand{\OII}{\ifmmode \left[{\rm O}\,\textsc{ii}\right]\,\lambda3727 \else [O\,{\sc ii}]\,$\lambda3727$\fi}
\newcommand{\oiii}{\ifmmode \left[{\rm O}\,\textsc{iii}\right] \else [O\,{\sc iii}]\fi}
\newcommand{  \OIIIbf   }{\ifmmode {\rm O}\,\textsc{iii}\,\lambda3133 \else O\,\textsc{iii}\,$\lambda3133$\fi}
\newcommand{  \OIIIuv   }{\ifmmode {\rm O}\,\textsc{iii}\,\lambda1663 \else O\,\textsc{iii}\,$\lambda1663$\fi}
\newcommand{  \oiv      }{\ifmmode {\rm O}\,\textsc{iv}  \else O\,\textsc{iv}\fi}
\newcommand{  \OIVuv    }{\ifmmode {\rm O}\,\textsc{iv}\,\lambda1402  \else O\,\textsc{iv}\,$\lambda1402$\fi}
\newcommand{  \OIVIR    }{\ifmmode {\rm O}\,\textsc{iv}\,25.9\,\mu {\rm m} \else O\,\textsc{iv}\,$25.9\,\mu$m\fi}
\newcommand{  \ovi      }{\ifmmode {\rm O}\,\textsc{vi}   \else O\,\textsc{vi}\fi}
\newcommand{  \Ovi      }{\ifmmode {\rm O}\,\textsc{vi}\,\lambda1035 \else O\,\textsc{vi}\,$\lambda1035$\fi}
\newcommand{  \nei      }{\ifmmode {\rm Ne}\,\textsc{i}   \else Ne\,\textsc{i}\fi}
\newcommand{  \neii     }{\ifmmode {\rm Ne}\,\textsc{ii}  \else Ne\,\textsc{ii}\fi}
\newcommand{  \NeiiIR   }{\ifmmode {\rm Ne}\,\textsc{ii}\,12.8\,\mu {\rm m} \else Ne\,\textsc{ii}\,$12.8\,\mu$m\fi}
\newcommand{  \neiii    }{\ifmmode {\rm Ne}\,\textsc{iii} \else Ne\,\textsc{iii}\fi}
\newcommand{  \neiv     }{\ifmmode {\rm Ne}\,\textsc{iv}  \else Ne\,\textsc{iv}\fi}
\newcommand{  \nev      }{\ifmmode {\rm Ne}\,\textsc{v}   \else Ne\,\textsc{v}\fi}
\newcommand{  \NevIR    }{\ifmmode {\rm Ne}\,\textsc{v}\,24.3\,\mu {\rm m} \else Ne\,\textsc{v}\,$24.3\,\mu$m\fi}
\newcommand{  \nevi     }{\ifmmode {\rm Ne}\,\textsc{vi}  \else Ne\,\textsc{vi}\fi}
\newcommand{  \mgi      }{\ifmmode {\rm Mg}\,\textsc{i} \else Mg\,\textsc{i}\fi}
\newcommand{  \mgii     }{\ifmmode {\rm Mg}\,\textsc{ii} \else Mg\,\textsc{ii}\fi}
\newcommand{  \MgII     }{\ifmmode {\rm Mg}\,\textsc{ii}\,\lambda2798 \else Mg\,\textsc{ii}\,$\lambda2798$\fi}
\newcommand{  \sii      }{\ifmmode {\rm S}\,\textsc{ii} \else S\,\textsc{ii}\fi}
\newcommand{  \siii     }{\ifmmode {\rm S}\,\textsc{iii} \else S\,\textsc{iii}\fi}
\newcommand{  \siv      }{\ifmmode {\rm S}\,\textsc{iv} \else S\,\textsc{iv}\fi}
\newcommand{  \sili     }{\ifmmode {\rm Si}\,\textsc{i}   \else Si\,\textsc{i}\fi}
\newcommand{  \silii    }{\ifmmode {\rm Si}\,\textsc{ii}  \else Si\,\textsc{ii}\fi}
\newcommand{  \Siliv    }{\ifmmode {\rm Si}\,\textsc{iv}  \else Si\,\textsc{iv}\fi}
\newcommand{  \SilIVuv  }{\ifmmode {\rm Si}\,\textsc{iv}\,\lambda1400  \else Si\,\textsc{iv}\,$\lambda1400$\fi}
\newcommand{  \AlIII   }{\ifmmode {\rm Al}\,\textsc{iii}\,\lambda1857 \else Al\,\textsc{iii}\,$\lambda1857$\fi}
\newcommand{  \Aliii   }{\ifmmode {\rm Al}\,\textsc{iii} \else Al\,\textsc{iii}\fi}
\newcommand{  \caii     }{\ifmmode {\rm Ca}\,\textsc{ii} \else Ca\,\textsc{ii}\fi}
\newcommand{  \feii     }{\ifmmode {\rm Fe}\,\textsc{ii} \else Fe\,\textsc{ii}\fi}
\newcommand{  \feiii    }{\ifmmode {\rm Fe}\,\textsc{iii} \else Fe\,\textsc{iii}\fi}
\newcommand{  \Kalpha   }{\ifmmode {\rm K}\alpha \else K$\alpha$\fi}

\newcommand{ \Lhb   }{\ifmmode L_{\hb} \else $L_{\hb}$\fi}
\newcommand{ \Lha   }{\ifmmode L_{\ha} \else $L_{\ha}$\fi}
\newcommand{ \fwhb  }{\ifmmode {\rm FWHM}\left(\hb\right) \else FWHM(\hb)\fi}
\newcommand{\sighb  }{\ifmmode \sigma\left(\hb\right) \else $\sigma\left(\hb\right)$\fi}
\newcommand{ \ewhb  }{\ifmmode {\rm EW}\left(\hb\right) \else EW(\hb)\fi}
\newcommand{ \fwha  }{\ifmmode {\rm FWHM}\left(\ha\right) \else FWHM(\ha)\fi}
\newcommand{ \ewha  }{\ifmmode {\rm EW}\left(\ha\right) \else EW(\ha)\fi}
\newcommand{ \Lmg   }{\ifmmode L\left(\mgii\right) \else $L\left(\mgii\right)$\fi}
\newcommand{ \fwmg  }{\ifmmode {\rm FWHM}\left(\mgii\right) \else FWHM(\mgii)\fi}
\newcommand{ \Lciv  }{\ifmmode L\left(\civ\right) \else $L\left(\civ\right)$\fi}
\newcommand{ \fwciv }{\ifmmode {\rm FWHM}\left(\civ\right) \else FWHM(\civ)\fi}
\newcommand{ \fwhm  }{\ifmmode {\rm FWHM} \else FWHM\fi} 
\newcommand{ \voff  }{\ifmmode v_{\rm off} \else $v_{\rm off}$\fi} 
\newcommand{ \vmax  }{\ifmmode v_{\rm max} \else $v_{\rm max}$\fi} 

\newcommand{ \mumg  }{\ifmmode \mu\left(\mgii\right) \else $\mu\left(\mgii\right)$\fi}
\newcommand{ \fmg   }{\ifmmode f\left(\mgii\right) \else $f\left(\mgii\right)$\fi}
\newcommand{ \muciv }{\ifmmode \mu\left(\civ\right) \else $\mu\left(\civ\right)$\fi}
\newcommand{ \fciv  }{\ifmmode f\left(\civ\right) \else $f\left(\civ\right)$\fi}

\newcommand{  \auvo     }{\ifmmode \alpha_{\nu,{\rm UVO}} \else $\alpha_{\nu,{\rm UVO}}$\fi}
\newcommand{  \Ledd     }{\ifmmode L_{\rm Edd} \else $L_{\rm Edd}$\fi}
\newcommand{  \lamLlam  }{\ifmmode \lambda L_{\lambda} \else $\lambda L_{\lambda}$\fi}
\newcommand{  \lLl      }{\ifmmode \lambda L_{\lambda} \else $\lambda L_{\lambda}$\fi}
\newcommand{  \nuLnu    }{\ifmmode \nu L_{\nu} \else $\nu L_{\nu}$\fi}
\newcommand{  \nLn      }{\ifmmode \nu L_{\nu} \else $\nu L_{\nu}$\fi}
\newcommand{  \Luv      }{\ifmmode L_{1350} \else $L_{1350}$\fi}
\newcommand{  \lLop     }{\ifmmode \log\left(\Lop/\ergs\right) \else $\log\left(\Lop/\ergs\right)$\fi}
\newcommand{  \Lthree   }{\ifmmode L_{3000} \else $L_{3000}$\fi}
\newcommand{  \lLthree  }{\ifmmode \log\left(\Lthree/\ergs\right) \else $\log\left(\Lthree/\ergs\right)$\fi}
\newcommand{  \Lsix      }{\ifmmode L_{6200} \else $L_{6200}$\fi}
\newcommand{  \lLisx     }{\ifmmode \log\left(\Lop/\ergs\right) \else $\log\left(\Lop/\ergs\right)$\fi}
\newcommand{  \Lxray    }{\ifmmode L_{\rm X} \else $L_{\rm X}$\fi}

\newcommand{  \Lhard    }{\ifmmode L_{\rm 2-10} \else $L_{\rm 2-10}$\fi}
\newcommand{  \Lsoft    }{\ifmmode L_{\rm 0.5-2} \else $L_{\rm 0.5-2}$\fi}

\newcommand{\Fthree}{\ifmmode F_{3000} \else $F_{3000}$\fi}
\newcommand{\fuv}{\ifmmode f_{\lambda}\left(1450{\rm \AA}\right) \else $f_{\lambda}\left(1450 {\rm \AA}\right)$\fi}
\newcommand{\fthree}{\ifmmode f_{\lambda}\left(3000{\rm \AA}\right) \else $f_{\lambda}\left(3000{\rm \AA}\right)$\fi}
\newcommand{\fH}{\ifmmode f_{\lambda}\left(1.65\micron\right) \else
$f_{\lambda}\left(1.65\micron\right)$\fi}

\newcommand{\fbol}{\ifmmode f_{\rm bol} \else $f_{\rm bol}$\fi}
\newcommand{\fbolwv}{\ifmmode f_{\rm bol}\left(\lambda\right) \else $f_{\rm bol}\left(\lambda\right)$\fi}
\newcommand{\fbolopt}{\ifmmode f_{\rm bol}\left(5100{\rm \AA}\right) \else $f_{\rm bol}\left(5100{\rm \AA}\right)$\fi}
\newcommand{\fbolthree}{\ifmmode f_{\rm bol}\left(3000{\rm \AA}\right) \else $f_{\rm bol}\left(3000{\rm \AA}\right)$\fi}
\newcommand{\fboluv}{\ifmmode f_{\rm bol}\left(1450{\rm \AA}\right) \else $f_{\rm bol}\left(1450{\rm \AA}\right)$\fi}

\newcommand{\fbolbat}{\ifmmode f_{\rm bol}\left(14-150\,\kev\right) \else $f_{\rm bol}\left(14-150\,\kev\right)$\fi}
\newcommand{\fbolhard}{\ifmmode f_{\rm bol}\left(2-10\,\kev\right) \else $f_{\rm bol}\left(2-10\,\kev\right)$\fi}

\newcommand{\fobs}{\ifmmode f_{\rm obs} \else $f_{\rm obs}$\fi}

\newcommand{  \mbh      }{\ifmmode M_{\rm BH} \else $M_{\rm BH}$\fi}
\newcommand{  \lmbh     }{\ifmmode \log\left(\mbh/\Msun\right) \else $\log\left(\mbh/\Msun\right)$\fi} 
\newcommand{  \lledd    }{\ifmmode L/L_{\rm Edd} \else $L/L_{\rm Edd}$\fi}
\newcommand{  \mmedd    }{\ifmmode \dot{m}/\dot{m}_{\rm \,Edd} \else $\dot{m}/\dot{m}_{\rm \,Edd}$\fi}
\newcommand{  \Lbol     }{\ifmmode L_{\rm bol} \else $L_{\rm bol}$\fi}
\newcommand{  \lbol     }{\ifmmode L_{\rm bol} \else $L_{\rm bol}$\fi}
\newcommand{  \lLbol    }{\ifmmode \log\left(\Lbol/\ergs\right) \else $\log\left(\Lbol/\ergs\right)$\fi} 
\newcommand{  \Lagn     }{\ifmmode L_{\rm AGN} \else $L_{\rm AGN}$\fi}
\newcommand{  \lagn     }{\ifmmode L_{\rm AGN} \else $L_{\rm AGN}$\fi}

\newcommand{  \tgrow     }{\ifmmode t_{\rm growth} \else $t_{\rm growth}$\fi}
\newcommand{  \tAD     }{\ifmmode t_{\rm acc} \else $t_{\rm acc}$\fi}
\newcommand{  \tacc    }{\ifmmode t_{\rm acc} \else $t_{\rm acc}$\fi}
\newcommand{  \tUni      }{\ifmmode t_{\rm Universe} \else $t_{\rm Universe}$\fi}

\newcommand{  \Mdotin	}{\ifmmode \dot{M}_{\rm infall} \else $\dot{M}_{\rm infall}$\fi}
\newcommand{  \Mdotbh	}{\ifmmode \dot{M}_{\rm BH} \else $\dot{M}_{\rm BH}$\fi}
\newcommand{  \Mdotad	}{\ifmmode \dot{M}_{\rm AD} \else $\dot{M}_{\rm AD}$\fi}
\newcommand{  \Mdotacc	}{\ifmmode \dot{M}_{\rm acc} \else $\dot{M}_{\rm acc}$\fi}
\newcommand{  \Mdotthin	}{\ifmmode \dot{M}_{\rm thin} \else $\dot{M}_{\rm thin}$\fi}
\newcommand{  \Mdotdisk	}{\ifmmode \dot{M}_{\rm disk} \else $\dot{M}_{\rm disk}$\fi}

\newcommand{  \Mindot	}{\ifmmode \dot{M}_{\rm infall} \else $\dot{M}_{\rm infall}$\fi}
\newcommand{  \Mbhdot	}{\ifmmode \dot{M}_{\rm BH} \else $\dot{M}_{\rm BH}$\fi}
\newcommand{  \Maddot	}{\ifmmode \dot{M}_{\rm AD} \else $\dot{M}_{\rm AD}$\fi}
\newcommand{  \Maccdot	}{\ifmmode \dot{M}_{\rm acc} \else $\dot{M}_{\rm acc}$\fi}
\newcommand{  \Mthdot	}{\ifmmode \dot{M}_{\rm thin} \else $\dot{M}_{\rm thin}$\fi}
\newcommand{  \Mdsdot	}{\ifmmode \dot{M}_{\rm disk} \else $\dot{M}_{\rm disk}$\fi}

\newcommand{  \as	}{\ifmmode a_{\rm *} \else $a_{\rm *}$\fi}
\newcommand{  \avec	}{\ifmmode \vec{a}_{\rm *} \else $\vec{a}_{\rm *}$\fi}
\newcommand{  \re	}{\ifmmode \eta      	 \else $\eta$\fi}
\newcommand{  \RISCO	}{\ifmmode R_{\rm ISCO}  \else $R_{\rm ISCO}$\fi}

\newcommand{  \mseed    }{\ifmmode M_{\rm seed} \else $M_{\rm seed}$\fi}
\newcommand{  \mbul     }{\ifmmode M_{\rm bulge} \else $M_{\rm bulge}$\fi} 
\newcommand{  \mstar    }{\ifmmode M_{*} \else $M_{*}$\fi} 
\newcommand{  \mgal     }{\ifmmode M_{*} \else $M_{*}$\fi} 
\newcommand{  \mhost    }{\ifmmode M_{\rm host} \else $M_{\rm host}$\fi}
\newcommand{  \mmsmall  }{\ifmmode M_{\rm BH}/M_{*} \else $M_{\rm BH}/M_{*}$\fi}
\newcommand{  \mmlarge  }{\ifmmode M_{*}/M_{\rm BH} \else $M_{*}/M_{\rm BH}$\fi}

\newcommand{  \mmdotlarge}{\ifmmode \dot{M}_*/\Mbhdot \else $\dot{M}_*/\Mbhdot$\fi}
\newcommand{  \mmdotsmall}{\ifmmode \Mbhdot/\dot{M}_* \else $\Mbhdot/\dot{M}_*$\fi}

\newcommand{  \mmwp     }{\ifmmode \left(M_{*}/M_{\rm BH}\right) \else $\left(M_{*}/M_{\rm BH}\right)$\fi}
\newcommand{  \ml       }{\ifmmode M_{*}/L_{*} \else $M_{*}/L_{*}$\fi}
\newcommand{  \mlwp     }{\ifmmode \left(M_{*}/L\right) \else $\left(M_{*}/L\right)$\fi}
\newcommand{  \mlk      }{\ifmmode \left(M_{*}/L_{K}\right) \else $\left(M_{*}/L_{K}\right)$\fi}
\newcommand{  \sigs     }{\ifmmode \sigma_{*} \else $\sigma_{*}$\fi}
\newcommand{  \Reff     }{\ifmmode R_{\rm e} \else $R_{\rm e}$\fi}
\newcommand{  \Rvir     }{\ifmmode R_{\rm vir} \else $R_{\rm vir}$\fi}
\newcommand{  \Rtwo     }{\ifmmode R_{200} \else $R_{200}$\fi}
\newcommand{  \Rfive    }{\ifmmode R_{500} \else $R_{500}$\fi}
\newcommand{  \Rgrp     }{\ifmmode R_{\rm grp} \else $R_{\rm grp}$\fi}
\newcommand{  \nser     }{\ifmmode n_{\rm s} \else $n_{\rm s}$\fi}
\newcommand{  \LSF      }{\ifmmode L_{\rm SF}  \else $L_{\rm SF}$\fi}
\newcommand{  \LFIR     }{\ifmmode L_{\rm FIR} \else $L_{\rm FIR}$\fi}
\newcommand{  \Lfir     }{\ifmmode L_{\rm FIR} \else $L_{\rm FIR}$\fi}
\newcommand{  \LTIR     }{\ifmmode L_{\rm TIR} \else $L_{\rm TIR}$\fi}
\newcommand{  \Ltir     }{\ifmmode L_{\rm TIR} \else $L_{\rm TIR}$\fi}

\newcommand{  \mdyn     }{\ifmmode M_{\rm dyn} \else $M_{\rm dyn}$\fi} 
\newcommand{  \mgas     }{\ifmmode M_{\rm gas} \else $M_{\rm gas}$\fi} 
\newcommand{  \mh       }{\ifmmode M_{\rm h} \else $M_{\rm h}$\fi}
\newcommand{  \mhalo    }{\ifmmode M_{\rm halo} \else $M_{\rm halo}$\fi}
\newcommand{  \sfr      }{\ifmmode {\rm SFR} \else SFR\fi}

\newcommand{ \Lcii     }{\ifmmode L_{\cii} \else $L_{\cii}$\fi}
\newcommand{ \fwcii  }{\ifmmode {\rm FWHM}\cii \else FWHM\cii\fi}

\newcommand{\bj}{\ifmmode b_{\rm J} \else $b_{\rm J}$\fi}

\newcommand{\iab}{\ifmmode i_{\rm AB} \else $i_{\rm AB}$\fi}

\newcommand{\jab}{\ifmmode J_{\rm AB} \else $J_{\rm AB}$\fi}
\newcommand{\hab}{\ifmmode H_{\rm AB} \else $H_{\rm AB}$\fi}
\newcommand{\kab}{\ifmmode K_{\rm AB} \else $K_{\rm AB}$\fi}

\newcommand{\jveg}{\ifmmode J_{\rm Vega} \else $J_{\rm Vega}$\fi}
\newcommand{\hveg}{\ifmmode H_{\rm Vega} \else $H_{\rm Vega}$\fi}
\newcommand{\kveg}{\ifmmode K_{\rm Vega} \else $K_{\rm Vega}$\fi}

\def\arcsec{\hbox{$^{\prime\prime}$}}

\newcommand{  \Chisq    }{\ifmmode \chi^{2} \else $\chi^{2}$}
\newcommand{  \nelec    }{\ifmmode n_{e} \else $n_{e}$\fi}     
\newcommand{  \nh       }{\ifmmode n_{\rm H} \else $n_{\rm H}$\fi}     
\newcommand{  \Ncol     }{\ifmmode N_{\rm col} \else $N_{\rm col}$\fi} 
\newcommand{  \NH       }{\ifmmode N_{\rm H} \else $N_{\rm H}$\fi}     

\def\sun{\hbox{$\odot$}}

\def\arcsec{\hbox{$^{\prime\prime}$}}
\DeclareRobustCommand{\ion}[2]{%
\relax\ifmmode
\ifx\testbx\f@series
{\mathbf{#1\,\mathsc{#2}}}\else
{\mathrm{#1\,\mathsc{#2}}}\fi
\else\textup{#1\,{\mdseries\textsc{#2}}}%
\fi} 

\usepackage[latin1]{inputenc}
\usepackage[T1]{fontenc}
\usepackage{soul}

\published{as ApJ, 966, 85}

\shorttitle{First year SDSS-V CL-AGNs}
\shortauthors{Zeltyn et al.}
\graphicspath{{./}{figures/}}

\begin{document}

\title{Exploring Changing-look Active Galactic Nuclei with the Sloan Digital Sky Survey V:\\ First Year Results}


\author[0000-0002-7817-0099]{Grisha Zeltyn}
\affiliation{School of Physics and Astronomy, Tel Aviv University, Tel Aviv 69978, Israel}

\author[0000-0002-3683-7297]{Benny Trakhtenbrot}
\affiliation{School of Physics and Astronomy, Tel Aviv University, Tel Aviv 69978, Israel}

\author[0000-0002-3719-940X]{Michael Eracleous}
\affiliation{Department of Astronomy \& Astrophysics, The Pennsylvania State University, University Park, PA 16802, USA}
\affiliation{Institute for Gravitation and the Cosmos, The Pennsylvania State University, University Park, PA 16802, USA}

\author[0000-0002-6893-3742]{Qian Yang}
\affiliation{Center for Astrophysics, Harvard \& Smithsonian, 60 Garden Street, Cambridge, MA 02138, USA}

\author[0000-0002-8179-9445]{Paul Green}
\affiliation{Center for Astrophysics, Harvard \& Smithsonian, 60 Garden Street, Cambridge, MA 02138, USA}

\author[0000-0002-6404-9562]{Scott F. Anderson}
\affiliation{Astronomy Department, University of Washington, Box 351580, Seattle, WA 98195, USA}

\author[0000-0002-5907-3330]{Stephanie LaMassa}
\affiliation{Space Telescope Science Institute, 3700 San Martin Drive, Baltimore, MD 21218, USA}

\author[0000-0001-8557-2822]{Jessie Runnoe}
\affiliation{Department of Physics and Astronomy, Vanderbilt University, VU Station 1807, Nashville, TN 37235, USA}

\author[0000-0002-9508-3667]{Roberto J. Assef} 
\affiliation{Instituto de Estudios Astrof\'isicos, Facultad de Ingenier\'ia y Ciencias, Universidad Diego Portales, Av. Ej\'ercito Libertador 441, Santiago, Chile} 

\author[0000-0002-8686-8737]{Franz E. Bauer}
\affil{Instituto de Astrof\'isica, Pontificia Universidad Cat\'olica de Chile, Casilla 306, Santiago 22, Chile}
\affil{Millennium Institute of Astrophysics (MAS), Nuncio Monse$\tilde{\rm n}$or S\'otero Sanz 100, Providencia, Santiago, Chile}
\affil{Space Science Institute, 4750 Walnut Street, Suite 205, Boulder, CO 80301, USA}

\author[0000-0002-0167-2453]{W.N. Brandt}
\affiliation{Department of Astronomy \& Astrophysics, The Pennsylvania State University, University Park, PA 16802, USA}
\affiliation{Institute for Gravitation and the Cosmos, The Pennsylvania State University, University Park, PA 16802, USA}
\affiliation{Department of Physics, 104 Davey Laboratory, The Pennsylvania State University, University Park, PA 16802, USA}

\author[0000-0001-9776-9227]{Megan C. Davis}
\affil{Department of Physics, 196A Auditorium Road, Unit 3046, University of Connecticut, Storrs, CT 06269, USA}

\author[0000-0001-9676-730X]{Sara E. Frederick}
\affiliation{Department of Physics and Astronomy, Vanderbilt University, VU Station 1807, Nashville, TN 37235, USA}

\author[0000-0001-8032-2971]{Logan B. Fries}
\affil{Department of Physics, 196A Auditorium Road, Unit 3046, University of Connecticut, Storrs, CT 06269, USA}

\author[0000-0002-3168-0139]{Matthew J. Graham}
\affil{California Institute of Technology, 1200 E. California Blvd, Pasadena, CA 91125, USA}

\author[0000-0001-9440-8872]{Norman A. Grogin}
\affiliation{Space Telescope Science Institute, 3700 San Martin Drive,
Baltimore, MD 21218, USA}

\author[0000-0002-5063-0751]{Muryel Guolo}
\affiliation{Department of Physics and Astronomy, Johns Hopkins University, 3400 North Charles Street, Baltimore, MD 21218, USA}

\author[0000-0002-8606-6961]{Lorena Hern\'andez-Garc\'ia}
\affil{Millennium Institute of Astrophysics (MAS), Nuncio Monse$\tilde{\rm n}$or S\'otero Sanz 100, Providencia, Santiago, Chile}

\author[0000-0002-6610-2048]{Anton M. Koekemoer}
\affiliation{Space Telescope Science Institute, 3700 San Martin Drive, Baltimore, MD 21218, USA}

\author{Mirko Krumpe}
\affiliation{Leibniz-Institut f\"ur Astrophysik Potsdam (AIP), An der Sternwarte 16, 14482 Potsdam, Germany}

\author[0000-0003-0049-5210]{Xin Liu}
\affiliation{Department of Astronomy, University of Illinois at Urbana-Champaign, Urbana, IL 61801, USA}
\affiliation{National Center for Supercomputing Applications, University of Illinois at Urbana-Champaign, Urbana, IL 61801, USA}
\affiliation{Center for Artificial Intelligence Innovation, University of Illinois at Urbana-Champaign, 1205 West Clark Street, Urbana, IL 61801, USA}

\author[0000-0002-7843-7689]{Mary Loli Mart\'{i}nez-Aldama}
\affiliation{Astronomy Department, Universidad de Concepci\'{o}n, Casilla 160-C, Concepci\'{o}n 4030000, Chile}

\author[0000-0001-5231-2645]{Claudio Ricci}
\affiliation{Instituto de Estudios Astrof\'isicos, Facultad de Ingenier\'ia y Ciencias, Universidad Diego Portales, Av. Ej\'ercito Libertador 441, Santiago, Chile} 
\affiliation{Kavli Institute for Astronomy and Astrophysics, Peking University, Beijing 100871, China}

\author[0000-0001-7240-7449]{Donald P.\ Schneider} 
\affiliation{Department of Astronomy \& Astrophysics, The Pennsylvania State University, University Park, PA 16802, USA}
\affiliation{Institute for Gravitation and the Cosmos, The Pennsylvania State University, University Park, PA 16802, USA}

\author[0000-0003-1659-7035]{Yue Shen}
\affiliation{Department of Astronomy, University of Illinois at Urbana-Champaign, Urbana, IL 61801, USA}
\affiliation{National Center for Supercomputing Applications, University of Illinois at Urbana-Champaign, Urbana, IL 61801, USA}

\author[0000-0003-2656-6726]{Marzena \'Sniegowska}
\affiliation{School of Physics and Astronomy, Tel Aviv University, Tel Aviv 69978, Israel}

\author[0000-0001-8433-550X]{Matthew J. Temple}
\affil{Instituto de Estudios Astrof\'isicos, Facultad de Ingenier\'ia y Ciencias, Universidad Diego Portales, Av. Ej\'ercito Libertador 441, Santiago, Chile}

\author[0000-0002-1410-0470]{Jonathan R. Trump}
\affil{Department of Physics, 196A Auditorium Road, Unit 3046, University of Connecticut, Storrs, CT 06269, USA}

\author[0000-0002-1935-8104]{Yongquan Xue}
\affiliation{CAS Key Laboratory for Research in Galaxies and Cosmology, Department of Astronomy, University of Science and Technology of China, Hefei 230026, China}

\author[0000-0002-8725-1069]{Joel R. Brownstein}
\affiliation{Department of Physics and Astronomy, University of Utah, 115 S. 1400 East, Salt Lake City, UT 84112, USA}

\author[0000-0002-4459-9233]{Tom Dwelly}
\affiliation{Max-Planck-Institut f{\"u}r extraterrestrische Physik, Giessenbachstra\ss{}e, D-85748 Garching, Germany}

\author[0000-0002-6770-2627]{Sean Morrison}
\affiliation{Department of Astronomy, University of Illinois at Urbana-Champaign, Urbana, IL 61801, USA}

\author[0000-0002-3601-133X]{Dmitry Bizyaev}
\affiliation{Apache Point Observatory and New Mexico State University, P.O. Box 59, Sunspot, NM, 88349-0059, USA}
\affiliation{Sternberg Astronomical Institute, Moscow State University, Moscow}

\author[0000-0002-2835-2556]{Kaike Pan}
\affiliation{Apache Point Observatory and New Mexico State University, P.O. Box 59, Sunspot, NM, 88349-0059, USA}

\author[0000-0001-9852-1610]{Juna A. Kollmeier}
\affiliation{The Observatories of the Carnegie Institution for Science, 813 Santa Barbara Street, Pasadena, CA 91101, USA}
\affiliation{Canadian Institute for Theoretical Astrophysics, 60 Saint George Street, Toronto, ON M5S 3H8, Canada}


\correspondingauthor{Grisha Zeltyn}
\email{grishazeltyn@tauex.tau.ac.il, benny@astro.tau.ac.il}

\vspace{1cm}
\begin{abstract}

``Changing-look'' active galactic nuclei (CL-AGNs) challenge our basic ideas about the physics of accretion flows and circumnuclear gas around supermassive black holes. 
Using first-year Sloan Digital Sky Survey V (SDSS-V) repeated spectroscopy of nearly 29,000 previously known AGNs, combined with dedicated follow-up spectroscopy, and publicly available optical light curves, we have identified 116 CL-AGNs where (at least) one broad emission line has essentially (dis-)appeared, as well as 88 other extremely variable systems. 
Our CL-AGN sample, with 107 newly identified cases, is the largest reported to date, and includes $\sim0.4\%$ of the AGNs reobserved in first-year SDSS-V operations.
Among our CL-AGNs, 67\% exhibit dimming while 33\% exhibit brightening.
Our sample probes extreme AGN spectral variability on months to decades timescales, including some cases of recurring transitions on surprisingly short timescales ($\lesssim 2$ months in the rest frame).
We find that CL events are preferentially found in lower-Eddington-ratio (\fedd) systems: Our CL-AGNs have a \fedd\ distribution that significantly differs from that of a carefully constructed, redshift- and luminosity-matched control sample (Anderson-Darling test yielding $p_{\rm AD}\approx 6\times10^{-5}$; median $\fedd\approx0.025$ vs. $0.043$).
This preference for low \fedd\ strengthens previous findings of higher CL-AGN incidence at lower \fedd, found in smaller samples.
Finally, we show that the broad \mgii\ emission line in our CL-AGN sample tends to vary significantly less than the broad \hb\ emission line.
Our large CL-AGN sample demonstrates the advantages and challenges in using multi-epoch spectroscopy from large surveys to study extreme AGN variability and physics. 

\end{abstract}


\section{Introduction}
\label{sec:intro}

The multiwavelength radiation emerging from active galactic nuclei (AGNs) has long been known to exhibit stochastic variability, of order of a few to tens of percents, and on timescales ranging from days to years \citep[e.g.,][and references therein]{Ulrich97,VB04,MacLeod12}. While the origin of this variability is not yet clearly understood, there has been a recent advancement in observations of yet more extreme and coherent-looking changes in AGNs, thanks to advances in time-domain astronomy. 
Among those dramatic variability events, ``changing-look'' AGNs (CL-AGNs) are those systems that are observed to transition between phenomenologically distinct spectral states, on rest-frame timescales as short as several months. Specifically, CL-AGNs identified in the UV-optical regime show the disappearance or appearance, or significant weakening or strengthening, of the broad emission line and/or the blue quasar-like continuum emission, components that are typically seen in unobscured AGNs (see, e.g., \citealt{RT23} for a recent review of these events).

While the earliest such changing-look event was reported nearly 50 yr ago \citep{Tohline76}, many more systems have been discovered over the past decade, starting from a few cases that expanded the study of CL-AGNs to higher, quasar-like luminosities \cite[i.e., $L>10^{45}\,\ergs$; see, e.g.,][]{LaMassa15,Runnoe16},\footnote{Throughout this work, we use the term ``quasars'' to describe AGN-dominated sources which, at least at some point, exhibit broad emission lines and a blue, power-law-like continuum.} followed by more sizable, spectroscopically-confirmed samples \cite[e.g.,][]{MacLeod16, MacLeod19, Yang18, Potts21, Green22} and higher-redshift sources \citep{Ross18,Guo20_hiz}.

In general, the physical mechanisms driving CL-AGN transitions remain uncertain. Most studies which have focused either on detailed analysis of individual objects \cite[e.g.,][]{Denney14,LaMassa15, Husemann16,Ruan16,Runnoe16, Sheng17, Stern18, Hutsemekers19, Trakhtenbrot19, Wang19, Hutsemekers20, Ricci20, Guolo21, Nagoshi21} or on larger samples \cite[e.g.,][]{MacLeod16, Yang18,Green22,Temple23} favor drastic changes to the accretion flows that power the AGNs as the source of observed variability, based mainly on evidence of the radiative response of various circumnuclear gas regions to accretion-generated, UV photons (e.g., the torus response in the IR regime), a lack of concurrent signatures of obscuration changes, and other evidence from multiwavelength observations \cite[e.g.,][]{Ruan19}. This explanation, however, poses a significant challenge to our understanding of accretion physics, as the observed changes occur on timescales that are much shorter than what is expected within the thin-accretion-disk paradigm (see, e.g., \citealt{Stern18,RT23} for discussions of the relevant timescales; but see also \citealt{Shen21} for a counterargument).

In contrast, evidence for variable obscuration being the driver of extreme-UV/optical spectral variations is much more limited. Such variations of the disk and the broad-line region (BLR) components have been inferred from the variable optical spectra of a few Seyfert 1.8 and 1.9 galaxies \citep[see][and references therein]{Goodrich95}. In addition, it has been proposed that some variations in the profiles of broad emission lines can be explained by partial obscuration of the BLR by outflowing dusty gas clumps \citep{GH18}. 
A sample of six CL-AGNs from the Sloan Digital Sky Survey II, assembled by \citet{Potts21}, revealed that the spectral changes in most of them can be accounted for by reddening due to dust. However, the authors also noted that the observed timescales of changes are too short for obscuration from torus clouds.
Recently, \citet[]{Zeltyn22} presented a CL-AGN which may be better explained by a variable, and perhaps dynamic, obscuring medium. 
Whichever the driving mechanisms of specific CL-AGN transitions may be, they allow us to probe the structure of AGNs and potentially test our current understanding of their accretion flows, as well as the BLR, the dusty torus, and other gas components that surround them.

Several models have been proposed to explain the extreme and fast transitions observed in CL-AGNs. These include, for example, rapid mass-accretion drops, enabled by a larger-than-fiducial sound speed \citep{ND18}, disk processes happening on the thermal and/or cooling-front timescales, such as thermal disk instabilities \citep{Stern18}, tidal disruptions of stars onto preexisting AGN disks \citep{Merloni15,Chan19}, semi-periodic effects that may explain recurring CL-AGNs \citep{Sniegowska20,Pan21}, and other effects in disks that deviate from the standard thin-disk paradigm \cite[e.g.,][]{DexterBegelman19,JiangBlaes20}. Some models even suggest a direct link between the formation and destruction of the BLR itself and variable accretion power, however this is only relevant for systems with very low accretion rates \cite[see][and references therein]{Elitzur14}.

Understanding these extreme phenomena requires studying large samples, preferably drawn from surveys with systematic and stable targeting and cadence strategies. While the searches for CL-AGNs and the emerging samples are indeed growing, there are still many challenges to overcome. The criteria used for finding CL-AGNs among large, multi-epoch spectroscopic datasets are complicated and vary between studies. Importantly, most large CL-AGN samples are based on repeated spectroscopy taken several years or even decades apart, as dictated by the corresponding survey strategies \cite[e.g.,][]{Yang18, Green22, Guo24_DESI}. Such low-cadence surveys prohibit a detailed analysis of the transitions themselves, and, in the absence of contemporary light curves, yield only upper limits on the transition timescales. 

While years-to-decades transition timescales are already challenging for accretion-disk models \cite[see, e.g., the discussion in][]{RT23}, some CL-AGNs were found to transition on even shorter timescales ($\lesssim1\,\rm{yr}$ rest frame), as seen either directly in spectroscopy \cite[e.g.,][]{Trakhtenbrot19, Ross20, Zeltyn22} or implied indirectly from photometric monitoring \cite[e.g.,][]{Gezari17, Yang18, Frederick19, Yan19, Green22}.

Another complementary approach for discovering CL-AGNs is to spectroscopically monitor those AGNs that exhibit dramatic photometric variability \cite[e.g.,][]{MacLeod16, MacLeod19, Rumbaugh18, Graham20, Senarath21,LopezNavas22,LopezNavas23}. While this photometry-driven approach benefits from large parent samples covered by wide-field, high-cadence surveys, it might miss host-dominated CL-AGNs, where extreme changes in the broad emission lines can occur without a significant continuum change. In this case, temporally resolving the spectral transition is also challenging.

An additional challenge for the identification of CL-AGN is that not all broad emission lines appear to vary in the same manner. Specifically, the broad \MgII\ line varies by a lesser degree in extremely variable quasars (EVQs) compared to the quasar continuum and does not exhibit the line `breathing' that is characteristic of broad Balmer emission lines \cite[e.g.,][and references therein]{Yang20_CLAGN}. These findings are consistent with the low degree of \mgii\ variability reported in some reverberation mapping (RM) campaigns \cite[e.g.,][]{Cackett15, Sun15,Homayouni20}, and may be linked to the particular excitation mechanisms, optical depths, and/or the BLR geometry of the \mgii\ emission line, or to some other, yet unknown BLR physics \cite[e.g.,][]{Sun15,Guo20_mgii}.

In this work, we present a large sample of CL-AGNs assembled from the first year of the Sloan Digital Sky Survey V (SDSS-V; \citealt{Kollmeier17,Almeida23}). Our work is part of a dedicated effort within SDSS-V to obtain repeated spectroscopy of $>30,000$ previously known SDSS quasars, some of which will be revisited several times within SDSS-V. This program will allow us to perform a systematic search for CL-AGNs and other dramatic variability events, directly in spectroscopy, over a multitude of timescales.

This paper is organized as follows. Section \ref{sec:sample} presents the main spectroscopic and the ancillary photometric data used to identify CL-AGNs, as well as the criteria used for selecting our final sample.
Section \ref{sec:control_and_qunatities} describes the reference AGN sample used to compare with our CL-AGN sample, and the spectral decomposition methods we use to analyze our CL-AGNs and to compare them with the reference sample. 
Section \ref{sec:analysis_results} presents the population properties of our CL-AGN sample as compared to the general AGN population, and discusses the implications of our findings. 
Finally, we conclude with Section~\ref{sec:conclusions}, where we summarize our key results.
Throughout this work, we adopt a flat $\Lambda$ cold dark matter cosmology with $H_0=70\,\kms\,\rm{Mpc}^{-1}$ and $\Omega_{\rm{m}}=0.3$.

\section{Data and Observations}
\label{sec:sample}

Our sample and analysis of strongly variable AGNs are based on optical spectroscopy obtained during the first year of SDSS-V of AGNs spectroscopically observed during the previous four SDSS generations (SDSS I-IV, DR16; \citealt{Ahumada20_DR16}). 

In this section, we describe the parent sample, the CL-AGN selection criteria, as well as the follow-up spectroscopic and ancillary photometric data used to corroborate our CL-AGN candidates. 
Table~\ref{tab:clagn_selection} summarizes this process and lists the number of sources selected or excluded in each of the steps described below.

\begin{deluxetable*}{lc}
\tablecaption{CL-AGN selection}
\label{tab:clagn_selection}
\tablewidth{\columnwidth}
\tabletypesize{\scriptsize}
\tablehead{
\colhead{Notes} & \colhead{Number}
}
\startdata
    Unique AGNs observed during first year of SDSS-V   & 44,761  \\ 
    AGNs with archival SDSS DR16 spectroscopy  &  29,631 \\ 
    Non-BHM-RM targets & 28,873 \\
    Sources flagged by the automated selection criterion (Section~\ref{subsec:spec_search}, Equation~\ref{eq:C_cut}) & 3338 \\
    Passed visual inspection (Section~\ref{subsec:spec_search}) & 130 \\
    Passed sample refinement through 
    spectral decomposition (Section~\ref{subsec:decomp_for_search}) & 123 \\
    Additional candidates that pass criterion after spectral decomposition (Section~\ref{subsec:decomp_for_search}) & +14 \\
    Passed sample refinement through 
    spectroscopic follow-up (Section~\ref{subsubsec:followup}) & 134 \\
    Final core sample: passed sample refinement through photometric cross-match (Section~\ref{subsubsec:crossmatch}) & 113 \\ 
    \hline
    BHM-RM CL-AGNs (Section~\ref{subsubsec:RM_CLAGN}) & 3 \\
    Other extremely variable sources (Section~\ref{subsubsec:EVQs_sample}) & 88 \\
\enddata
\end{deluxetable*}

\subsection{SDSS-V Repeated Spectroscopy}
\label{subsec:sdssv}

Our main dataset consists of medium-resolution ($R\sim2000$) spectra obtained through the Black Hole Mapper (BHM) program within the first year of operations of SDSS-V, covering 2020 October through 2021 June. These spectra were acquired using the Baryon Oscillation Spectroscopic Survey \cite[BOSS;][]{Smee13} spectrograph, mounted on the Sloan Foundation 2.5 m telescope \citep{Gunn06} at the Apache Point Observatory, and utilizing plates to position fibers on preselected science targets (as well as standard stars and sky regions). 
The new, robotic Focal Plane System that constitutes a key upgrade of SDSS-V \citep{Pogge20} compared with previous SDSS projects was \emph{not} operational during the relevant observation period.

One of the key goals of the BHM program is to investigate spectral AGN variability using repeated spectroscopy of previously known broad-line AGNs with archival SDSS spectroscopy (see \citealt{Almeida23} for a detailed account of SDSS-V/BHM targeting). 
These core AGN targets with previous Sloan spectra can be divided into two main subsets: 

\begin{enumerate}
    \item All-Quasar Multi-Epoch Spectroscopy (AQMES) targets. About 20,000 $i\leq19.1$ broad-line AGNs drawn from SDSS-I--IV, that will be observed 2--12 times throughout the entire duration of SDSS-V, with medium-to-low cadence (months to years).
    \item Reverberation-mapping targets (BHM-RM). About 1400 $i\lesssim20$ broad-line AGNs that will be observed over 100 times throughout the entire duration of SDSS-V with high cadence (down to $\sim1-2$ days in the observer frame).
\end{enumerate}
Additional sets of BHM sources that may also provide repeated spectroscopy include targets that were observed as part of the SPectroscopic IDentfication of eRosita Sources \cite[SPIDERS;][]{Clerc16} and the Chandra Source Catalog \cite[CSC;][]{Evans20} programs, which target X-ray point-like sources identified with the eROSITA and Chandra telescopes, respectively.

During its first year of operations, the SDSS-V/BHM program obtained at least one new spectrum for 44,761 unique AGNs. For 29,631 of these sources, there was at least one archival SDSS spectrum available in SDSS DR16 \citep{Ahumada20_DR16}. 
We note that 28,804 of these (97.2\%) are part of the DR16 quasar catalog compiled by \citet[][hereafter DR16Q]{Lyke20_DR16Q}.
We further restrict this sample of 29,631 sources by excluding 758 BHM-RM targets, for which the large number of (high-cadence) spectra was found to challenge our automatic CL-AGN search methodology (discussed further in Section \ref{subsubsec:RM_CLAGN}). 
Excluding the BHM-RM targets, we are left with 28,873 unique AGNs.
We use this latter sample of non-BHM-RM AGNs with repeated spectroscopy, covering $0<z<5.1$ (median: $z=1.49$; see Figure~\ref{fig:z_L_dist}) as the parent sample for our search for CL-AGNs and other extremely variable AGNs.

The archival DR16 spectra we use were obtained following a variety of selection criteria (targeting various types of galaxies) and with two different instrumental setups \citep{Smee13}. Specifically, the spectra collected up to 2008 July (and published in Data Release 7 of SDSS-I--II; \citealt{Abazajian09_DR7}) were obtained with the SDSS spectrographs and 3\arcsec diameter fibers, and focused on a large, flux-limited sample of quasars, selected through a rather homogeneous color-based criterion \citep{Richards02,Schneider10}.
All subsequent spectra were obtained with the BOSS spectrographs and 2\arcsec diameter fibers, and added a large number of quasars selected through a wider variety of criteria, including probabilistic methods, variability, and other sets (see DR16Q).
We discuss the implications of the aperture differences for our work in Section \ref{subsec:decomposition}. 
The various generations of SDSS surveys also slightly differ in their wavelength coverage, with the red end of the spectra changing from $\approx$9200\,\AA\ for the SDSS spectrograph to $\approx$10300\,\AA\ in the BOSS configuration used for SDSS-IV and the first year of SDSS-V.
For completeness, we note that our work used products of version v6.0.9 of the BOSS pipeline.

\subsection{Searching for Changing-look Active Galactic Nuclei candidates}
\label{subsec:spec_search}

Given the large size of the parent SDSS-V sample, the first step in our search employed automated and relatively simple methods. Our approach avoids detailed spectral decomposition, which relies on a large set of assumptions (see Section \ref{subsec:decomposition} below) that may not be appropriate for all observed, multi-epoch spectra of all sources. 

Specifically, we searched for extreme variations of broad-line emission. The broad emission lines considered in our search are (depending on the target's redshift): \ha, \hb, \MgII, \CIII, and \CIV. The search was conducted in the following way. We first shifted each spectrum to its rest frame, and cleaned it by removing bad pixels (marked with $\texttt{ivar} = 0$) and pixels with particularly low signal-to-noise ratios (S/N$<1$). Each spectrum was smoothed by applying a rolling median with a window size of 9 pixels.\footnote{Each spectral pixel covers 69\,\kms.}
For each spectrum, the continuum level was measured near each of the aforementioned broad emission lines based on adjacent relatively line-free spectral bands. Specifically, we calculated the median flux density in two such bands, located blueward and redward of the line in question \cite[see Table 1 in][]{MejiaRestrepo16}, and linearly interpolated between the two median flux densities. This simple linear continuum model was subtracted from the total spectrum to obtain the emission-line spectrum. To quantify the emission-line flux, we integrated the continuum-subtracted flux density over a wavelength band corresponding to $\pm4000\,\kms$ around the expected central wavelength of the emission line of interest.
The rest-frame wavelength ranges used for continuum placement and line flux integration are listed in Table~\ref{tab:waves}.
This process allowed a measurement of the broad emission-line fluxes without the need to decompose the continuum emission (into host and AGN) or to model the line profiles.

\begin{deluxetable}{lccc}
\tablecaption{Wavelength ranges used in the automated CL-AGN candidate search}
\label{tab:waves}
\tablewidth{\columnwidth}
\tabletypesize{\scriptsize}
\tablehead{
\colhead{Emission Line} & \multicolumn{2}{c}{Continuum\tablenotemark{\footnotesize a}} & \colhead{Line Flux\tablenotemark{\footnotesize b}}\\
\colhead{} & \colhead{Blue} & \colhead{Red}  & \colhead{}
}
\startdata
    \ha\   & (6150--6250) & (6950--7150) & (6477.1--6652.1) \\ 
    \hb\   & (4670--4730) & (5080--5120) & (4797.8--4927.5) \\ 
    \mgii\ & (2650--2670) & (3030--3070) & (2761.4--2836.1) \\ 
    \ciii\ & (1680--1720) & (1960--2020) & (1883.3--1934.2)\\ 
    \civ\  & (1420--1470) & (1680--1720) & (1528.4--1569.7)\\ 
\enddata
\tablenotetext{a}{Wavelength bands used for placing the simple, linear continuum model.}
\tablenotetext{b}{Wavelength range used for the simple line flux integration.}
\tablecomments{All wavelengths are given in vacuum, rest frame, and in units of \AA.}
\end{deluxetable}

The line flux variability for each source was estimated by comparing all the pairs of line measurements available from our multi-epoch spectroscopy. 
To evaluate the significance of the line variability, we used the following quantity:
\begin{equation}
\label{eq:criterion_final}
    \rm C \left(\rm line \right) \equiv F_2/F_1 - \Delta\left(F_2/F_1\right),
\end{equation}
where $F_2/F_1$ is the ratio between the (continuum-subtracted) line fluxes for the two specific epochs, with $F_2$ defined to be the higher flux of the two, and $\Delta\left(F_2/F_1\right)$ is the 1 $\sigma$ equivalent uncertainty on the line flux ratio, propagated from the two error spectra. 
\footnote{We note that the spectrophotometric calibration of the data is designed to be accurate at the $\sim5\%$ level \cite[e.g.,][]{Shen15_SDSS_RM,Almeida23}.}

The quantity defined in Equation~\ref{eq:criterion_final} is designed to estimate the fractional flux variability of each emission line while taking into account the uncertainties in the flux measurements. We note that this quantity, which measures changes in emission-line fluxes based on simple integration of observed flux densities, may often underestimate the real change in the line flux. This may happen because the integrated flux encompasses not only the variable broad emission component but also some essentially constant narrow emission line flux. It is thus possible that a number of objects that did experience a significant change in their broad emission-line flux would be missed by our search. One way to address this would involve applying our search criteria to narrow-line-subtracted spectra. However, this approach would necessitate a full and reliable spectral decomposition for all epochs of the 28,873 AGNs in our parent sample, a task that requires a nontrivial set of assumptions to be used in the spectral decomposition procedure. 
This is beyond the scope of the present work. We will return to this point in Section \ref{subsec:decomp_for_search}.

After calculating $\rm C \left(\rm line \right)$ for each emission line, for all pairs of spectra for a given source we define our initial CL-AGN candidates to be any source for which the broad emission-line flux changed by at least a factor of two. That is, a CL-AGN candidate is a source for which the maximal fractional emission-line flux variability among all available pairs is greater than two, that is, 
\begin{equation}
    \max\left[ \rm C\left(\rm line\right)\right]>2  \, .
    \label{eq:C_cut}
\end{equation}
This universal cut was designed to focus only on those systems that show (line) flux variations far above the typical rest-frame UV-optical variability seen in AGNs.
For example, \cite{MacLeod12} find that only $\sim1\%$ of quasars vary in flux by more than a factor of two over periods of 8--10 years (in the observed frame). 
For comparison, the recent study by \cite{Guo24_DESI} adopts a criterion that corresponds to line variation by a factor of $2.5$, however with a different treatment of uncertainties. 
We note that some previous CL-AGN samples were selected based on the significance, but not necessarily the amplitude, of their emission-line variations (see discussions in \citealt{MacLeod19} and \citealt{Guo24_DESI}).
In addition, we visually inspected all the spectra obtained using a  randomly selected plate, including well over 100 AGNs, and verified that our chosen automatic search criterion did not overlook obvious CL-AGN candidates using our search criterion.

In total, our selection criteria flagged 3338 candidates (among the 28,873 unique AGNs in our parent sample), all of which were then visually inspected to identify CL-AGNs or other extreme-variability events. 
Performing such a visual inspection on a dataset as large as our parent sample was not practical, which again demonstrates the utility of our automatic search.
This visual inspection focused on identifying calibration issues mistakenly flagged by the automatic search procedure. We mainly inspected narrow emission lines, whenever they were present, as these are not expected to vary significantly on the relevant timescales (see, e.g., \citealt{Peterson82} but also \citealt{Peterson13}). Additionally, we examined the consistency between multiple SDSS-V spectra of the same source, whenever more than one SDSS-V spectrum was available (obtained within months). Furthermore, many sources were incorrectly flagged as showing significant spectral variations due to incorrect estimation of the continuum and/or broad-line emission due to noise in one of the epochs---a situation that is relatively easy to identify visually. 
Two examples of CL-AGN candidates, flagged by our automated search but not passing visual inspection, are shown in Figure~\ref{fig:selection_examples} (in Appendix~\ref{app:selection}).

Many of the flagged objects that were visually confirmed as showing genuine, extreme line changes were found to display prominent emission in all their (accessible) broad lines, even in their dimmest states. 
We chose not to designate these objects as CL-AGN candidates, as  we instead prefer to focus on those  objects which show the (dis-)appearance of at least one broad emission line. We further discuss the objects removed in this step in Section~\ref{subsubsec:EVQs_sample} below.

In total, 130 AGNs passed visual inspection as viable CL-AGN candidates. The number of flagged candidates, the number of candidates passing the visual inspection, and the true-positive rate (TPR) are listed in Table~\ref{tab:criterion}, which further breaks down these numbers based on the emission line(s) used for selection.
Notably, the drastic reduction from the $>$3300 initially flagged sources to the 130 viable candidates is driven by high-redshift sources, which are unfortunately prone to several selection effects, which we further discuss in Section~\ref{subsec:sample_props} (including our inability to rule out flagging due to flux-calibration issues). Indeed, our automatic search is rather efficient (TPR roughly between $10\%$ and $20\%$) in the lower-redshift regime, where \ha\ and/or \hb\ are accessible.

We note that, as the quantity we use as a selection criterion (Equations~\ref{eq:criterion_final} and \ref{eq:C_cut}) employs flux ratios, it is meaningless if either one of the fluxes is negative. In such cases, we decided to ignore the specific measurement of the particular line in the corresponding particular epoch, and did not consider it when looking for (drastic) line variability. We note that considering objects which otherwise did not pass our criterion, but where one of the emission-line flux measurements yielded a negative value as CL-AGN candidates, greatly enlarged the number of flagged candidates (from 3338 to 5329). However, a visual inspection of a subset of these newly flagged sources revealed that a vast majority of them were flagged due to an incorrect estimation of the emission line and/or continuum flux levels, typically driven by low-$S/N$ data in the relevant spectral regions. Therefore, the exclusion of such objects from our search may lead us to miss only a few genuine CL-AGNs, at a cost of much higher search efficiency.

We acknowledge that our search for extreme spectral AGN variability may be incomplete, as most SDSS-V spectra were not inspected visually, and some variable AGNs might have been missed by our search. However, we stress that our primary focus was to ensure the purity of our large sample, rather than its completeness. This is further detailed in Section \ref{subsec:ancillary_data} below, where we present our efforts to corroborate our candidates through follow-up spectroscopy and photometric light curves.

Another, broader issue that affects our sample completeness is inherent to the SDSS-V survey design. Since the core repeated spectroscopy programs within SDSS-V focus on previously known, luminous broad-line AGNs (i.e., quasars) identified in archival SDSS data, any search for dramatic variability events in (early) SDSS-V data is expected to favor the identification of dimming events of these previously known AGNs, rather than of their further brightening. This bias is discussed in detail, and quantified, by \cite{ShenBurke21}. We will return to this point in Sections \ref{subsec:final_sample} and \ref{subsec:line_comparison}.

We have also experimented with alternative search procedures and criteria, specifically ones based on changes in the flux of the lines relative to those in the (adjacent) continuum.
Several studies have examined CL-AGNs samples selected through this kind of approach (e.g., the criterion defined in \citealt{MacLeod19}, which is also employed in \citealt{Green22} and \citealt{Guo24_DESI}).
Specifically, we looked into using the equivalent widths (EWs) of the broad emission lines, rather than their integrated absolute fluxes, as an alternative selection path. 
In principle, one would expect that searching for large EW variations would be robust to calibration issues. However, we found such methods to be less suitable for our needs, for several reasons. First, they would flag cases where the continuum changed while the broad line roughly retained its flux. Although such events could be considered as showing extreme spectral variability, these variations are different from the CL-AGNs we chose to focus on in the present work (i.e. systems with dramatic broad emission-line variations). 
Second, given the possible driving mechanisms of CL-AGNs, one may expect the broad emission lines and the AGN continuum to vary (roughly) in unison, either due to the emission lines being reprocessed seed ionizing continuum radiation\footnote{Note that the ionizing continuum itself is far beyond the observed spectral range.}, or due to a variable obscurer attenuating both emission components. In such situations, the EW can remain roughly constant even in genuine CL-AGNs.

\begin{deluxetable*}{lc|cc|cc|c}
\tablecaption{Automated CL-AGN candidate search}
\label{tab:criterion}
\tablewidth{\columnwidth}
\tabletypesize{\scriptsize}
\tablehead{
\colhead{} & \colhead{} & \multicolumn{2}{c}{Before Refinement} & \multicolumn{2}{c}{After Refinement} & \colhead{Core Sample}\\[-1ex]
\colhead{Emission Line} & \colhead{\# Flagged} & \colhead{\# Candidates} & \colhead{TPR\tablenotemark{\footnotesize a}} & \colhead{\# Candidates} & \colhead{TPR\tablenotemark{\footnotesize b}} & \colhead{\# CL-AGNs}
}
\startdata
    \ha\   &  173 & 38 & 22\% & 37 & 21\% & 38 \\ 
    \hb\   &  637 & 81 & 13\% & 74 & 12\% & 79 \\ 
    \mgii\ & 1228 & 44 & 3.6\% & 43 & 3.5\% & 43 \\ 
    \ciii\ & 1065 &  2 & 0.19\% & 2 & 0.19\% & 1\\ 
    \civ\  & 1086 &  3 & 0.28\% & 3 & 0.28\% & 2\\ 
    \hline
    All lines & 3338 & 130 & 3.9\% & 123 & 3.7\% & 113 \\
\enddata
\tablenotetext{a}{True-Positive Rate, i.e., the percentage of candidates confirmed through visual inspection, out of all the sources flagged based on the corresponding line.}
\tablenotetext{b}{True-Positive Rate, i.e., the percentage of candidates confirmed through visual inspection and spectral refinement, out of all the sources flagged based on the corresponding line.}
\tablecomments{Some sources were flagged by multiple lines and thus result in a total count of flagged sources and candidates per emission line that exceeds the total counts.}
\end{deluxetable*}

\subsection{Sample Refinement Through Spectral Decomposition}
\label{subsec:decomp_for_search}

To ensure the robustness of the broad emission-line measurements for our CL-AGN sample, we performed a full spectral decomposition for the 130 CL-AGN candidates, using the dedicated and widely used \texttt{PyQSOFit} package \citep{QSOFit}. We discuss this procedure in detail in Section \ref{subsec:decomposition}, and here only briefly mention the key steps related to the sample refinement. 
Specifically, for each CL-AGN candidate where this was applicable, we rescaled the brighter spectra to yield the \OIII\ narrow-line flux observed in the dimmer relevant spectrum. This is motivated by the expectation that narrow emission lines do not vary significantly over the timescales probed by our data. 
We then subtracted the best-fitting (rescaled) narrow-line profiles from all epochs of the CL-AGN candidate. 
Subsequently, we reevaluated each CL-AGN candidate using the rescaled, narrow-line-subtracted multi-epoch spectra. This step led to the removal of seven candidates, for which these refined spectra no longer satisfied our CL selection criterion\footnote{That is, the re-scaled spectra resulted in $\rm C({\rm line})\leq2$.}, leaving a total of 123 candidates.

As previously discussed in Section \ref{subsec:spec_search}, our automated search, which was applied to the observed SDSS spectra, might have missed some CL-AGNs due the inclusion of nonvariable narrow-line emission in our line flux estimators. 
We thus used the detailed spectral decomposition procedure mentioned above (and discussed in Section~\ref{subsec:decomposition}) in an attempt to identify additional, robust CL-AGNs. These objects, which initially did not meet our criterion, were (sporadically) identified during our experimentation with various search methods and were confirmed as viable CL-AGN candidates through visual inspection, and---importantly---met the criterion only when applied to the narrow-line-subtracted and renormalized spectra. 
This process identified 14 additional robust CL-AGN candidates, increasing our sample to 137 candidates. 

\subsection{Ancillary Observations and Data}
\label{subsec:ancillary_data}

With an initial sample of CL-AGN candidates in hand, we acquired several sets of additional, ancillary observations and publicly available data to further refine the sample, and to gain further insights regarding the nature of the variable emission from our candidates. This included additional optical spectroscopy, optical light curves, and mid-IR light curves, as detailed below.

\subsubsection{Follow-up Spectroscopy}
\label{subsubsec:followup}

In addition to the SDSS-V spectroscopy, we performed an extensive spectroscopic effort to corroborate the nature of the identified CL-AGN candidates, using various facilities. This follow-up effort is important for two reasons: (i) to rule out false-positive candidates caused by calibration issues in the new SDSS-V spectra, which may be mistaken for real flux changes in our automated search and visual inspection (particularly for spectra lacking narrow emission lines); and (ii) to follow up on variable sources to see whether they exhibit further dramatic spectral variability \cite[see][for an example of such a case]{Zeltyn22}. 

Additional optical spectroscopy was obtained with the 2 m Faulkes Telescope North (FTN) and South (FTS) facilities, which are part of the Las Cumbres Observatory network \cite[LCOGT;][]{Brown13}; the Low-Resolution Spectrograph 2 \cite[LRS2;][]{Chonis16} on the 10 m Hobby--Eberly Telescope (HET; \citealt{Hill21}) at McDonald Observatory; and the Double Spectrograph \cite[DBSP;][]{OG82} on the 5.1 m Hale Telescope at the Palomar Observatory. 
Spectra were obtained through either long slits with widths of either 2\arcsec\ (FTN and FTS) or 1.5\arcsec\ (Hale), or with dynamic, seeing-matched apertures of $1.7-2.5$
\arcsec (HET), and calibrated using standard stars observed during the corresponding nights. 
Table~\ref{tab:all_followups} in Appendix \ref{app:follow-up} lists the targets observed through this effort, as well as the spectral setups used in these observations.
The spectra were reduced using standard procedures and the well-tested pipelines of the corresponding instruments (see Appendix~\ref{app:follow-up}).

In total, we obtained 109 spectra for 74 unique sources of interest, including 52 spectra for 33 unique sources from our sample of 137 CL-AGN candidates, with no particular prioritization.  
Among the latter set of 33 CL-AGN candidates, in 22 cases the additional spectra confirmed the spectral changes seen in the first-year SDSS-V spectra, while three sources were identified as false candidates. For the remaining eight CL-AGN candidates, the follow-up spectroscopy could neither confirm nor refute their CL nature. 
The three refuted candidates are all $z>1.5$ sources, where the SDSS-V spectroscopy indicated a significant dimming of both continuum and broad emission lines, but the follow-up spectroscopy is highly consistent with the archival SDSS spectra. Given the high redshifts of the source, we could not use the \oiii\ line to account for such effects, which are most likely driven by flux-calibration issues.
An example of how follow-up spectroscopy helped to identify one of these three false CL-AGN candidates is presented in Figure~\ref{fig:selection_examples} (in Appendix~\ref{app:selection}).

Consequently, after these spectroscopic follow-up efforts, we retained 134 of our CL-AGN candidates for further analysis.
Interestingly, the follow-up spectra of six targets displayed various degrees of variability compared with the (rather recent) SDSS-V spectra, with some targets showing a continuation of the trends discovered in SDSS-V (i.e., the AGN continued to dim or brighten), while others showed a reversal of the SDSS-V trend (e.g., a brightening AGN began to fade). The number of candidates observed, confirmed, and refuted with each observational facility are listed in Table~\ref{tab:follow-up}.

\begin{deluxetable*}{lc|ccccc}
\tablecaption{Follow-up observations and photometric cross-match of CL-AGN candidates}
\label{tab:follow-up}
\tablewidth{\columnwidth}
\tablehead{
\colhead{Facility} & 
\multicolumn{1}{c}{All\tablenotemark{\footnotesize a}} & \multicolumn{4}{c}{CL-AGN Candidates\tablenotemark{\footnotesize b}} \\[-1ex]
\colhead{} & \colhead{Observed} & \colhead{Observed} & \colhead{Confirmed\tablenotemark{\footnotesize c}}
& \colhead{Refuted\tablenotemark{\footnotesize c}} & \colhead{Inconclusive\tablenotemark{\footnotesize c}} & \colhead{Further} 
\\[-3ex]
\colhead{} & \colhead{} & \colhead{} & \colhead{}  & \colhead{} & \colhead{} & \colhead{Variability} 
\\[-5ex]
}
\startdata
    \hline
    LCO     &  54 & 24 & 14 & 3 & 7 & 5 \\ 
    HET     &  25 & 15 & 10 & 0 & 5 & 3 \\ 
    Palomar &  11 &  2 &  2 & 0 & 0 & 0  \\ 
    \hline
    Total spec.\tablenotemark{\footnotesize d} & 74 & 33 & 22 & 3 & 8 & 6 \\
    \hline
    ZTF   & - & 116 &  7 & 10 &  99 & 4 \\ 
    ATLAS & - & 128 & 13 & 11 &  104 & 3 \\
    CRTS  & - &  98 &  4 &  6 &  88 & - \\
    PS1   & - &  115 &  5 &  6 &  104 & - \\
    \hline
    Total phot.\tablenotemark{\footnotesize d} & - & 133 & 19 & 21 & 93 & 4 \\
    \hline
\enddata
\tablenotetext{a}{Spectroscopic follow-up observations conducted for all types of extreme-variability candidates flagged within the scope of our work.}
\tablenotetext{b}{Spectroscopic follow-up observations and photometric light-curve data for CL-AGN candidates that satisfied our automatic CL selection criterion and passed visual inspection.}
\tablenotetext{c}{The unique number of CL-AGN candidates that the ancillary data (spectroscopy or light curves) confirmed their CL nature, refuted it, or did not provide conclusive evidence.}
\tablenotetext{d}{The total number of unique sources with spectroscopic follow-up observations and/or with photometric cross-matches.}
\end{deluxetable*}

\subsubsection{Optical photometric light curves}
\label{subsubsec:crossmatch}

To further corroborate and examine our candidates, we used publicly available optical photometric light curves obtained through the Zwicky Transient Facility \cite[ZTF;][]{Masci19,ZTF}, Asteroid Terrestrial-impact Last Alert System \cite[ATLAS;][]{Tonry18}, Catalina Real-time Transient Survey \cite[CRTS;][]{Drake09}, and Pan-STARRS1 \cite[PS1;][]{Kaiser10} datasets and compared these with synthetic photometry derived from the candidates' (SDSS) spectra.
Specifically, we generated ZTF and ATLAS light curves using the corresponding forced-photometry services and the DR16 coordinates of our AGNs,\footnote{Instructions for producing ZTF and ATLAS light curves can be found at \href{https://irsa.ipac.caltech.edu/data/ZTF/docs/ztf_forced_photometry.pdf}{https://irsa.ipac.caltech.edu/data/ZTF/docs/} and \href{https://fallingstar-data.com/forcedphot/}{https://fallingstar-data.com/forcedphot/}, respectively.} which employ forced point-spread function fit photometry on images to create the light curves \cite[see, e.g.,][for more details on the forced-photometry procedures]{Tonry18, Smith20, Masci23}. 
The CRTS and PS1 light curves were obtained by cross-matching our sources with the publicly available catalogs, allowing for a search radius of 1\arcsec and 2\arcsec, respectively.

Performing such photometric cross-matches was particularly useful for identifying flux-calibration issues in SDSS-V spectra, i.e., cases where the synthetic photometry derived from these spectra showed large discrepancies compared with (pseudo-)concurrent photometry (i.e., $\Delta m \gtrsim 1$ mag). 

Inspecting the photometric data led to a removal of 21 candidates from our sample, retaining 113. 
Indeed, all the refuted sources are at relatively high redshift ($z>1.2$), where we could not use \oiii\ to assess the robustness of their selection as CL-AGN candidates.
For 19 of the remaining candidates, the photometry was consistent with the spectral variations seen in the SDSS-V data, which could allow to further tighten the constraints on the timescales of extreme spectral variability. 
Moreover, the light curves of four targets showed prolonged variability after the SDSS-V spectra were taken, with some targets continuing the trends identified between the archival and the new SDSS-V spectra (i.e., the target keeps brightening or dimming), while others showed a reversal of the SDSS-V trend (i.e., the spectra suggested dimming, but the light curve showed rebrightening). 

Table ~\ref{tab:follow-up} summarizes the number of candidates observed, confirmed, and refuted by each facility.
Figure~\ref{fig:selection_examples_phot} (in Appendix~\ref{app:selection}) shows two examples of how optical light curves were used to refute or retain CL-AGN candidates.

\subsubsection{WISE infrared light curves}

We used publicly available data from the Wide-field Infrared Survey Explorer \cite[WISE;][]{Wright10} to create IR light curves for our candidates in the W1 ($\sim$3.4\,\mic) and W2 ($\sim$4.6\,\mic) bands, with a cadence of about 6 months.\footnote{WISE light curves were extracted using the Gator catalog list at \href{https://irsa.ipac.caltech.edu/applications/Gator/}{https://irsa.ipac.caltech.edu/applications/Gator/}} Specifically, we searched for counterparts within radii of  2\arcsec\ and 3\arcsec, for the AllWISE (covering 2010--2011; \citealt{ALLWISE}) and the NEOWISE
reactivation mission (NEOWISE-R, covering 2013 to the present; \citealt{NEOWISE}) datasets, respectively \citep{Mainzer11_NEOWISE,Mainzer14_NEOWISER}.\footnote{We have decided to be more permissive with NEOWISE-R matches, as NEOWISE-R sources are not prematched between frames.}
We obtained matches for all 113 of our candidates in the NEOWISE-R dataset. In the AllWISE dataset, we found matches for 109 out of 113 candidates.

These data were collected in order to examine the mechanism behind the changing-look transitions in our sample, as the IR emission covered by WISE is not expected to be significantly affected by variable obscuration. In contrast, dramatic variations to the accretion flow are expected to be echoed by the IR emission, as the latter is thought to be driven by reprocessing the former on timescales of weeks to months \cite[e.g.,][and references therein]{Minezaki19,RT23}.

We show two examples of WISE light curves in Figure~\ref{fig:wise_examples} (in Appendix~\ref{app:wise}).

\subsection{Final Core Sample}
\label{subsec:final_sample}

Our final core sample consists of 113 robust CL-AGN candidates, listed in Table~\ref{tab:cands}. 
Spectra of several examples of CL-AGN from our final sample are presented in Figure \ref{fig:example}. 
We discuss some of these and other noteworthy cases in Appendix~\ref{app:individual}, and show the spectra of \emph{all} the systems in our final sample in Figure~\ref{fig:all_spec} there (available online).

Of the 113 CL-AGNs in our final core sample, 38 were selected based on drastic changes in their broad \ha\ line emission, 79 based on \hb, 43 based on \mgii, two based on \civ, and one based on \ciii\ (see Table ~\ref{tab:criterion}). 
There is significant and complex overlap between these subsets, as demonstrated in the line-pairwise Venn diagrams shown in Figure~\ref{fig:venn_lines}.
Of the various overlapping subsets, we particularly note that
(i) of the two \civ-selected CL-AGNs, one is also  \ciii-selected; 
(ii) 30 of the 38 (79\%) \hb-selected CL-AGNs for which the \ha\ line is accessible are also \ha-selected CL-AGNs;
and 
(iii) 22 of the 37 (59\%) \hb-selected CL-AGNs for which the \mgii\ line is accessible are also \mgii-selected CL-AGNs.

Our CL-AGNs span a redshift range of $0.06<z<2.4$, and the SDSS-V spectroscopy was capable of probing extreme variability on rest-frame timescales in the range of 1 year $\lesssim \Delta t_{\rm rest} \lesssim$ 19 yr. 
In some cases, our subsequent follow-up spectroscopy revealed dramatic variability on yet shorter timescales (as short as $\Delta t_{\rm rest} \approx 2$ months; see \citealt{Zeltyn22}).
Of these 113 systems, 67\% (76 systems) show dimming in their recent SDSS-V spectroscopy compared to previous data, while 33\% (37 systems) show brightening.  
\begin{figure*}
    \centering
    \includegraphics[width=1\textwidth,clip,trim={2cm 0 1cm 1cm}]{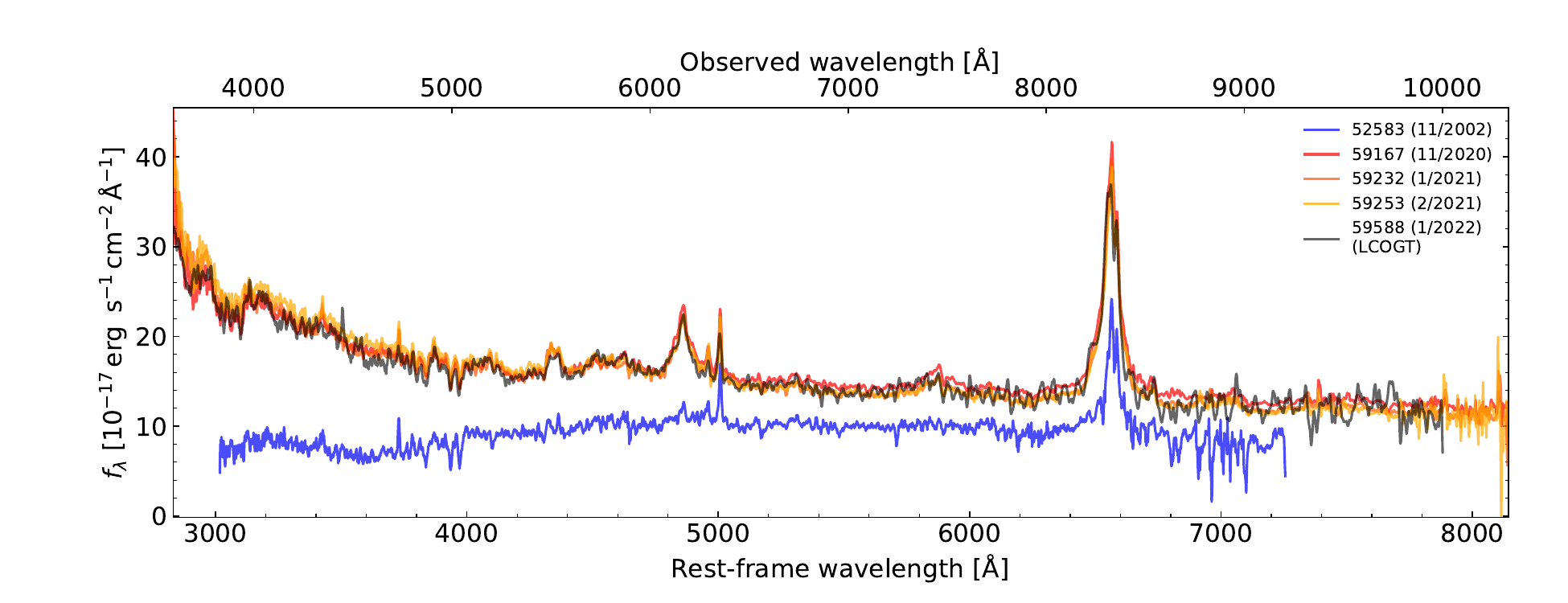}
    \includegraphics[width=1\textwidth,clip,trim={2cm 0 0.7cm 1cm}]{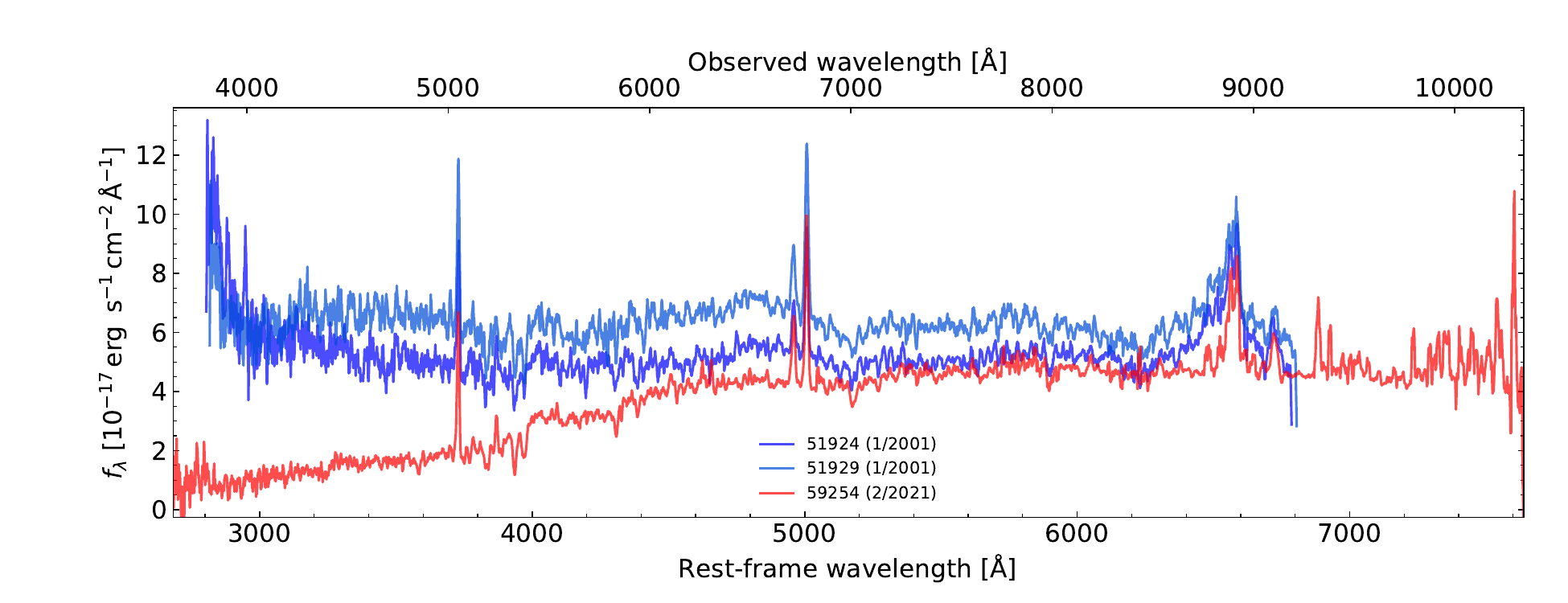}
    \includegraphics[width=1\textwidth,clip,trim={2cm 0 0.8cm 1cm}]{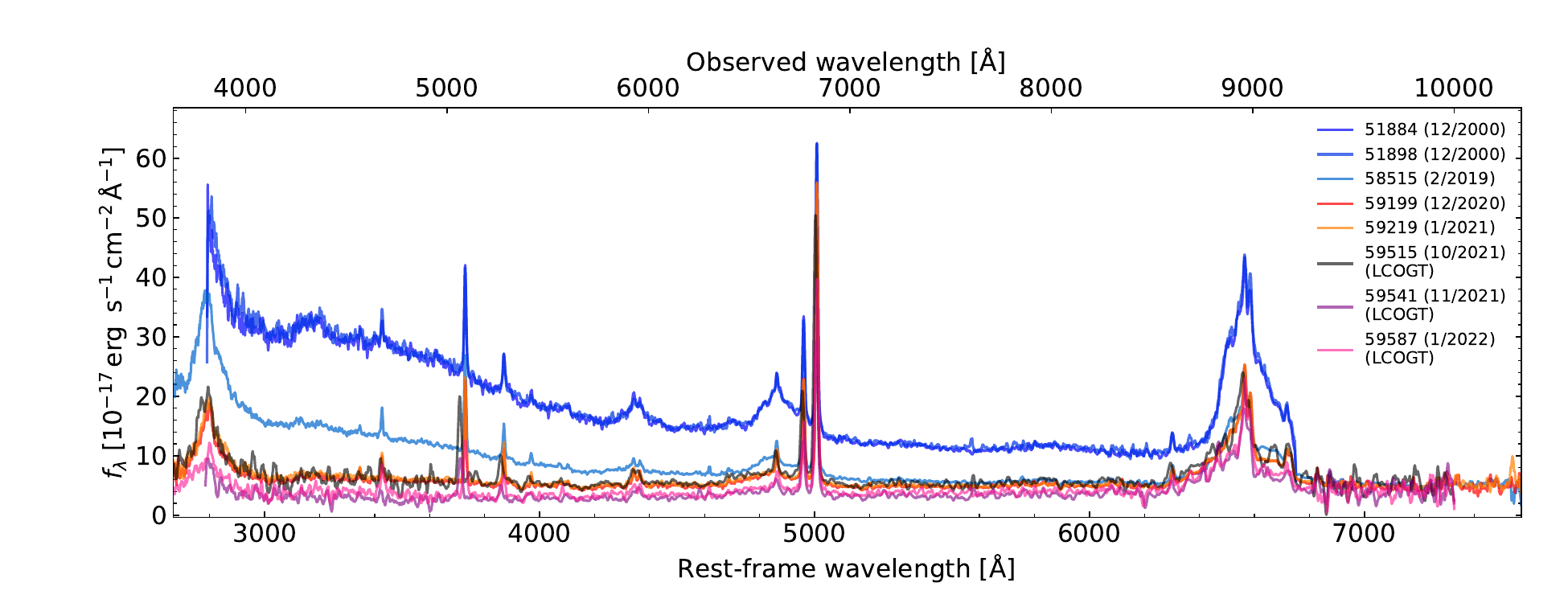}
    \caption{Three CL-AGN candidates observed in SDSS-V. 
    In all panels, the colors of the spectra mark the observing epoch, moving from early, SDSS-I--IV spectroscopy (blue-cyan) to the recent SDSS-V spectroscopy (red-orange-yellow), and our follow-up observations (with LCOGT; black, purple, magenta).
    All the spectra were boxcar-smoothed over 7 pixels.
    \textit{Top -- J075934.95+322143.3 ($z=0.26888$)}: The archival 2002 spectrum classifies this source as a Type 1.9 quasar, while the more recent 2021 SDSS-V and 2022 follow-up LCOGT spectra show the appearance of the broad \hb\ and a quasar-like continuum. 
    \textit{Middle -- J090550.30+003948.1 ($z=0.354041$)}: The archival SDSS spectra show a typical quasar, while the more recent 2021 SDSS-V spectrum shows the disappearance of the broad \hb\ and the quasar-like continuum, and a significant weakening of the broad \ha.
    \textit{Bottom -- J012946.72+150457.3 ($z=0.364851$)}: The archival SDSS spectra reveal a dimming between 2000 and 2019. Recent (2020--2021) SDSS-V spectra show the continuation of the dimming, with the broad \hb\ almost completely disappearing. Additional LCO spectra taken in late 2021 and early 2022 (black, purple, magenta) confirm the SDSS-V dimming.
    }
    \label{fig:example}
\end{figure*}

The nonunity ratio of brightening to dimming events we find (i.e., 1:2), is expected given the nature of our sample, which focuses on revisiting previously known broad-line AGNs. 
Specifically, \cite{ShenBurke21} quantified the expected bias in flux-limited quasar variability studies. They find that if all AGNs intrinsically vary in a symmetrical manner, the ratio of significant brightening to dimming events is indeed expected to be \~1:2 for an AQMES-like survey (i.e., $i\lesssim19$ sources revisited after $\sim$10--20 yr).

Of the 113 CL-AGNs in our final core sample, 87 appear in the DR16Q. We stress that the remaining 26 CL-AGNs do have optical spectra in the SDSS DR16 dataset, but were not included in DR16Q.
These 26 sources were reobserved in SDSS-V as part of the SPIDERS and/or CSC targets.

Our final sample of 113 CL-AGNs includes 104 newly discovered CL-AGNs. The remaining nine objects were already noted in prior studies, including 
five CL-AGNs that are part of the sample presented in \citet{MacLeod19}, 
three in \citet{Green22},
two in \citet{MacLeod16}, 
two in \citet{Guo24_DESI},
and one in each of \citet{Ruan16} and \citet{Yang18}.\footnote{There is some overlap between the samples reported in all these studies.}
In addition, our sample includes the archetypal CL quasar studied in detail by \citet[][J0159+0033]{LaMassa15}, and another intriguing CL-AGN which was published earlier by the SDSS-V collaboration \cite[J1628+4329;][]{Zeltyn22}.
The relevant references for these nine previously known CL-AGNs are also included in Table \ref{tab:cands}.

Our first-year SDSS-V-based sample of 113 spectroscopically identified CL-AGNs greatly increases the number of such sources, and is the largest spectroscopically selected CL-AGN sample reported to date (particularly when considering the 88 additional extreme spectral variability events we discuss in Section~\ref{subsubsec:EVQs_sample}). 
For comparison, a study looking for CL-AGNs in SDSS-IV found 19 CL-AGNs, of which 15 were newly discovered \cite[][]{Green22}. Another SDSS-IV-based study, focusing on high-redshift CL-AGNs, identified 23 systems at $z>1.5$ \citep{Guo20_hiz}.
Other large-sample studies carried out by \citet{MacLeod19} and \citet{Yang18} discovered 17 and 21 new CL-AGNs, respectively. 
Most recently, \cite{Guo24_DESI} published a sample of 56 CL-AGNs, along with 44 additional objects that exhibit significant broad-line variability, identified in the early data collected with The Dark Energy Spectroscopic Instrument (DESI) spectroscopic survey \citep{DESI23_EDR}, which reobserved over 82,000 sources that already had spectra in SDSS DR16. 
While DESI has clear advantages over SDSS-V in terms of the total number of AGNs with repeated spectroscopy, of sensitivity and spectral resolution, it does not (currently) plan to further revisit these sources, for example to study recurring and/or short-timescale spectral transitions---which are part of the design drivers of SDSS-V (see Section~\ref{subsec:sdssv} above). Specifically, the AQMES-Medium component of SDSS-V is designed to obtain up to 10 new spectra of $\sim2000$ quasars over the entire duration of SDSS-V \cite[see][]{Almeida23}, which will allow to study in detail, and indeed temporarily resolve, changing-look transitions and other extreme-variability events.

\begin{figure*}
    \centering
    \includegraphics[width=0.32\textwidth,clip, trim={0.5cm 0 0.5cm 0}]{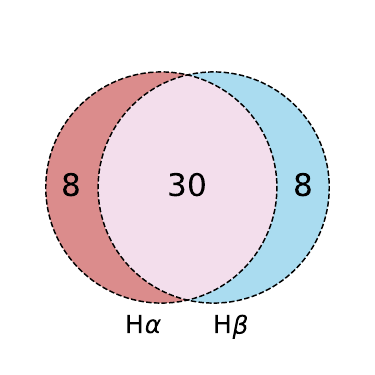}   
    \includegraphics[width=0.32\textwidth,clip, trim={0.5cm 0 0.5cm 0}]{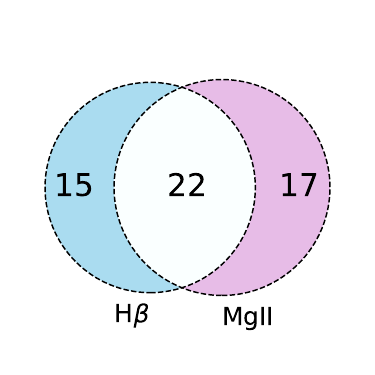}
    \includegraphics[width=0.32\textwidth,clip, trim={0.5cm 0 0.5cm 0}]{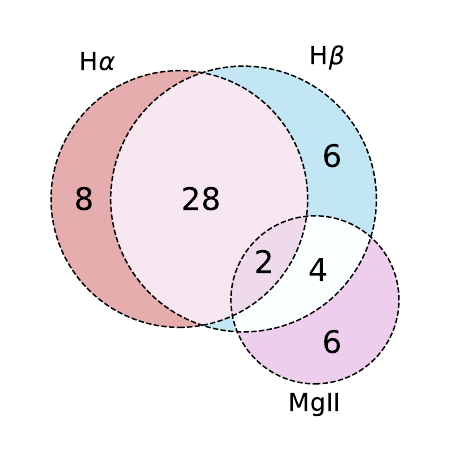}
    \caption{Venn diagrams showing the overlap between CL-AGNs in our final core sample that were selected based on the extreme variability of different emission lines. The panels show the subsets of objects that have spectral coverage of both \ha\ and \hb\ (left); both \hb\ and \mgii\ (center); or \ha, \hb, and \mgii\ (right).}
    \label{fig:venn_lines}
\end{figure*}

\subsection{Additional Extremely Variable Sources}
\label{subsec:additional_sources}


In addition to our core sample of 113 CL-AGNs, we have identified two subsets of extremely variable sources during our work: three CL-AGN candidates drawn from BHM-RM, and 88 sources showing extremely variable but persistent broad emission lines; 
these subsets are described below.
We stress that none of these sources is included in our detailed analyses (described in Sections \ref{sec:control_and_qunatities} and \ref{sec:analysis_results}), which focus solely on the 113 candidates in our core CL-AGN sample.

\subsubsection{Additional Changing-look Active Galactic Nuclei from BHM-RM}
\label{subsubsec:RM_CLAGN}

Our CL-AGN candidate selection criterion (Section \ref{subsec:spec_search}) initially yielded an exceptionally large number of candidates among BHM-RM targets, due to the pairwise nature of our search and the large number of (high-cadence) pairs of spectra for BHM-RM targets. To make our automatic search efficient, we decided to exclude BHM-RM targets from our primary, complete sample. 

However, during our initial experimentation with various search methods, we identified three additional, robust CL-AGN candidates based on the spectra collected within the BHM-RM effort.
The multi-epoch spectra and light curves of these three sources make them irrefutable CL-AGNs (also by the criterion in Equation~\ref{eq:C_cut}), which could be of particular interest for follow-up studies, given the continued spectroscopic monitoring within SDSS-V/BHM-RM.
We therefore chose to list them in Table~\ref{tab:cands} (with appropriate demarcation). 
The multi-epoch spectra of these three BHM-RM CL-AGNs are also part of Figure~\ref{fig:all_spec} (available online).

We stress that our search for CL-AGNs within BHM-RM data is incomplete, and that this dataset could indeed include many more CL-AGNs. A more nuanced search should be conducted as part of a future analysis of SDSS-V data.

\subsubsection{Extremely Variable Quasars}
\label{subsubsec:EVQs_sample}

In the course of our work, we identified 88 additional sources that satisfied our automatic selection criterion (Equation~\ref{eq:C_cut}), and were confirmed by visual inspection as presenting genuine, dramatic changes in their broad-line emission. However, unlike in our final, core CL-AGN sample, in these sources all the observable broad lines remained prominent and robustly detected, in all epochs, even in their dimmer states.\footnote{Including in our follow-up spectroscopy, when available.}
Thus, we deemed these EVQs to be qualitatively different from our working definition of CL-AGNs, i.e., systems where at least one of the broad lines essentially (dis-)appears. 

There are two important, and potentially interlinked, caveats regarding the identification of these extremely variable sources. 
First, the apparent distinction between them and the ``proper'' CL-AGNs we retain in our core sample may be driven by our visual inspection process, which in some cases may miss the presence of broad emission lines concealed in low-$S/N$, dim-state spectra. 
Moreover, a high fraction of starlight in the fiber, whether due to a lower AGN luminosity (see the host galaxy) and/or aperture and distance effects, can make the identification of broad emission lines harder in some systems.
Second, the physical distinction may also be artificial, as we cannot rule out that all types of (extreme) spectral variability are driven by the same physical mechanisms, particularly by dramatic changes to the accretion-driven (ionizing) continuum radiation.
It is indeed possible that some of the dimming events among these 88 EVQs will lose their broad emission lines, and thus become proper CL-AGNs, some time in the (near) future. We hope to be able to monitor them within SDSS-V and/or using dedicated observations with other telescopes. 

Notwithstanding these caveats, these 88 sources are noteworthy examples of extreme spectral variations, and are part of the more general class of extremely variable or flaring AGNs \cite[see, e.g.,][]{Lawrence16, Graham17, Assef18, Rumbaugh18}. 
We tabulate all 88 such sources in Table~\ref{tab:cands} and present two examples in Figure~\ref{fig:selection_examples} (in Appendix \ref{app:selection}).
The full, statistical analysis of these EVQs is beyond the scope of the present paper.

\begin{deluxetable*}{ccccccccccccc}
\tablecaption{Robust CL-AGNs and EVQs identified in this work}
\label{tab:cands}
\tablewidth{\textwidth}
\tabletypesize{\scriptsize}
\tablehead{
\colhead{Name} & \colhead{$z$} & \colhead{Epochs} & \colhead{Class} & \multicolumn{5}{c}{Emission-Line Change Parameter \tablenotemark{\footnotesize a}} & \multicolumn{3}{c}{Ancillary Data} & \colhead{Notes\tablenotemark{\footnotesize e}}
\\[-2ex]
\colhead{} & \colhead{} & \colhead{(MJD)} & \colhead{} & \colhead{C(\Ha)} & \colhead{C(\hb)} & \colhead{C(\mgii)} & \colhead{C(\ciii)} & \colhead{C(\civ)} & \colhead{Spec. \tablenotemark{\footnotesize b}} & \colhead{Opt. phot. \tablenotemark{\footnotesize c}} & \colhead{WISE \tablenotemark{\footnotesize d}} & \colhead{}
\\[-2ex]
}
\startdata
J000719.90$+$253128.6 & 0.559321 & 56543, 59217 & EVQ & - &2.1 & 1.3 & - & - & 0 & 1 & 2 & new \\
J000750.51$+$005815.2 & 0.345113 & 52903, 59202 & CL-AGN & - &2.9 & - & - & - & 0 & 1 & 0 & new \\
J001103.49$+$010032.6 & 0.485892 & 55478, 59202 & CL-AGN & - &-2 & 2.8 & - & - & 0 & 1 & 1 & new \\
J001133.95$-$000001.1 & 0.44707 & 52518, 59202 & EVQ & - &1.7 & 2.2 & - & - & 0 & 1 & 0 & new \\
J001333.96$+$014506.8 & 0.881555 & 55511, 59202 & CL-AGN & - &- & 2.3 & - & - & 0 & 1 & 0 & new \\
J002956.65$+$010537.2 & 0.373015 & 55447, 59167 & EVQ & 2 &2.2 & - & - & - & 0 & 1 & 1 & new \\
J003242.74$+$003110.9 & 0.360592 & 51782, 59167 & EVQ & - &2.3 & - & - & - & 0 & 2 & 2 & new \\
J003841.53$+$000226.7 & 0.649649 & 55182, 59187 & CL-AGN & - &4.1 & 1.4 & - & - & 0 & 1 & 2 & new \\
J003849.27$-$011722.4 & 0.500226 & 55827, 59187 & EVQ & - &1.3 & 2.1 & - & - & 0 & 1 & 2 & new \\
J004459.09$-$010629.3 & 0.228215 & 52531, 59187 & CL-AGN & 3.2 &5.1 & - & - & - & 2 & 1 & 2 & new \\
\enddata
\tablenotetext{a}{The line variation parameter, as defined by Equation~\ref{eq:criterion_final}. All the sources in this table have (at least) one emission line with $C({\rm line})>2$ (Equation~\ref{eq:C_cut}). For CL-AGNs from our core sample (``Class'': CL-AGN), these values are calculated using the decomposed, narrow-line-subtracted spectra. For EVQs and RM CL-AGNs (``Class'': EVQ and CL-AGN RM, respectively), these values are calculated directly from the observed spectra.}
\tablenotetext{b}{Follow-up spectroscopy flag. ``0'': no follow-up spectroscopy taken, ``1'': follow-up spectroscopy is inconclusive, ``2'': follow-up spectroscopy confirms the SDSS-V spectroscopy.}
\tablenotetext{c}{Optical photometry flag. ``1'': the available photometric data are irrelevant or inconclusive for our work, ``2'': photometric data confirm concurrent spectral changes.}
\tablenotetext{d}{WISE-based, IR photometry flag. ``0'': no relevant WISE data, ``1'': trend in WISE light curve does not match concurrent spectral changes, ``2'': trend in WISE light curve matches concurrent spectral changes.}
\tablenotetext{e}{Additional information regarding each source. ``new'': object was not mentioned in previous  works (to the best of our knowledge). Otherwise, we mark the previous studies that noted some of the CL-AGNs in our sample. ``L15'': object noted by \citet{LaMassa15}, M16: \citet{MacLeod16}, R16: \citet{Ruan16}, Y18: \cite{Yang18}, H19: \citet{Hutsemekers19}, M19: \citet{MacLeod19}, G20: \citet{Graham20}, G22: \citet{Green22}, Z22: \citet{Zeltyn22}, G24: \citet{Guo24_DESI}.}
\tablecomments{Table~\ref{tab:cands} is published in its entirety in machine-readable format. A portion is shown here for guidance regarding its form and content.}
\end{deluxetable*}

\section{Comparison Samples and Key Measurements}
\label{sec:control_and_qunatities}

\subsection{The Control Sample(s)}
\label{subsec:control_samples}

To contextualize our sample of CL-AGNs and compare it to the broader population of (SDSS-observed) broad-line AGNs, we relied on the catalog of spectral measurements for the SDSS DR16Q catalog, as measured and compiled by \citet[][hereafter WS22]{WS22}. Those spectral measurements were (also) obtained using the \texttt{PyQSOFit} spectral-fitting package \citep[][see more details below]{QSOFit}. 
Only 87 of the 113 CL-AGNs in our final sample appear in the WS22 catalog. The remaining 26 sources are not part of DR16Q, which is the ``parent'' catalog for the WS22 work (as noted in Section~\ref{subsec:final_sample}).

In Section~\ref{subsec:prop_dist} we compare our sample of CL-AGNs to this control sample of quasars, in terms of key properties such as AGN luminosity, black hole (BH) mass, and accretion rate (\Lbol, \mbh, and \fedd, respectively; see Section~\ref{subsec:calc_ang_prop} below for how these quantities are derived). 
These comparisons necessitate the construction of control subsamples that are subsets of the large WS22 control sample, and which are \emph{matched} to our CL-AGNs sample in terms of redshift and certain other properties.

The matched control samples are constructed as follows:
\begin{enumerate}[label=\roman*.]

\item 
Each control sample is meant to match our CL-AGNs in terms of redshift and one other property, \Lbol, \mbh, or \fedd, as measured in (earlier) SDSS-I-IV spectroscopy.

\item
For each such pair of properties, we first split the two-dimensional space covered by our CL-AGN sample into $10\times10$ bins. Importantly, we map the fraction of CL-AGNs among our sample that reside in these bins, i.e., the two-dimensional fractional distribution of our CL-AGNs.

\item
For each nonempty bin in our CL-AGN two-dimensional (fractional) distribution, we select those quasars from the WS22 catalog that correspond to the same bin in redshift and the other relevant property. 

\item
We then randomly downselect only a subset of the bin-matched WS22 quasars so that the resulting two-dimensional fractional distribution of the matched control sample will be identical to that of our CL-AGN sample, i.e., if only 4\% of the CL-AGNs in our core sample reside in a certain $(z,\Lbol)$ bin, the WS22 quasars in all bins are downselected such that 4\% of them will reside in that specific bin.

\end{enumerate}

We stress that, due to the random nature of the downselection process in the last step above, each property-matched control sample is pseudo-random, unique, and not entirely reproducible. Thus, in our statistical analyses that rely on these control samples, we repeat the process of control sample construction 1000 times and base our insights on the collective emergent statistical results.

\subsection{Spectral Measurements of Changing-look Active Galactic Nuclei}
\label{subsec:decomposition}

Throughout this work, we have used three different sets of spectral measurements for our CL-AGN sample, as outlined below. 
Each set of measurements has certain merits and limitations, making it useful for various aspects of our analysis, which focuses on continuum and (broad) emission-line properties (luminosities, line widths) as well as a few key derived AGN/SMBH properties (BH mass and accretion rate).

\subsubsection{\citet[]{WS22} Catalog Values}

We take the subset of our CL-AGN sample whose earlier SDSS spectral epoch (i.e., pre-SDSS-V) appears in the WS22 catalog. As noted above, this is applicable for 87 out of our 113 CL-AGN candidates, for which we simply adopt the measurements tabulated in WS22. 
This approach ensures the consistency between the measurements for the CL-AGN and comparison samples measurements, but is obviously limited to only a subset ($\sim$77\%) of our entire CL-AGN core sample.

This approach was used for Figure~\ref{fig:z_L_dist} and for the analysis shown in Figures~\ref{fig:rates} and \ref{fig:histograms_ws22}.

\subsubsection{\citet[]{WS22}-like Decomposition}

We decompose the spectra of all our CL-AGN candidates following the methodology outlined in WS22, using \texttt{PyQSOFit} \citep{QSOFit}. A detailed account of the fitting procedure can be found in WS22 and references therein. In short, the fitting procedure employed in WS22 de-reddens the spectra using Milky-Way dust maps \citep[]{SFD98}, and models the continuum emission using a combination of a power law, a third-order polynomial (introduced to account for peculiar continuum shapes, e.g. due to reddening), and \ion{Fe}{ii} UV and optical empirical templates \citep[]{BG92,VW01}. Each of the emission lines in the continuum-subtracted spectra are then fitted with combinations of Gaussians, as detailed in WS22. The uncertainties in the continuum and line parameters were obtained through a Monte Carlo refitting approach, using 25 realizations for each spectrum, relying on the corresponding error spectra.

This approach allows us to take advantage of the entire sample of CL-AGN candidates (113 in total) and enhance the statistical significance of our results, while maintaining high consistency with the WS22 catalog values. We verified that the properties obtained from our decomposition are highly consistent with those derived by WS22, in terms of both accuracy and precision. We also verified that the usage of our spectral decomposition (rather than the WS22 catalog values) does not affect the primary findings presented through the rest of this work (see discussion in Section~\ref{subsec:prop_dist}).

This approach was used for Figure~\ref{fig:histograms_all} (in Appendix~\ref{app:alter_dists}) and the corresponding analysis, and for deriving the values tabulated in Table~\ref{tab:spec_measurements}.

\subsubsection{Two-epoch Decomposition}

For comparisons between multi-epoch spectra of the same source (i.e., for line variability measurements), as well as to determine the representative flux ratio value for each source (as defined in Equation \ref{eq:criterion_final}), we conducted detailed spectral decompositions that are better suited to our needs. As these measurements focus only on the core CL-AGN sample, the (strict) requirement to be fully consistent with WS22 is lifted. 

For each CL-AGN in our sample, we chose two spectra to represent the `dim' and `bright' AGN states. Many of our candidates displayed complex temporal behavior that was captured in multiple spectra, such as gradual transitions or alternating changes between dimmer and brighter states; however, for the sake of simplicity, we only used two spectra to represent each CL transition. Specifically, for each object, we used an archival SDSS spectrum (pre-SDSS-V) and a more recent SDSS-V spectrum, focusing only on those pairs of spectra that satisfy our CL-AGN selection criterion (Equation~\ref{eq:C_cut}), we chose the brightest and dimmest spectra available within these distinct survey-defined datasets. Whenever several spectra of comparable brightness level were available, we chose the latest spectrum from the SDSS archive and the earliest from new SDSS-V observations. This choice is motivated by our desire to probe, or constrain, the shortest possible CL transitions between archival SDSS and new SDSS-V spectroscopy. It is possible that yet faster transitions will be identified within SDSS-V, which would be extremely interesting given how poorly understood these phenomena are and the challenges they pose to AGN models. We leave searching for transitions occurring within SDSS-V observations to a future work.

Our decomposition also relied on the \texttt{PyQSOFit} package, however with a somewhat different setup compared with that used in WS22. 
The continuum was decomposed using a power law, a polynomial, and empirical UV and optical iron emission templates \citep{BG92,VW01}. 
We modeled each broad emission line using two Gaussian components, while the narrow lines (of \hb, \oiii, \ha, \nii, and $\left[\sii\right]$) were modeled with one Gaussian, each. 
We further tied the widths and shifts of all narrow lines in each spectral complex, and forced the ratio between the [\ion{O}{iii}]$\,\lambda\lambda4959,5007$, [\ion{N}{ii}]$\,\lambda\lambda6549,6585$, and [\ion{S}{ii}]$\,\lambda\lambda6718,6732$ doublets to be 1:3, 1:3, and 1:1 (respectively).\footnote{For comparison, WS22 used three Gaussians for the \ha, \hb, and \civ\ broad lines; included an additional broadened component for each of the two \oiii\ lines, and did not force these lines to have any specific intensity ratio.} 

As we are mainly interested in emission-line flux measurements, we did not include any host decomposition in our modeling. We visually verified that this does not negatively affect the quality of the continuum (shape) fits of our spectra.
This choice also means that we cannot account for stellar absorption features underlying the (broad) Balmer lines--a common issue in nearly all studies of broad-line AGNs (see e.g. \citealt{Oh15} for an exception).

In order to minimize the effects of calibration issues on our measurements, whenever possible and deemed justified (see below), we matched the narrow \oiii\ emission flux measurements in the two epochs, as follows. For each CL-AGN, we first fitted the two epochs separately with both broad and narrow emission lines. Next, we renormalized the bright-state spectrum using the ratio between the narrow \oiii\ line fluxes in the bright and dim states, so that the narrow-line emission is consistent between the two epochs.
The \oiii-based normalization factors we applied are in the range $0.72-1.87$, with a median value of $1.03$.
Furthermore, to maximize the consistency between the (broad) emission-line measurements of the two epochs, we subtracted the narrow emission-line profiles deduced from the dim-state spectrum from the (scaled) bright-state spectrum, so that the latter remains essentially free of narrow-line emission. Finally, the scaled, narrow-line-subtracted bright-state spectrum was fitted through another run of \texttt{PyQSOFit}, but now without any narrow-line components. 
This procedure enhances the quality of the broad-line profile fitting for the bright-state spectra, particularly for the rather complex and blended \ha\ profiles.

For 50 of our CL-AGN spectra, the two epochs were obtained using different fiber sizes (i.e., 3\arcsec-diameter fibers for the SDSS spectrographs and 2\arcsec-diameter fibers for the BOSS spectrographs, for the earlier and later epochs, respectively). For such sources, a normalization based on matching narrow emission-line fluxes may yield inconsistent results, since more narrow-line emission from the (extended) narrow-line region may enter the larger apertures used for the earlier epochs. For these sources, and only when the narrow emission-line flux difference between the epochs could (qualitatively and, roughly, quantitatively) be explained by the difference in fiber sizes, we did not perform any renormalization, but rather carried out our final spectral measurements on the original (pipeline produced) spectra.
For yet other candidates, visual inspection of the decomposed spectra revealed that the normalization procedure produced unphysical results, such as a mismatch between host components in the two epochs (i.e., a drastic difference between the appearances of host stellar absorption features). In such cases, which can be perhaps explained by seeing differences between the two epochs, we also chose to not perform any renormalization and instead relied on the original spectra.

The spectral measurements resulting from this approach were used for Figures~\ref{fig:df_vs_dt} and \ref{fig:df1_vs_df2}, and the corresponding analysis.

\subsection{Calculation of Active Galactic Nuclei Properties}
\label{subsec:calc_ang_prop}

The various spectral decomposition and measurement approaches allow us to derive estimates of bolometric luminosities (\Lbol), black hole masses (\mbh), and Eddington ratios (\fedd). 
The only purpose of these derived quantities for the present work is to compare our CL-AGNs to the general quasar population (i.e., to the DR16Q-based WS22 catalog). As these extensive catalogs relied on SDSS-I-IV spectra for their spectral measurements, we also chose to use the spectral measurements of the earlier, SDSS-I-IV spectra of our CL-AGNs in what follows.
We acknowledge that all these derived quantities are based on a set of assumptions, and carry large systematic uncertainties. Our derived quantities should thus be treated with the same caution as do other compilations of \Lbol, \mbh\ and \fedd\ for broad-line AGNs with rest-frame UV-optical spectroscopy \cite[see, e.g.,][and references therein]{Shen13,MejiaRestrepo22}.

Following the methods of WS22, we estimate the bolometric luminosity, \Lbol, using the measured monochromatic continuum luminosities, \lamLlam, at three specific wavelengths: 1350, 3000, and 5100 \AA\ (\Luv, \Lthree, and \Lop, respectively). The corresponding bolometric corrections are 3.81, 5.15, and 9.26, respectively \citep{Richards06}. We prioritize using the estimate based on \Lthree, as it is less affected by host contamination compared to \Lop\ and less affected by reddening than \Luv. For sources where \Lthree\ is unavailable, we used whichever of the other two continuum measurements that was available (no single source had both \Luv\ and \Lop\ measurements available).
We acknowledge that estimating the AGN continuum emission using \Lthree\ may also be inaccurate. Specifically, contamination from hot (young) stars and/or from (blended) Fe transitions within the relevant spectral band may be mistaken for AGN continuum. We note, however, that \Lthree\ is the preferred estimate in WS22, from which we draw our parent sample.

We estimate \mbh\ using single-epoch, virial BH mass prescriptions \citep[see][and references therein]{MejiaRestrepo22}. Contrary to WS22, who used only the broad \hb, \mgii, and \civ\ emission lines to estimate BH masses, we also utilized the broad \Ha\ emission line. This choice is explained in detail in what follows. To minimize potential systematic errors due to host contamination, we adopted a hierarchical approach. Whenever possible, we used mass estimates based on the luminosity and width of the broad component of the \Ha\ emission line, following the prescription of \cite{GreeneHo05} but rescaling it by $\times 4/3$ (or $+0.125$ dex; see \citealt{MejiaRestrepo22} for details). As this estimate does not rely on continuum measurements, it is the least affected by host contamination. Whenever \Ha\ was not accessible, we relied on width of the broad \mgii\ line and \Lthree\ (again: due to host contamination and reddening concerns). When neither the broad \Ha\ nor the broad \mgii\ lines were available, we deferred to prescriptions that use either the broad \hb\ line and \Lop, or the broad \civ\ line and \Luv, depending on the source redshift.

We estimate \fedd\ following 
\begin{equation}
\label{eq:fedd}
    \fedd\equiv\frac{\Lbol}{L_{\rm
    Edd}}=\frac{\Lbol}{1.5\times10^{38} \times (\mbh/\Msun)\,\ergs} \, ,
\end{equation}
which is appropriate for solar-metallicity gas.
We note that the BH mass prescriptions we use, and hence the derived Eddington ratios, rely on standard and universal relations between BLR size and AGN continuum luminosity derived from RM campaigns. In practice, every AGN may be offset from these relations with an intrinsic scatter of ${\lesssim}0.15$ dex \cite[e.g.,][]{Bentz13}, and sources with particularly high accretion rates  may show more pronounced and systematic offsets \cite[see, e.g.,][]{DW19}. The latter effect should not affect our analysis and results, as the bulk of our CL-AGN sample, as well as the control samples, show relatively low accretion rates.

Throughout our analysis (described below), we have utilized two distinct sets of the quantities \Lbol, \mbh, and \fedd.
First, we employed the aforementioned mass prescriptions to calculate \mbh\ and \fedd\ for all the quasars in the WS22 catalog. This calculation relied on the spectral measurements in that catalog (luminosity and line widths), rather than using the \mbh\ values provided in the catalog.
Second, we also calculated these quantities based on our own ``WS22-like'' spectral decomposition of the earliest SDSS-I--IV epochs of our candidates (see Section \ref{subsec:decomposition} for details). As these values are not part of any public catalog, we have compiled them along with the relevant spectral measurements (luminosity and line widths) in Table~\ref{tab:spec_measurements}. 
Also, as noted in Section \ref{subsec:decomposition}, we verified that the spectral measurements and derived properties stemming from our ``WS22-like'' decomposition are highly consistent with the ones available through the WS22 catalog, with differences much smaller than 0.1 dex.

\begin{deluxetable*}{lccccccc}
\tablecaption{Spectral measurements of CL-AGNs}
\label{tab:spec_measurements}
\tablewidth{\textwidth}
\tabletypesize{\scriptsize}
\tablehead{
\colhead{Name} & \colhead{$z$} & \colhead{Spec.} & \colhead{$\log L$} & \colhead{FWHM} & \colhead{$\log\mbh$} & \colhead{$\log\Lbol$} & \colhead{$\log\fedd$} 
\\
[-2ex]
\colhead{} & \colhead{} & \colhead{complex} & \colhead{$\left(\ergs\right)$} & \colhead{$\left(\kms\right)$} & \colhead{$\left(\Msun\right)$} & \colhead{$\left(\ergs\right)$} & \colhead{} 
}
\startdata
J000750.51$+$005815.2 & 0.345113 & \ha & 42.3 & 7203 & 8.37 & 44 & -2.56 \\
J001103.49$+$010032.6 & 0.485892 & \ha & 42.4 & 3774 & 7.82 & 44.7 & -1.34 \\
J001333.96$+$014506.8 & 0.881555 & \mgii & 44.2 & 3747 & 8.2 & 44.9 & -1.43 \\
J003841.53$+$000226.7 & 0.649649 & \mgii & 44.7 & 5690 & 8.83 & 45.4 & -1.62 \\
J004459.09$-$010629.3 & 0.228215 & \ha & 42.3 & 5228 & 8.07 & 44.8 & -1.43 \\
J004616.68$+$000249.1 & 0.4531 & \mgii & 43.3 & 4177 & 7.74 & 44.1 & -1.87 \\
J011536.11$+$003352.4 & 0.364029 & \ha & 42.4 & 7371 & 8.44 & 44.4 & -2.26 \\
J011851.07$-$010417.3 & 0.573171 & \mgii & 44.1 & 4515 & 8.28 & 44.8 & -1.64 \\
J012025.45$+$142727.9 & 0.263752 & \ha & 41.9 & 5850 & 7.97 & 43.8 & -2.32 \\
J012256.19$-$000252.6 & 0.340502 & \ha & 42.2 & 4633 & 7.9 & 44.4 & -1.7 \\
\enddata
\tablecomments{Table~\ref{tab:spec_measurements} is published in its entirety in the machine-readable format.
A portion is shown here for guidance regarding its form and content.}
\end{deluxetable*}

\section{Analysis and Results}
\label{sec:analysis_results}

In what follows, we examine various properties of the core CL-AGN sample, as well as some (multi-epoch) emission-line measurements. 
We stress that the analyses described hereafter focus only on the core sample of 113 CL-AGNs (Section~\ref{subsec:final_sample}), and do \emph{not} include the three additional CL-AGNs identified within the BHM-RM data (Section~\ref{subsubsec:RM_CLAGN}).

\subsection{Changing-look Active Galactic Nuclei Sample Properties}
\label{subsec:sample_props}

To ensure consistency with the fitting procedures in WS22, we initially examine only the subset of our core CL-AGN sample that is present in the DR16Q catalog (87 out of 113 CL-AGNs), and use the catalog values from WS22 for both our CL-AGN subsample and the WS22 comparison sample.
This means that we are using only measurements based on the earlier (archival) spectra for the following analysis.

\begin{figure}
    \centering
    \includegraphics[width=1.0\columnwidth]{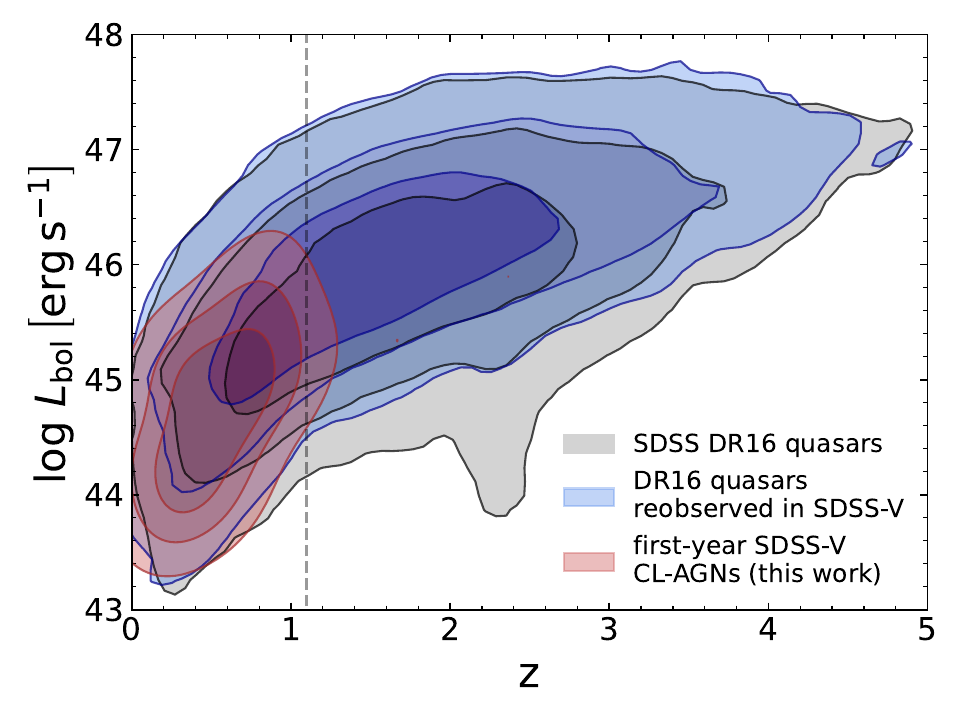}
    \caption{The distributions of SDSS DR16 quasars \cite[black contours;][]{WS22}, SDSS DR16 quasars that have been reobserved in the first year of SDSS-V and that also appear in the WS22 quasar catalog (blue contours), and SDSS-V CL-AGNs that appear in the WS22 catalog (red contours), in the luminosity-redshift plane. For each distribution, the enclosed regions correspond to the 68\%, 95\%, and 99.7\% percentiles. The dashed vertical line marks the redshift where the broad \Hb\ line moves out of the SDSS-V/BOSS spectral coverage.}
    \label{fig:z_L_dist}
\end{figure}

Figure \ref{fig:z_L_dist} compares the redshift--luminosity distribution of three different samples: the entire DR16Q catalog, the subset of the DR16 quasar catalog that has been reobserved during the first year of SDSS-V operations (not including BHM-RM targets), and our CL-AGN sample (which was drawn from the first-year SDSS-V data).
As can be seen in the figure, CL-AGNs occupy a lower redshift and luminosity range compared to all known SDSS DR16 quasars, even when accounting for the partial coverage of SDSS-V repeated spectroscopy. 
Specifically, 94\% of the subset of CL-AGNs that are present in the WS22 catalog reside at $z\lesssim0.9$, compared with 14\% of the entire WS22 quasar catalog and 19\% of the quasars reobserved in the first year of SDSS-V operations.

This difference may be explained by the longer rest-frame timescales that are probed for low-redshift targets, and hence the higher probability of detecting a significant spectral change at low $z$.
An additional explanation may lie in the known anticorrelation between AGN luminosity and variability amplitude in general \cite[e.g.,][]{VB04, Wilhite08, Kozlowski10, Macleod10, MW13, Morganson14, Simm16, Caplar17} and extreme variability in particular \cite[e.g.,][]{Rumbaugh18}, as well as the occurrence of CL-AGNs  \cite[e.g.,][]{MacLeod19, Green22, Temple23}.

Furthermore, part of the difference between the distributions could be attributed to our method of searching for CL-AGNs, which primarily relied on detecting variability in specific emission lines. Beyond $z \sim 1.1$, the \Hb\ line is shifted out of the BOSS spectra. 
At higher redshifts, our CL-AGN search algorithm relies instead primarily on the broad \mgii\ line. Several studies reported that this line displays less variability than the broad Balmer lines \cite[e.g.,][and references therein]{Cackett15, Sun15, Homayouni20,Yang20_CLAGN}. In addition, beyond $z\approx1$, the BOSS spectra lack the prominent narrow \OIII\ emission line, which was used to robustly identify significant changes in the broad \hb\ line (i.e., as a flux-calibration anchor). 
We were not able to employ such an approach at yet higher redshifts, thus leading to the apparent bias of our sample toward $z \lesssim 1$ sources.

\begin{figure*}
  \centering
  \includegraphics[width=1.0\textwidth, clip, trim={2cm 0.5cm 1.9cm 1.5cm}]{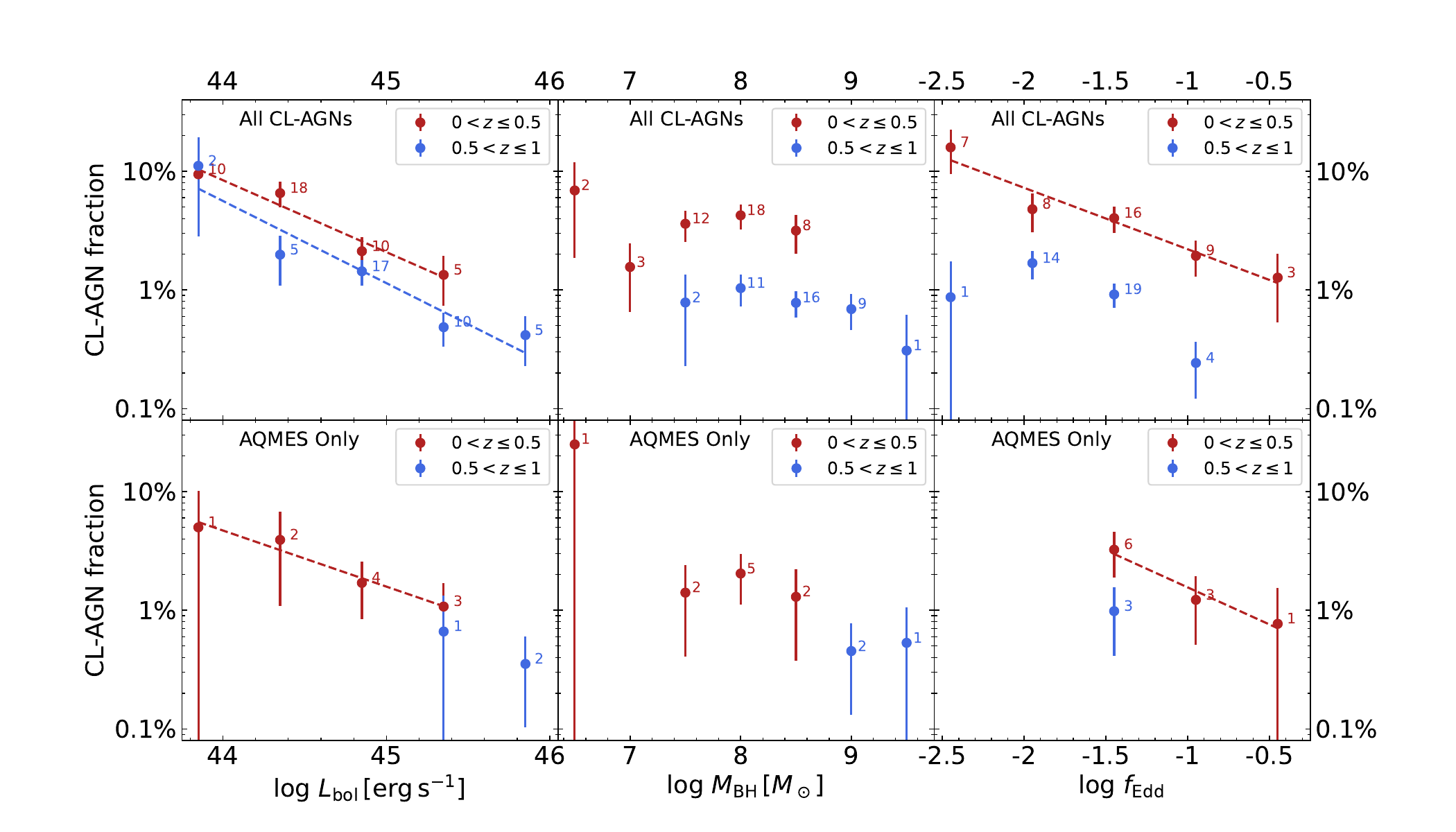}
  \caption{\textit{Top row:} the CL-AGN occurrence rate as a function of \Lbol\ (left), \mbh\ (middle), and \fedd\ (right), for $0<z\le0.5$ (red) and $0.5<z\le1$ (blue), for the 82 CL-AGN at $z\leq1$ in our core sample that also appear in the WS22 catalog. 
  The rates were calculated in 0.5 dex wide bins, and the number of CL-AGNs in each such bin is indicated next to it. 
  To illustrate some of the trends in these panels and to guide the eye, we also show (as dashed lines) simple best linear fits (in the log-log space) to the binned data points, neglecting the uncertainties, for the cases where there is a statistically significant correlation (i.e., a Spearman correlation test that resulted in $p_{\rm S}<0.01$). 
  \textit{Bottom row:} same as top panels, but for the 13 AQMES CL-AGN at $z\leq1$.}
  \label{fig:rates}
\end{figure*}

In Figure \ref{fig:rates} we examine the occurrence rates of CL-AGNs, \fcl, and its dependence on redshift, luminosity, BH mass, and Eddington ratio (all derived, again, from the earlier SDSS spectra). For this analysis, we only consider the 82 CL-AGNs at $z\leq1$ in our core sample that also appear in the WS22 catalog, and further split them to those located at $0<z\leq0.5$ and $0.5<z\leq1$ (43 and 39 sources, respectively). 

The top row of panels of Figure \ref{fig:rates} presents the occurrence rates of our (subset of) CL-AGNs, calculated by dividing the number of detected CL-AGNs by the total number of WS22 quasars with repeated spectroscopy in SDSS-V, in each of the two redshift ranges, and for each bin of AGN bolometric luminosity, BH mass, and Eddington ratio. 
We recall that the DR16Q quasars, and thus the WS22 quasars, constitute a heterogeneous superset constructed with various selection criteria (and related biases).
In each panel, and for each of the redshift ranges, we split the corresponding sample of CL-AGNs to evenly spaced bins of width 0.5 dex, such that each bin would contain at least one CL-AGN (i.e., the bins cover the range of properties observed in our CL-AGN sample).
As can be seen in Figure~\ref{fig:rates},
\fcl\ increases with decreasing \lbol\ and \fedd, across all redshifts. For example, for $0<z\leq0.5$, the CL-AGN occurrence rate increases from $\fcl\sim1\%$ for $\log(\Lbol/\ergs)\approx45.4$ to $\sim 10\%$ for $\log(\Lbol/\ergs)\approx 43.9$. Similarly, the occurrence rate increases from $\fcl\sim1\%$ for $\log\fedd\approx-0.4$ to $\sim15\%$ for $\log\fedd\approx-2.5$. Conversely, we find an apparently flat dependence of \fcl\ on \mbh\ ($p_{\rm S}=0.62$). For the higher-redshift bin of $0.5<z\leq1$ the trends are qualitatively similar.

The bottom row of panels of Figure~\ref{fig:rates} presents the occurrence rates of CL-AGNs discovered within the AQMES program. 
As mentioned in Section~\ref{subsec:sdssv}, AQMES provides the largest and most homogeneously selected sample of AGNs with repeated spectroscopy within SDSS-V, with a systematically programmed cadence of repeated observations.
While the focus on AQMES-observed AGNs allows us to provide rates based on a clear selection criteria, it naturally comes at the cost of a smaller sample (14 CL-AGNs).
Here, \fcl\ is calculated by dividing the number of detected AQMES CL-AGNs by the total number of AQMES targets revisited during the first year of SDSS-V operations, within the corresponding bin (in \Lbol, \mbh, \fedd) and redshift range.
The results for the AQMES subsample are qualitatively similar to the results for the entire sample: \fcl\ decreases with increasing luminosity and Eddington ratio, and does not depend on \mbh. These AQMES-only panels are currently sparsely populated, but we envision that they will provide the most statistically sound results as SDSS-V progresses, given the clean target selection criteria, and the high purity of the AQMES program.

It is important to emphasize that, as the calculation of \fcl\ relies on a straightforward division of the number of CL-AGNs by the total number of AGNs, it does not properly account for the observational bias stemming from the relation between luminosity and redshift, inherent biases in the CL-AGN search, and degeneracies between the different BH properties (Equation~\ref{eq:fedd}). Therefore, the trends seen in Figure \ref{fig:rates} are only suggestive of a link between fundamental AGN properties and the changing-look phenomena, and should be regarded with caution. Instead, they serve as predictions for the occurrence rates of CL-AGNs in an SDSS-V--like survey based on \Lbol, \mbh, and \fedd. Additionally, as our search methods may have missed some CL-AGNs (i.e., our search is pure according to our definition of CL-AGNs, but incomplete), these rates should be viewed as lower limits.

\subsection{What Active Galactic Nuclei Properties are Intrinsically Related with the Changing-look Phenomenon?}
\label{subsec:prop_dist}

The AGN luminosity, BH mass and Eddington ratio are closely related through $\fedd\propto \Lbol/\mbh$. To examine their relevance to our CL-AGNs while accounting for potential degeneracies, we constructed three control samples drawn from the WS22 catalog. 
All three control samples are first matched to our CL-AGN sample in terms of redshift distribution, to counteract any redshift-related selection biases. Importantly, each of the control samples is further matched to our CL-AGN sample in the distribution of one of the properties we examine (\Lbol, \mbh, or \fedd), so we can compare the distributions of the other two properties between the control and CL-AGN samples (for a detailed description of the matched control samples construction procedure, see Section~\ref{subsec:control_samples}). 
Moreover, to ensure that our control samples and CL-AGNs are indeed drawn from the same parent sample, the following analysis focuses on the subset of 87 CL-AGNs that are found in the DR16Q and WS22 catalogs.  
We further verified that our conclusions do not depend on this choice to focus on this subset of 87 CL-AGNs, by performing the same analysis also on the entire core sample of 113 CL-AGNs.
To obtain the spectral measurements needed for this latter analysis of the entire CL-AGN sample, and the corresponding derived AGN properties, we used the WS22-like spectral decomposition of the earlier (DR16) spectra of our CL-AGNs, which allowed us to take advantage of our entire sample of 113 CL-AGNs, and thus to increase the statistical significance of our results. 

\begin{figure*}
  \centering
  \raisebox{70pt}{\parbox[b]{.1\textwidth}
  {\large $\Lbol\,\&\,z$ matched}}\hfill
  \includegraphics[width=.40\textwidth,trim={0 0 16.5cm 4cm}, clip]{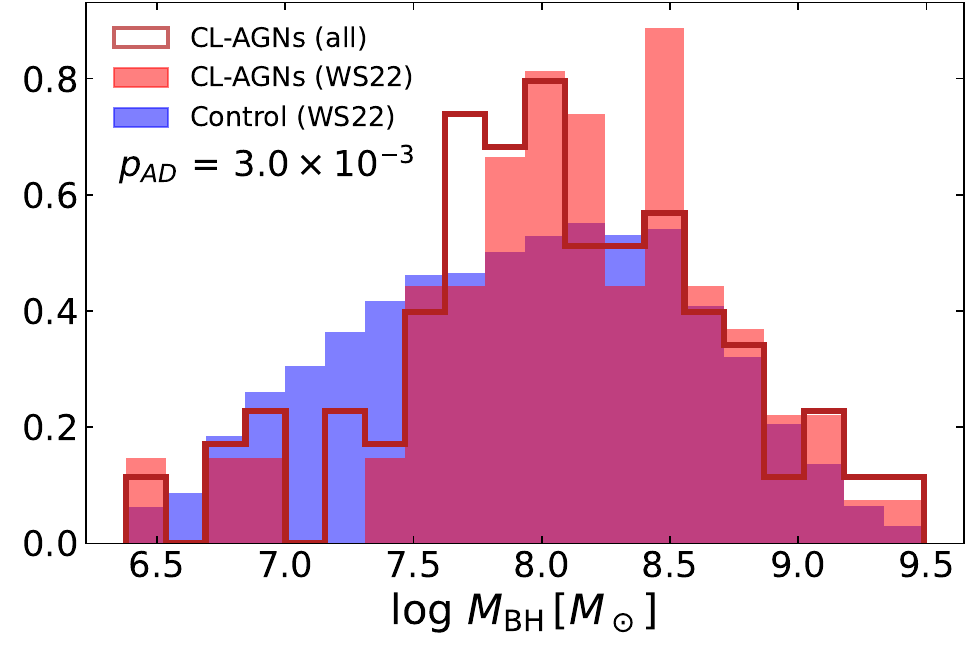}\hfill
  \includegraphics[width=.40\textwidth, trim={0 0 16.5cm 4cm}, clip]{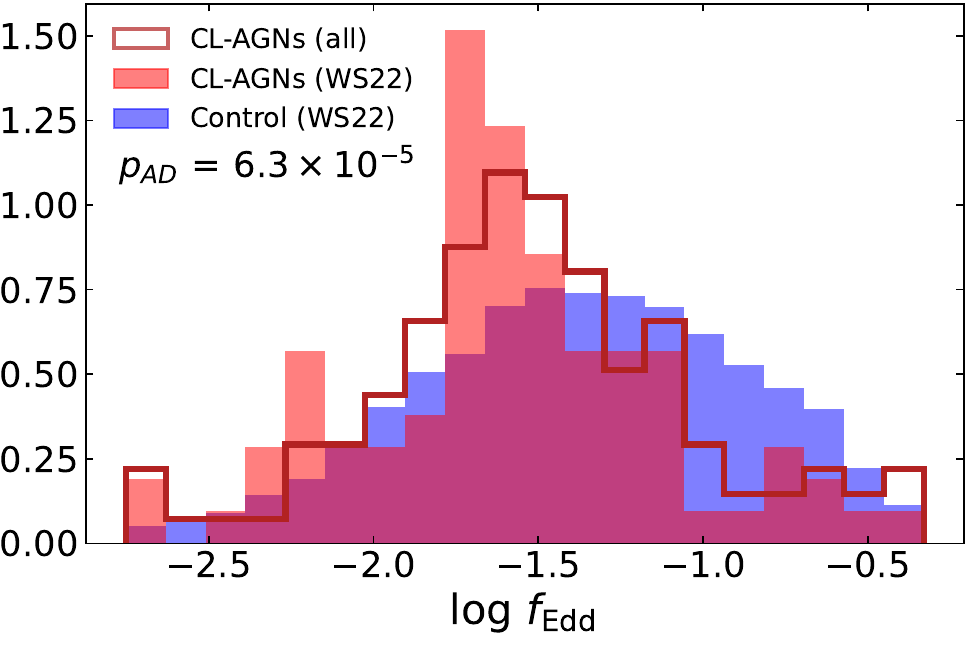}\\
  \raisebox{70pt}{\parbox[b]{.1\textwidth}
  {\large $\mbh\,\&\,z$ matched}}\hfill
  \includegraphics[width=.40\textwidth, trim={0 0 16.5cm 4cm}, clip]{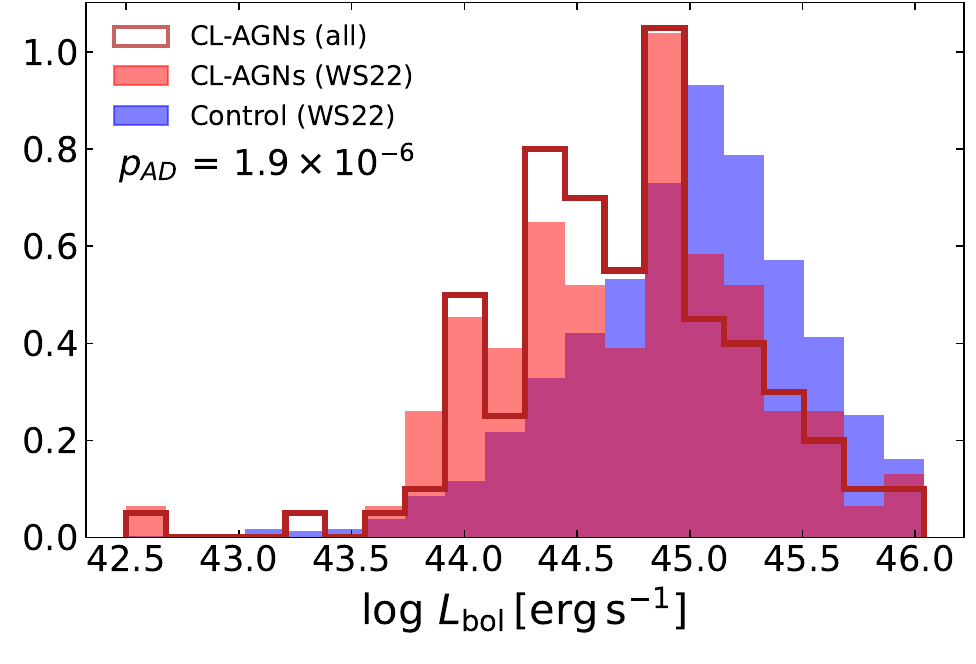}\hfill
  \includegraphics[width=.40\textwidth, trim={0 0 16.5cm 4cm}, clip]{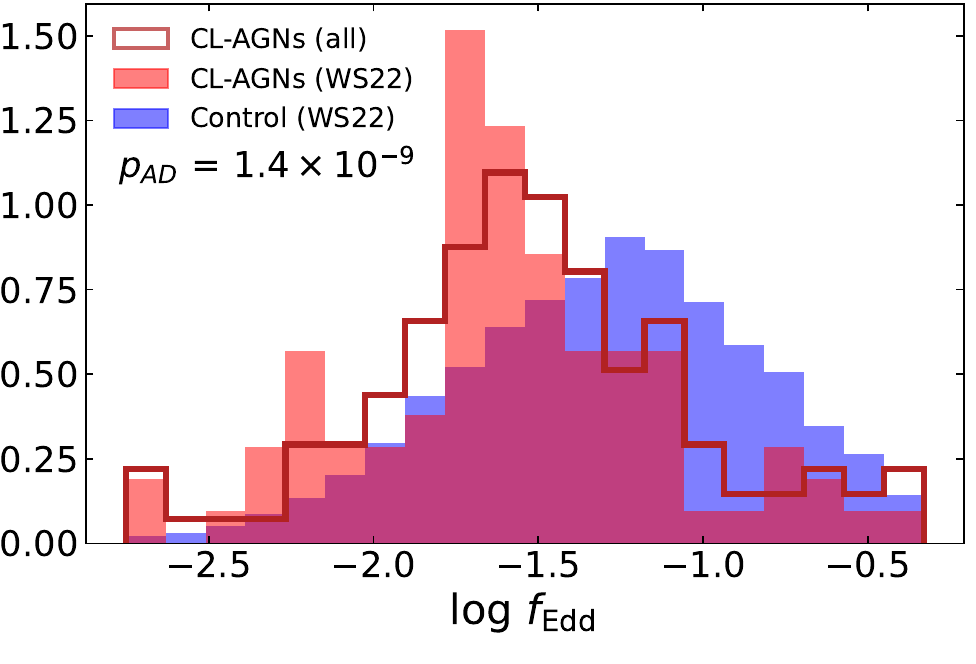}\\
  \raisebox{70pt}{\parbox[b]{.1\textwidth}{\large $\fedd\,\&\,z$ matched}}\hfill
  \includegraphics[width=.40\textwidth, trim={0 0 16.5cm 4cm}, clip]{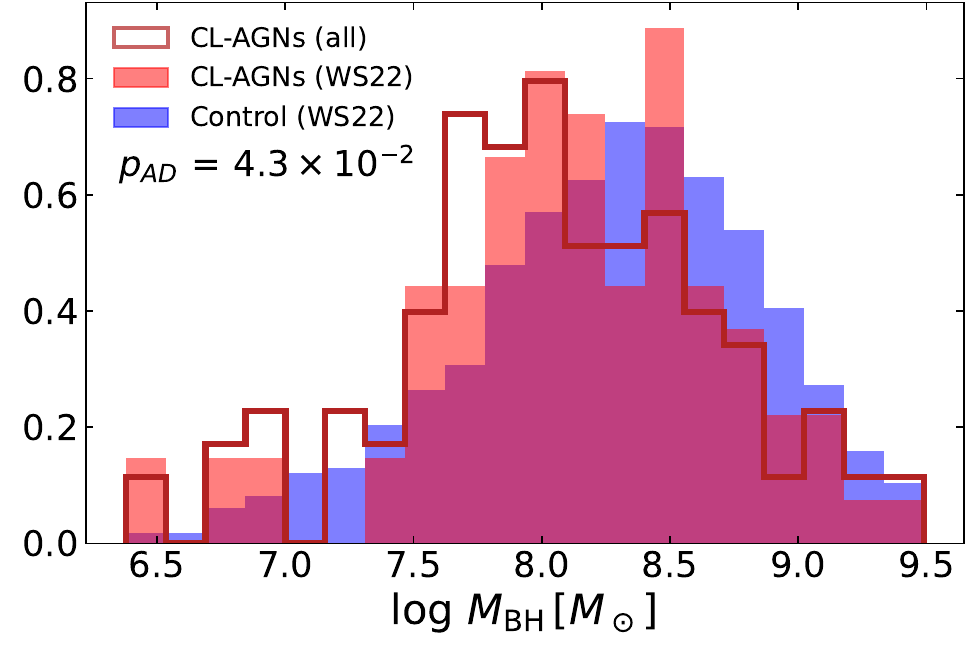}\hfill
  \includegraphics[width=.40\textwidth, trim={0 0 16.5cm 4cm}, clip]{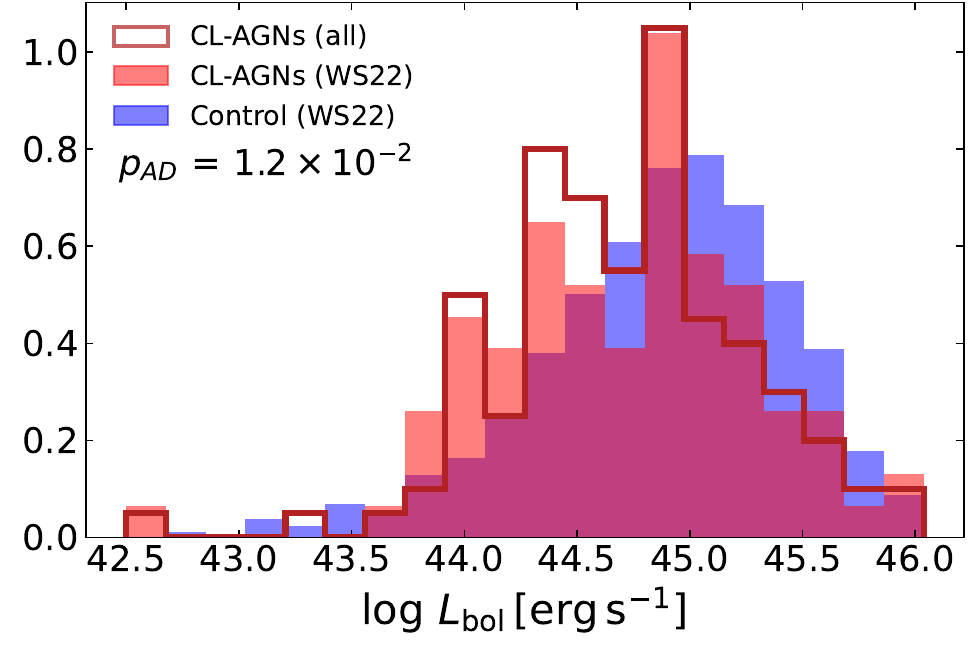}
\caption{Distributions of \Lbol, \mbh, and \fedd for our core sample of 113 CL-AGNs (red lines), for the subset of 87 CL-AGNs that appear in the WS22 catalog (red shaded bars), and for the matched control samples drawn from the WS22 catalog (blue shaded bars). 
Each row of panels shows control samples that are matched based on redshift and either \Lbol, \mbh, or \fedd\ (top, middle, and bottom panels, respectively).
Each panel includes the median $p$-value derived from Anderson-Darling tests between the WS22 subset of CL-AGNs and 1000 realizations of the corresponding control sample.}
\label{fig:histograms_ws22}
\end{figure*}

Figure~\ref{fig:histograms_ws22} displays the distributions of \Lbol, \mbh, and \fedd\ for CL-AGNs from our sample and the corresponding control samples  (red and blue shaded bars, respectively), 
where the control samples are drawn from the WS22 catalog and are matched to the subset of 87 CL-AGNs which appear in that catalog based on a combination of redshift and either \Lbol, \mbh, or \fedd (top, middle, and bottom panels, respectively). 
For each of the properties examined, we compare the distributions for our CL-AGN sample and the corresponding control sample using the Anderson--Darling (AD) test.\footnote{We have verified that our results and conclusions remain unchanged if we instead apply the Kolmogorov--Smirnov test. However, the AD test is more appropriate for comparing the samples under study (see discussion in \href{https://asaip.psu.edu/articles/beware-the-kolmogorov-smirnov-test/}{https://asaip.psu.edu/articles/beware-the-kolmogorov-smirnov-test/}).}
Appendix~\ref{app:alter_dists} and Figure~\ref{fig:histograms_all} present a complementary analysis conducted for our entire core sample of 113 CL-AGNs, based on our own ``WS22-like'' spectral decomposition.

Given that the control samples are constructed through a semi-random process (i.e., the control samples are tailored to match our CL-AGN sample in various properties but are otherwise drawn randomly from the WS22 quasar catalog; see Section~\ref{subsec:control_samples}), the histograms in Figure~\ref{fig:histograms_ws22} (and in Figure~\ref{fig:histograms_all}) are unique to each realization of the control sample. Therefore, in order to obtain robust results for the statistical tests discussed below, we repeated the construction of the control samples and the statistical tests that compare them to our CL-AGNs 1000 times. All the $p$-values mentioned below are the median values of the resulting distributions.\footnote{We verified that the distributions of the $p$-values are unimodal and well centered (approximately) around these representative values.}

As can clearly be seen in the top panels of Figure \ref{fig:histograms_ws22}, there is a statistically significant difference in the distributions of both \mbh\ and \fedd\ for the \Lbol-matched samples ($p_{\rm AD} = 3\times10^{-3}$ and $6\times10^{-5}$, respectively).
For the \mbh-matched samples (middle panels), there is a statistically significant difference in the distributions of both \Lbol\ and \fedd\ ($p_{\rm AD}=2\times10^{-6}$ and $1\times10^{-9}$, respectively).
The significantly lower $p_{\rm AD}$ of the latter comparison suggests that the \fedd\ distributions are different to a greater degree than the \Lbol\ ones. 
Most importantly, for the \fedd-matched samples, there is no statistically significant difference in the distributions of both \mbh\ and \Lbol\ ($p_{\rm AD}>0.01$).

Replacing the subset of 87 CL-AGNs that are part of the WS22 catalog with the full sample of 113 CL-AGNs yields generally consistent results, with some changes in the $p_{\rm AD}$ values and---importantly---with a more significant difference between the (\Lbol-matched) \fedd\ distributions than between the (\fedd-matched) \Lbol\ distributions (see Appendix~\ref{app:alter_dists} and Figure~\ref{fig:histograms_all}). 
However, unlike for the WS22 subset, the entire CL-AGN sample appears to significantly differ in \Lbol\ compared with the control sample, when matching for \fedd\ ($p_{\rm AD}=2\times10^{-3}$).
We suspect that this difference is the direct result of the fact that the 26 CL-AGNs in our sample that are not part of WS22 catalog were not part of the (parent) DR16Q catalog, and are therefore ``diluting'' the sample with sources that cannot, and should not, be compared with the control sample we use for our analysis.

Our analysis suggests that, among the properties we investigate here, it is most likely the Eddington ratio that differentiates our CL-AGN from the general quasar population and thus that the Eddington ratio---rather than BH mass or AGN luminosity---may have an intrinsic relation to the changing-look phenomenon.
This conclusion is supported by the fact that the most statistically significant difference between the CL-AGN sample and the various control samples we have assembled is in the \fedd\ distribution.
We thus postulate that any trends between CL-AGNs and either BH mass or AGN luminosities are likely caused by their intrinsic links to the Eddington ratio.

To further quantify the difference we find between the \fedd\ distributions of our CL-AGNs and of the control sample, we note that for the \Lbol\ matched sample, the median \fedd\ for the CL-AGN WS22 subset is $(\fedd)_{\rm med}=0.025$, while for the corresponding control sample it is higher, at $(\fedd)_{\rm med}=0.043$.
When considering the entire core sample of 113 CL-AGNs, and the corresponding control sample, the medians are $(\fedd)_{\rm med}=0.030$ and $0.050$, respectively.
This difference is in qualitative (and roughly quantitative) agreement with what was found in other studies of CL-AGNs (e.g., Figure~8 in  \citealt{Rumbaugh18}, Figure~6 in \citealt{MacLeod19}, Figure~13 in \citealt{Green22}; see also \citealt{Temple23}).
The two \fedd\ distributions have significant overlap, with median absolute deviations (MADs) of 0.27 and 0.36 dex, for the WS22 subset of CL-AGNs and the control sample, respectively. However, the difference between them goes beyond the aforementioned AD tests: for example, while $\approx$25\% of the control sample lies at $\fedd>0.1$, only $\approx$10\% of our CL-AGNs  are found in that range.\footnote{The MAD values and percentiles we quote reflect the subset of 87 CL-AGNs that are in the WS22 catalog, and the corresponding control sample. For the latter, the quoted quantities are themselves the median values obtained from the 1000 realizations of the control samples.}

These results have several important implications. 
First, the lack of a sharp drop in the mass distribution of CL-AGNs around the value of $10^{8}\,\Msun$, as seen in tidal disruption event (TDE) rates \cite[e.g.,][]{vV18,Yao23}, disfavors TDEs as the main driving mechanism of CL-AGNs, contrary to suggestions in some studies (see \citealt{Merloni15} for a specific case and detailed discussion).
The drop in TDE rates for $\mbh\gtrsim10^{8}\,\Msun$ is commonly interpreted in the context of basic TDE physics; a $1\,\Msun$ main-sequence star can be disrupted outside of the BH horizon (and produce an observable TDE) only if $\mbh < 10^{8}\,\Msun$ (for a nonspinning BH), or $\mbh < 7\times 10^{8}\,\Msun$ (for a maximally spinning BH; see \citealt{Gezari21} and \citealt{Rossi21_SSRv} for recent reviews). 
Among our sample of CL-AGNs, 57 ($\sim$50\%) and 12 ($\sim$10\%) sources exceed these two limiting BH masses, respectively.

Another implication relates to the possible link between the extreme variability observed in CL-AGNs and the more commonly seen stochastic optical variability seen in the general AGN population \cite[see][and references therein]{VB04}. One possibility is that the extreme and apparently coherent changes observed in CL-AGN (and extremely variable AGNs in general) are simply the rarest cases at the tail of the stochastic AGN variability distribution. 
Testing this idea directly using (currently available) SDSS-V data is well beyond the scope of the present work, however we note that the (final) SDSS-V dataset would be ideal for this kind of analysis.
The source of typical AGN variability remains uncertain, and may be attributed to accretion-disk instabilities, Poisson processes, and/or gravitational microlensing. Irrespective of the underlying mechanism, essentially all relevant studies have found the variability amplitude to be anticorrelated with luminosity \cite[e.g.,][]{VB04, Macleod10, Simm16, Caplar17,Arevalo23}. Some of these studies have also investigated a possible link between variability amplitude and Eddington ratio, finding conflicting results \cite[e.g.,][]{Simm16, Caplar17,Arevalo23}. Our work, however, demonstrates that when controlling for the Eddington ratio, CL-AGNs do not have a significantly lower luminosity as compared to the AGN population. 

\subsection{Comparing the Variability of Different Emission Lines}
\label{subsec:line_comparison}

\begin{figure}
\includegraphics[width=0.5\textwidth, clip, trim={0 1.7cm 0 0}]{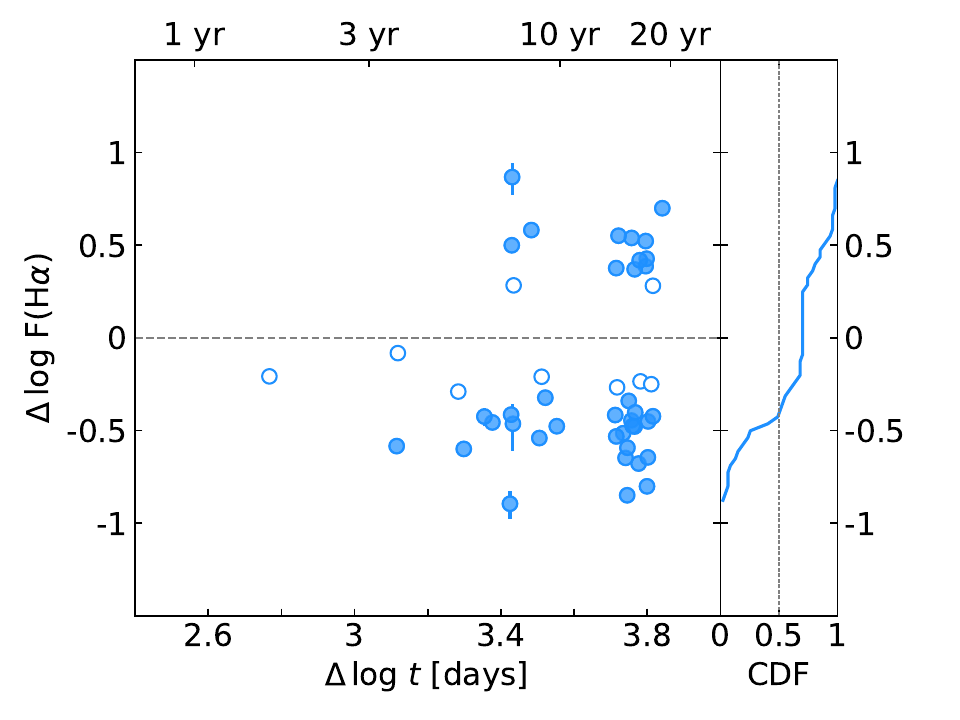}\hfill
\includegraphics[width=0.5\textwidth, clip, trim={0 1.7cm 0 1.19cm}]{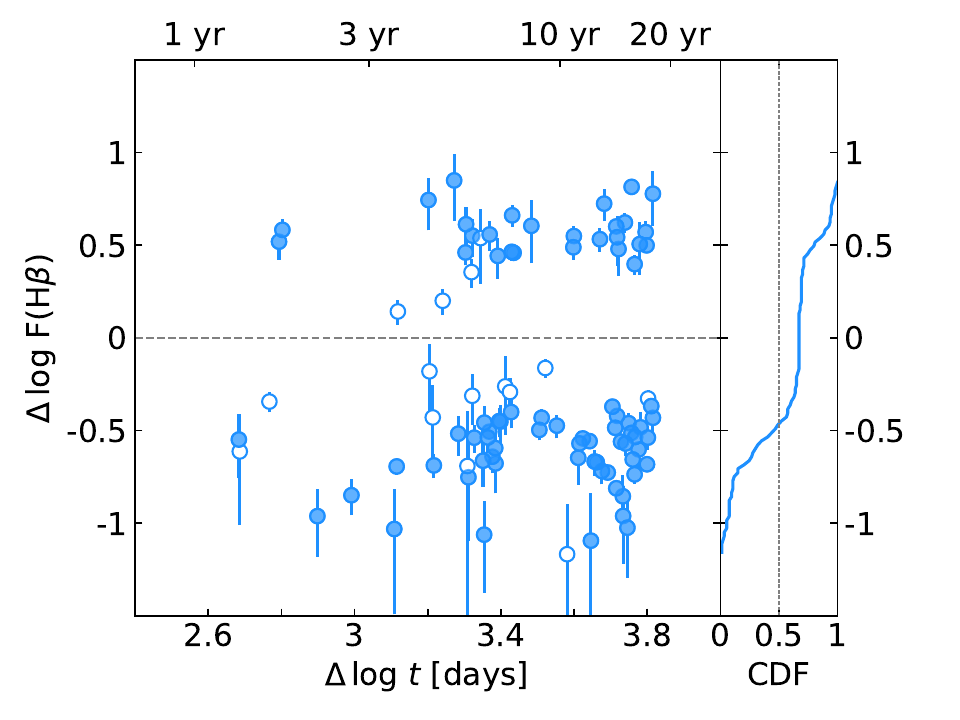}\hfill
\includegraphics[width=0.5\textwidth, clip, trim={0 0 0 1.17cm}]{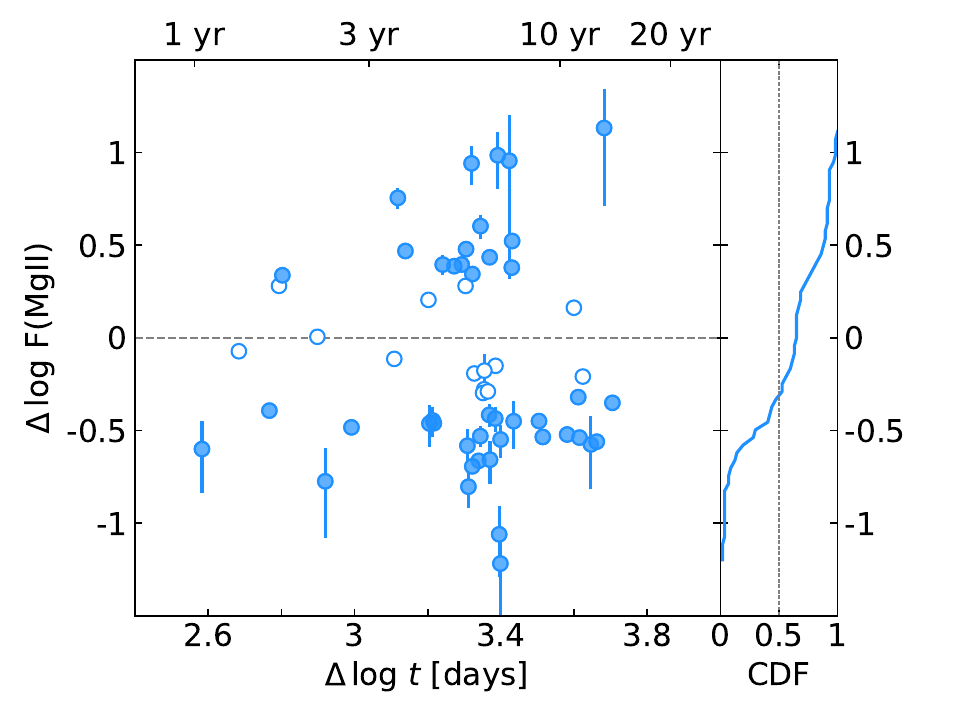}
\caption{The distributions of line flux changes between the SDSS-V and DR16 epochs for the \ha\ (top), \hb\ (middle), and \mgii\ (bottom) broad emission lines, and the corresponding rest-frame time differences, for our CL-AGN sample. Objects that were selected through the relevant emission line in each panel are shown as filled circles, while objects that were selected through one of the other emission lines are represented by empty circles. Displayed to the right of each panel is the corresponding CDF of the emission line flux ratio for all the CL-AGNs in each panel.}
\label{fig:df_vs_dt}
\end{figure}

We now examine how various broad emission lines have changed in our CL-AGNs, as generally not all lines are expected to vary by the same amount between any two spectral epochs.
This analysis relies on our two-epoch spectral decomposition approach, which provides spectral measurements for the `dim' and `bright' states of each of the CL-AGNs in our sample (see Section~\ref{subsec:decomposition}).

The three panels in Figure \ref{fig:df_vs_dt} present the distributions of line flux changes between the SDSS-V and DR16 epochs for the \ha, \hb, and \mgii\ broad emission lines, and the corresponding (rest-frame) time differences for our CL-AGN sample. Each panel displays the objects that were selected through changes in the corresponding emission line (filled circles; see Section \ref{subsec:spec_search}) and those that were selected through one of the other emission lines (empty circles).
To the right of each panel, we show the cumulative distribution functions (CDF) of the emission-line flux ratios for all the CL-AGNs in that panel. As can be seen, the median $\Delta\log F\left(\rm line\right)$ in all panels are negative, indicating that a majority of our candidates are dimming or ``turn-off'' events. This outcome aligns with our expectations, given that our data primarily rely on repeated spectroscopy of sources that were previously identified as 
broad-line AGNs in archival SDSS data (see Section~\ref{subsec:final_sample}).

\begin{figure*}
\includegraphics[width=0.32\textwidth,clip,trim={0.4cm 0.5cm 0.4cm 0.5cm}]{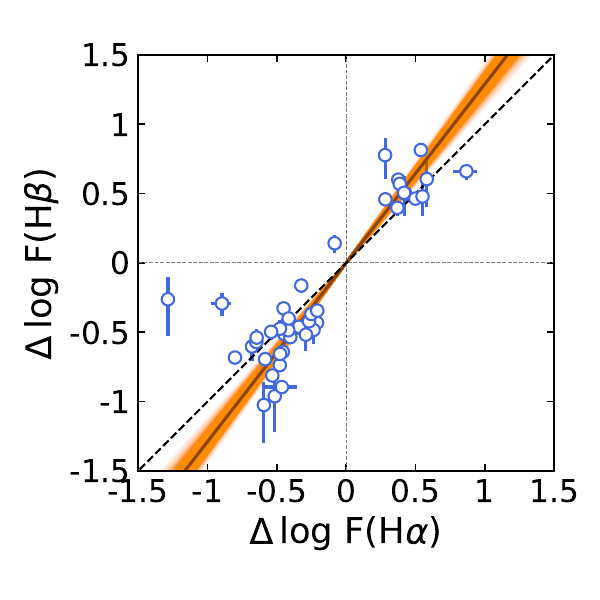}\hfill
\includegraphics[width=0.32\textwidth,clip,trim={0.4cm 0.5cm 0.4cm 0.5cm}]{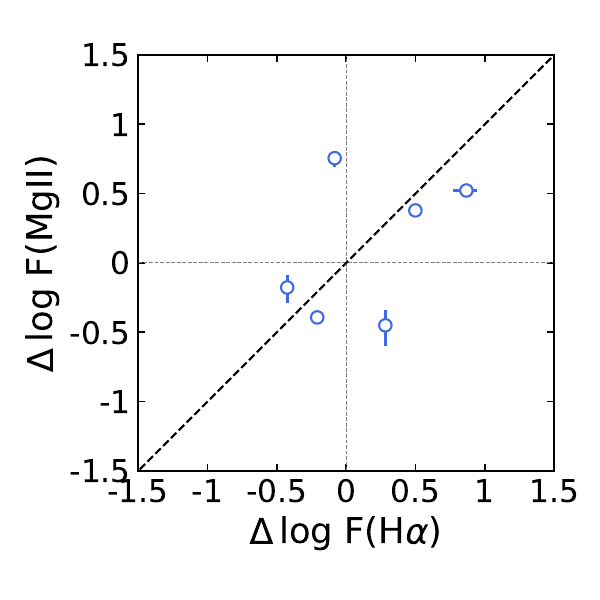}\hfill
\includegraphics[width=0.32\textwidth,clip,trim={0.4cm 0.5cm 0.4cm 0.5cm}]{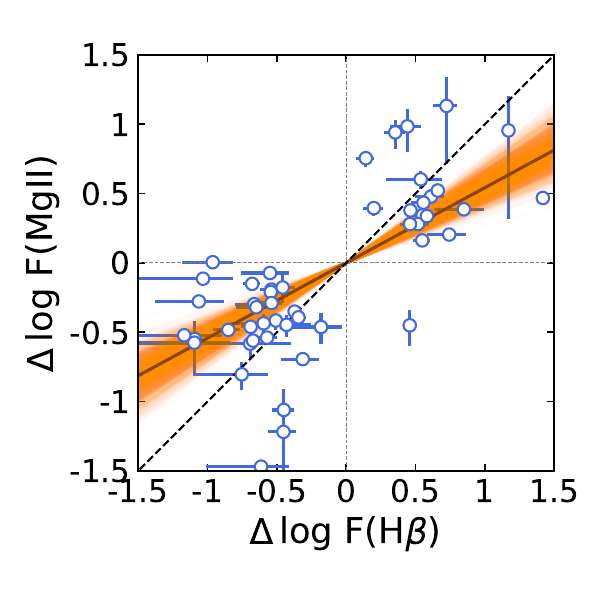}
\caption{A comparison of the flux ratios between the SDSS-V and DR16 epochs for different emission lines within our CL-AGN sample: \ha\ vs \hb\ (left), \ha\ vs \mgii\ (middle), and \hb\ vs \mgii\ (right). The dashed lines represent the hypothetical scenario in which the broad emission lines would change at the same rate (i.e., a 1:1 relation), while the solid lines denote the best power-law fits to the data in each panel, where the intercept is manually set to zero during the fitting procedure. The orange regions surrounding the solid lines indicate the associated uncertainty for each fit, estimated using the pair bootstrap method.}
\label{fig:df1_vs_df2}
\end{figure*}

The three panels in Figure~\ref{fig:df1_vs_df2} compare the ratios of various emission-line fluxes, between the SDSS-V and earlier (DR16) epochs for our sample of CL-AGNs.\footnote{Since our final core sample includes only three CL-AGNs where \civ\ is observed, and two \civ-selected CL-AGNs, we chose to exclude this line from this analysis.} The dashed lines represent the hypothetical scenario in which the broad emission lines would change at the same rate (i.e., a 1:1 relation), while the solid lines denote the best power-law fits to the data in each panel.\footnote{The fits were obtained through a standard orthogonal distance regression.} The uncertainties of each fit were estimated using the pair bootstrap method.
For the fits shown in Figure~\ref{fig:df1_vs_df2}, we have set the intercept to zero. This choice minimizes the bias introduced by our CL-AGN sample having more dimming events compared to brightening ones (67\% and 33\% of the CL-AGN core sample, respectively), which is reflected in the way the first and third quadrants in Figure~\ref{fig:df1_vs_df2} are populated. Moreover, a nonzero intercept would suggest that some CL-AGNs occupy peculiar regions in the two-line-changes phase space, where one emission line strengthens while the other one weakens, including in an apparently self-inconsistent way and irregardless of whether the CL-AGN is dimming or brightening. 
Such a scenario seems physically implausible. 
As can be seen in all panels of Figure~\ref{fig:df1_vs_df2}, almost all objects exhibit highly correlated line variability, in the sense that various emission lines either brighten or dim simultaneously.

The left panel, which presents the flux variability of \ha\ vs. that of \hb, shows the tightest correlation among the three pairs of lines we examine, which is expected given that both lines originate from the same species (and gas). 
All but one of the 46 CL-AGNs included in that panel (98\%) are found in the 1st and 3rd quadrants, i.e. both lines brighten or dim synchronously (a Spearman correlation test yields $\rho_{\rm S}=0.79$).
There is, however, a significant deviation from the 1:1 relation, as the best power-law fit suggests that \hb\ exhibits greater variation than \ha.
We first note that 32 out of the 46 CL-AGNs in that panel ($\approx$70\%) have $| \Delta\log(\hb)| > |\Delta\log(\ha)|$.
Moreover, the best-fit slope we obtain for the correlation of $\Delta\log(\hb)$ vs. $\Delta\log(\ha)$ is $1.29ֿ\pm0.05$, which suggests a significant deviation from the 1:1 relation
.

The trend of more variability in \hb\ than in \ha\ is consistent with what is expected from a variable obscuration scenario, in which shorter wavelengths would be affected more than longer ones \cite[see, e.g.,][and references therein]{RAR17,Zeltyn22}. We discuss this possibility further in Section~\ref{subsec:acc_vs_obsc} below.

The middle panel, which presents the flux variability of \ha\ vs. that of \mgii, is sparsely populated, as the (redshift-dependent) spectral coverage of most of our candidates does not allow simultaneous measurement of these two emission lines. Consequently, we refrain from fitting these data and drawing any conclusions.

The right panel presents the flux variability of \hb\ vs. \mgii, where the two lines also appear to vary in a qualitatively synchronous manner, with all but one of the 51 CL-AGNs in this panel (98\%) located in the 1st and 3rd quadrants ($\rho_{\rm S}=0.70$).
There is again a significant deviation from a purely linear (1:1) relation, with \hb\ typically exhibiting significantly more variation than \mgii. 
Of the 51 CL-AGNs in this panel, 37 (73\%) have  $| \Delta\log(\hb)| > |\Delta\log(\mgii)|$.
The best-fitting slope for these data is determined to be $0.54^{+0.08}_{-0.07}$, a significant deviation from the 1:1 relation.

This result is consistent with previous works which showed that the \mgii\ emission line is less variable compared to the Balmer emission lines \cite[e.g.,][]{Cackett15, Sun15,Homayouni20}. 
As discussed in these works (and references therein), there are several possible factors contributing to this observed trend, including the specific excitation mechanisms, optical depth, and the BLR geometry of the \mgii\ emission line, or---alternatively---some other, not-yet-understood aspect of BLR physics \cite[e.g.][]{Sun15}. 
In addition, this trend, unlike the \hb\ vs. \ha\ trend seen in the left panel of Figure~\ref{fig:df1_vs_df2}, is inconsistent with variable obscuration driving (most of) the CL-AGN events in our sample, as in this scenario we would expect the bluer \mgii\ line to show higher variability than \hb\ (assuming they are emitted from roughly the same region within the BLR; see, e.g., \citealt{Homayouni20}).
It is important to note, however, that our data display considerable scatter around the best-fit line, with some objects showing perfectly correlated (1:1) variability, while other objects show greater variability in \mgii\ than in \hb.

In any case, the limited variability in \mgii\ relative to \hb\ (right panel of Figure \ref{fig:df1_vs_df2}) underscores the limitations facing any search for CL-AGNs at $1\lesssim z \lesssim 2$ (including our own). Investigating CL-AGNs in this redshift region is further complicated by the inability to use (strong) narrow emission lines to overcome potential spectral calibration issues, which presents a considerable challenge when searching for CL-AGNs at high $z$ in general.

\subsection{Accretion Versus Obscuration}
\label{subsec:acc_vs_obsc}

To date, most studies of CL-AGNs identified through (rest-frame) UV-optical spectroscopy have favored the hypothesis that fluctuations in the ionizing continuum, caused by changes to the accretion flow, are the (primary) driver of the CL-AGN phenomena, as opposed to attributing them to variable obscuration, which is indeed the favorable explanation for most of the CL-AGNs identified in the X-ray regime \cite[see, e.g.,][and references therein]{RT23}. This view has led to the usage of the term ``changing-state'' AGN to describe the former type of CL-AGN events.
One of the key lines of evidence in favor of this explanation is the coincident and correlated IR variability observed in several CL-AGNs \cite[e.g.,][]{Sheng17, Stern18, Yang18, Wang19}. Within the widely accepted picture of AGN structure, the mid-IR emission is not expected to be substantially influenced by variable obscuration, however it is expected to respond to (dramatic) UV variability from the central engine. 
Mid-IR light curves could thus serve as a relatively efficient, if incomplete, way to separate changing accretion from changing obscuration events in AGNs.

To examine the mechanism behind the CL-AGNs in our sample, we acquired WISE light curves for all of our candidates. 
Out of the total 113 candidates, 46 exhibited variability in the WISE light curve that was consistent with the observed optical changes, whereby a significant IR brightening (dimming) occurred within the period between the archival SDSS and the new, brighter (dimmer) SDSS-V spectra, while 14 other CL-AGNs displayed no such concurrent IR variations. 
The example WISE light curves in Figure~\ref{fig:wise_examples} (in Appendix~\ref{app:wise}) demonstrate these two scenarios.
For the remaining 53 candidates, the light curves were either inconclusive or had missing data points during the optical transition.

This simple analysis suggests that the majority of CL-AGN are likely linked to variations in the accretion flow, in agreement with the prevalent interpretation in the literature, although there may also be a considerable fraction of CL-AGNs driven by variable obscuration. 
We note in particular that one of the CL-AGNs in our sample (J1628+4329) indeed shows substantial evidence for an obscuration-driven, transient changing-look event, as discussed in detail in \cite{Zeltyn22}.
The absence of observed IR variability in some of our CL-AGNs, however, does not necessarily point to variable obscuration being the source of variability for these sources. 
To directly determine the source of variability for (each of) the sources identified following the approach presented here (and in similar studies), we would ideally need to conduct a coordinated multiwavelength campaign, including X-ray observations \cite[e.g.,][]{LaMassa15,Ruan19,Temple23,Yang23}. This effort is a challenging one due to the extensive data volume of SDSS-V. Alternatively, a careful and detailed spectral analysis may hint at the source of variability \cite[e.g.,][]{Ruan14,Zeltyn22}; such an endeavor, however, is beyond the scope of this work.

\section{Conclusions}
\label{sec:conclusions}

This study used the data accumulated during the first year of SDSS-V spectroscopic observations to identify AGNs that experienced dramatic variations to their broad emission-line strength, and to investigate their nature.
The main outcomes of this study are as follows:

\begin{enumerate}
    
    \item We have constructed a high-purity sample totaling 116 CL-AGNs (113 in our core sample and three from BHM-RM), the largest such sample to date, and 88 other dramatic emission-line variations, drawn from a survey that is (partially) designed to identify such events directly through wide-field, repeated spectroscopy. 
    See Sections~\ref{subsec:final_sample} and \ref{subsec:additional_sources}, Table~\ref{tab:cands}, and Figure~\ref{fig:all_spec}.
    
    \item In our experience, the selection of CL-AGNs and other (dramatic) spectral variations is far from straightforward, given the need to quantify dimmed spectra with limited $S/N$, and other limitations. 
    Additional spectroscopic observations and cross-match to (pseudo-)concurrent photometric data are important. 
    See Sections~\ref{subsubsec:followup} and \ref{subsubsec:crossmatch}, and Figures~\ref{fig:selection_examples} and \ref{fig:selection_examples_phot} in Appendix~\ref{app:selection}.
    
    \item 
    Our core sample of 113 CL-AGNs covers the redshift range of $0.06<z<2.4$. The SDSS-V spectroscopy of this core sample probes spectral transition timescales of 1 yr $\lesssim \Delta t_{\rm rest} \lesssim$ 19 yr, with yet faster transitions revealed by our additional spectroscopy (as short as $\Delta t_{\rm rest} \approx 2$ months; see \citealt{Zeltyn22}). Of the 113 systems in our core sample, 67\% (76 systems) show dimming in their recent SDSS-V spectroscopy compared to previous data, while 33\% (37 systems) show brightening.
    See Sections~\ref{subsec:sample_props} and \ref{subsec:line_comparison}, and Figures~\ref{fig:z_L_dist} and \ref{fig:df_vs_dt}.

    \item 
    Roughly $0.4\%$ of the quasars reobserved in the first year of SDSS-V operations, and $1.25\%$ among those at $z<1$, showed CL-AGN--like behavior over a timescale of $<20$ yr. We stress that this is a lower limit on the intrinsic occurrence rate of such phenomena, as SDSS-V observations are designed to focus on previously known quasars and miss inactive galaxies in which quasar-like features have recently appeared.
    See Section~\ref{subsec:sample_props} and Figure~\ref{fig:rates}.
    
    \item 
    Our sample of CL-AGNs has lower Eddington ratios compared with a ($z$ and \Lbol- or \mbh-matched) control sample. The corresponding median values for the subset of our CL-AGNs where this comparison is most robust are $\fedd\approx0.025$ vs. $0.043$ (i.e., a factor of $>1.7$ lower). 
    This finding is in agreement with previous studies, which were based on much smaller spectroscopic samples. 
    See Section~\ref{subsec:prop_dist} and Figure~\ref{fig:histograms_ws22}, and Figure~\ref{fig:histograms_all} in Appendix ~\ref{app:alter_dists}.
    
    \item 
    The available WISE IR data suggest that most---but not all---of the sources in our sample are driven by changes to the radiation that emerges from their accretion flows. This is based on the identification of IR flux changes that  (qualitatively) agree with those seen in optical spectroscopy (or lack thereof).
    A more detailed investigation of this issue would require detailed analysis of the optical spectra and of responsive, multiwavelength follow-up observations.
    See Section~\ref{subsec:acc_vs_obsc} and Figure~\ref{fig:wise_examples}.
    
\end{enumerate}

We stress that the present study is the first sample study of CL-AGNs (and related phenomena) based on SDSS-V data.
We envision several near-future pathways for expanding the study of such systems.
The ongoing SDSS-V/BHM program provides a growing number of repeated spectroscopy of (previously known) AGNs, thus enabling us to further enlarge the sample of CL and extremely variable AGNs and to better understand how such systems are related to more subtle and typical AGN variability.
Obtaining responsive, multiwavelength targeted observations, particularly for CL-AGNs identified in the higher-cadence SDSS-V observations, would allow us to constrain the transition timescales and to further test ideas regarding the  mechanisms that drive these phenomena (i.e. accretion vs. obscuration).
Finally, studying the host galaxies of CL-AGNs through dedicated observations (e.g., understanding the stellar populations in host-dominated spectra of recently dimmed sources) could not only provide more insights regarding CL-AGNs, but indeed be utilized to study the AGN--host connection in general.

The SDSS-V survey design and progress guarantee that yet larger and richer samples of extremely variable AGNs, and specifically CL-AGNs, will become available in the near future for in-depth analyses.

\begin{acknowledgments}

We thank the anonymous referee and statistics editor for their insightful and constructive comments, which helped us improve the paper.
We thank Jamie Burke, Joseph Farah, Daichi Hiramatsu, Estefania Padilla Gonzalez, Craig Pellegrino, Megan Newsome, and Giacomo Terreran, for their assistance with the acquisition and reduction of LCOGT data.
We acknowledge support from:
the European Research Council (ERC) under the European Union's Horizon 2020 research and innovation program (grant agreement number 950533; G.Z., B.T., M.\'S); 
the Israel Science Foundation (grant number 1849/19; G.Z., B.T., M.\'S);
FONDECYT Regular \#123171 (R.J.A.), \#1200495 (F.E.B.), and \#1230345 (C.R.); 
ANID BASAL project FB210003 (R.J.A., F.E.B., C.R.);
ANID Millennium Science Initiative ICN12\_009 (F.E.B., L.H.G.);
DLR grant FKZ 50 OR 2307 (M.K.);
NSF grants AST-1715579, AST-2009947 (Y.S.), and AST-2206499 (X.L.); 
FONDECYT Postdoctorado \#3220516 (M.J.T.); 
FONDECYT Iniciaci\'on \#11241477 (L.H.G.);
NSFC grants 12025303, 12393814, and 11890693 (Y.Q.X.);
Millenium Nucleus NCN19-058 TITANs (M.L.M.-A.);
NSF grant AST-2106990 and the Penn State Eberly Endowment (W.N.B).

Funding for the Sloan Digital Sky Survey V has been provided by the Alfred P. Sloan Foundation, the Heising-Simons Foundation, the National Science Foundation, and the Participating Institutions. SDSS acknowledges support and resources from the Center for High-Performance Computing at the University of Utah. SDSS telescopes are located at Apache Point Observatory, funded by the Astrophysical Research Consortium and operated by New Mexico State University, and at Las Campanas Observatory, operated by the Carnegie Institution for Science. The SDSS web site is \url{www.sdss.org}.

SDSS is managed by the Astrophysical Research Consortium for the Participating Institutions of the SDSS Collaboration, including Caltech, The Carnegie Institution for Science, Chilean National Time Allocation Committee (CNTAC) ratified researchers, The Flatiron Institute, the Gotham Participation Group, Harvard University, Heidelberg University, The Johns Hopkins University, L'Ecole polytechnique f\'{e}d\'{e}rale de Lausanne (EPFL), Leibniz-Institut f\"{u}r Astrophysik Potsdam (AIP), Max-Planck-Institut f\"{u}r Astronomie (MPIA Heidelberg), Max-Planck-Institut f\"{u}r Extraterrestrische Physik (MPE), Nanjing University, National Astronomical Observatories of China (NAOC), New Mexico State University, The Ohio State University, Pennsylvania State University, Smithsonian Astrophysical Observatory, Space Telescope Science Institute (STScI), the Stellar Astrophysics Participation Group, Universidad Nacional Aut\'{o}noma de M\'{e}xico, University of Arizona, University of Colorado Boulder, University of Illinois at Urbana-Champaign, University of Toronto, University of Utah, University of Virginia, Yale University, and Yunnan University.



The ZTF forced-photometry service was funded under the Heising-Simons Foundation grant No. 12540303 (PI: Graham). The Hobby--Eberly Telescope (HET) is a joint project of the University of Texas at Austin, the Pennsylvania State University, Ludwig-Maximillians-Universitaet Muenchen, and Georg-August Universitaet Goettingen. The HET is named in honor of its principal benefactors, William P. Hobby and Robert E. Eberly.

\end{acknowledgments}

\facilities{Sloan (SDSS and BOSS), ING:Herschel (ISIS), LCOGT (FTN: FLOYDS), ARC (KOSMOS), HET (LRS-2).}

\software{{\tt AstroPy} \citep{Astropy13,Astropy18,Astropy22},
{\tt Matplotlib} \citep{Matplotlib07}, 
{\tt NumPy} \citep{NumPy20}, {\tt SciPy} \citep{SciPy20}, {\tt PyQSOFit} \citep{QSOFit, Shen19}, {\tt floyds\_pipeline}, {\tt PypeIt} \citep{Prochaska20_pypeit}, {\tt Panacea}}

\clearpage
\bibliography{sdssv_clagns}{}
\bibliographystyle{aasjournal}

\begin{appendix}

\section{Follow-up Observations}
\label{app:follow-up}

Table \ref{tab:all_followups} lists the follow-up spectroscopic observations carried out for each target, as well as the spectral setups used in each observation.
The follow-up spectra were obtained and reduced with a variety of facilities and corresponding pipelines.
Spectra obtained with the LCOGT/FLOYDS instruments were reduced using a custom version of the \texttt{floyds\_pipeline} PyRAF-based pipeline, developed by the LCOGT SN team.\footnote{ \url{https://github.com/LCOGT/floyds_pipeline}}
Spectra obtained with the HET/LRS2 instrument were reduced using the \texttt{Panacea} pipeline.\footnote{\url{https://github.com/grzeimann/Panacea}}
Spectra obtained with the Hale/DBSP instrument were reduced using the \texttt{PypeIt} spectral-reduction platform (versions 1.4.1 and 1.5.0).\footnote{\url{https://github.com/pypeit/PypeIt}}

\begin{deluxetable}{lcc}[h!]
\tablecaption{Follow-up Observations}
\label{tab:all_followups}
\tablewidth{\textwidth}
\tabletypesize{\scriptsize}
\tablehead{
\colhead{Name} & \colhead{Facility} & \colhead{MJD} 
\
}
\startdata
J000715.14$+$260413.4 & LCOGT & 59517 \\
J001153.87$+$015912.2 & LCOGT & 59495 \\
J004149.62$+$243050.5 & LCOGT & 59514 \\
J004459.09$-$010629.3 & LCOGT & 59495 \\
J011536.11$+$003352.4 & LCOGT & 59501 \\
J012946.72$+$150457.3 & LCOGT & 59515 \\
 & LCOGT & 59541 \\
 & LCOGT & 59587 \\
 & LCOGT & 59888 \\
 & LCOGT & 59915 \\
J015854.50$+$001307.3 & LCOGT & 59593 \\
 & HET & 59927 \\
J020428.59$-$031013.7 & LCOGT & 59501 \\
 & LCOGT & 59932 \\
J020649.48$-$041452.7 & LCOGT & 59501 \\
J021359.08$-$025352.8 & LCOGT & 59514 \\
 & LCOGT & 59560 \\
\enddata
\tablecomments{Table~\ref{tab:all_followups} is published in its entirety in the machine-readable format. A portion is shown here for guidance regarding its form and content.}
\end{deluxetable}

\clearpage

\section{Demonstrating Changing-look Active Galactic Nuclei Selection and Rejection}
\label{app:selection}

\begin{figure}
\centering
\includegraphics[width=0.475\textwidth]{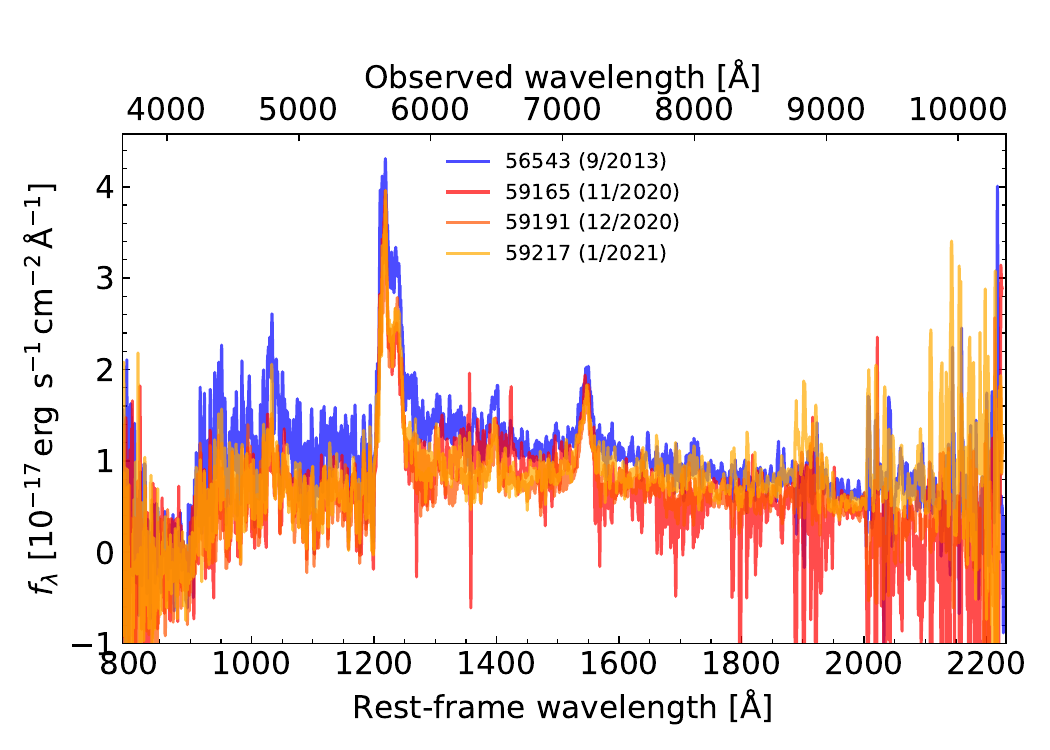}
\includegraphics[width=0.475\textwidth]{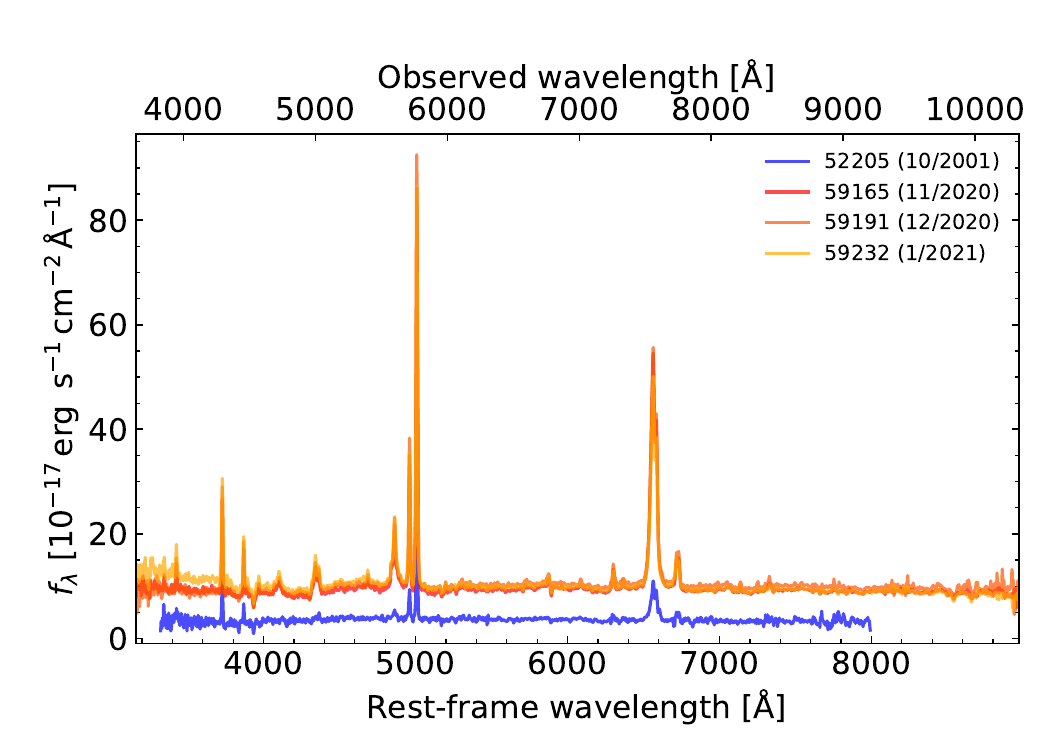}
\includegraphics[width=0.475\textwidth]{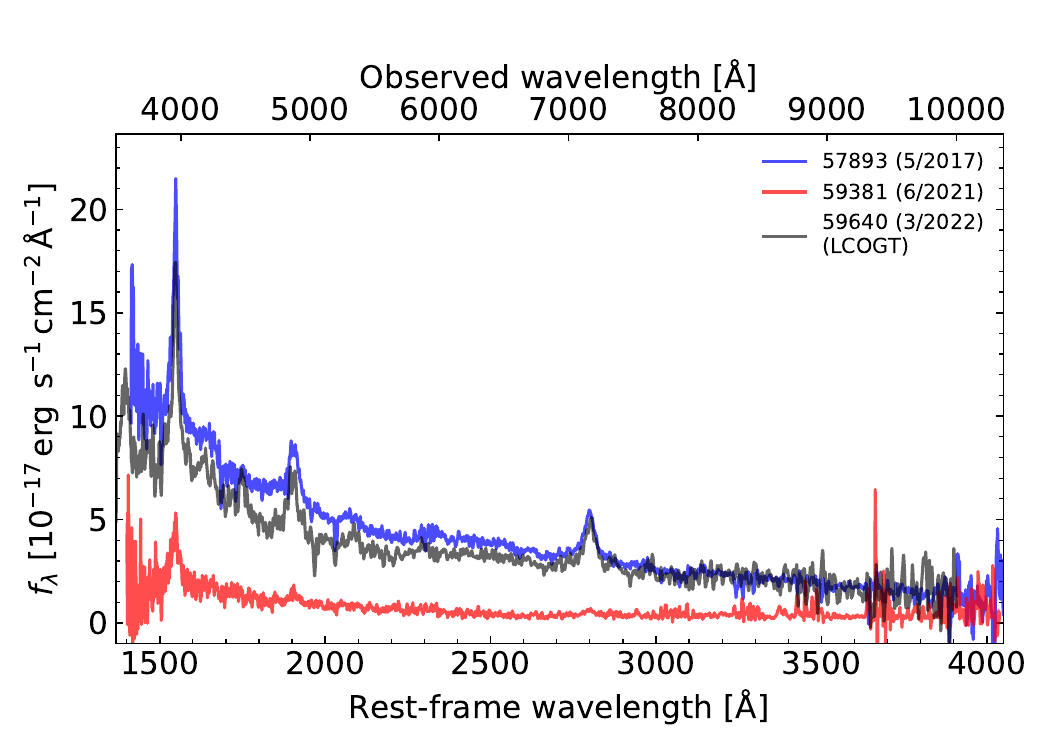}
\includegraphics[width=0.475\textwidth]{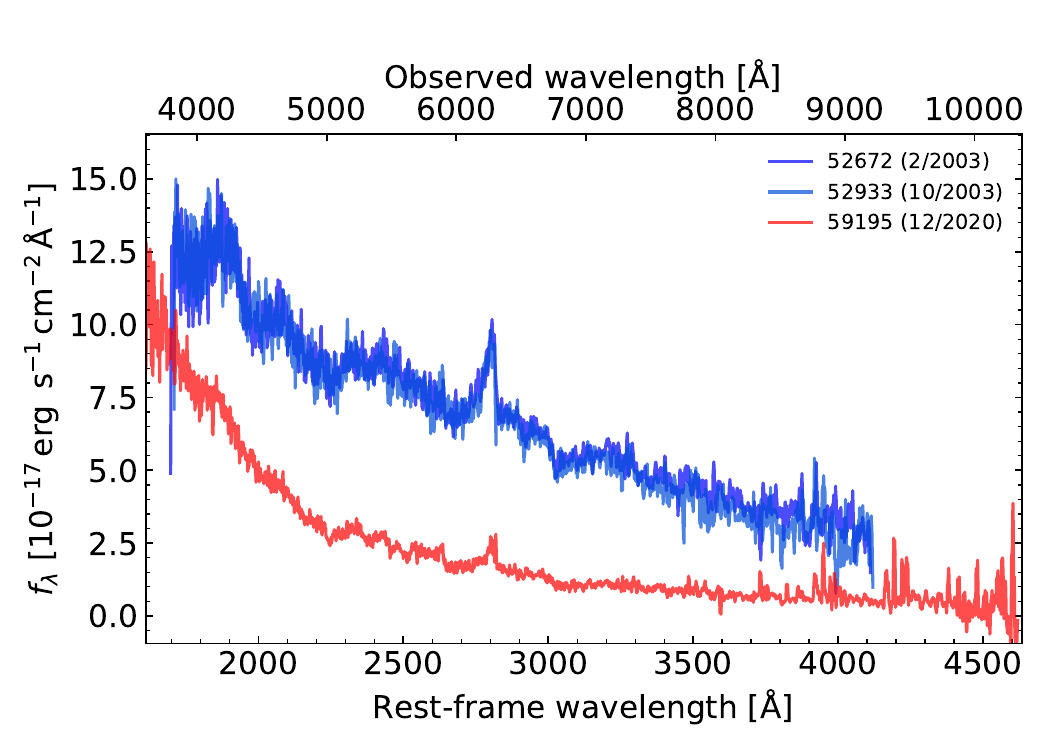}
\includegraphics[width=0.475\textwidth]{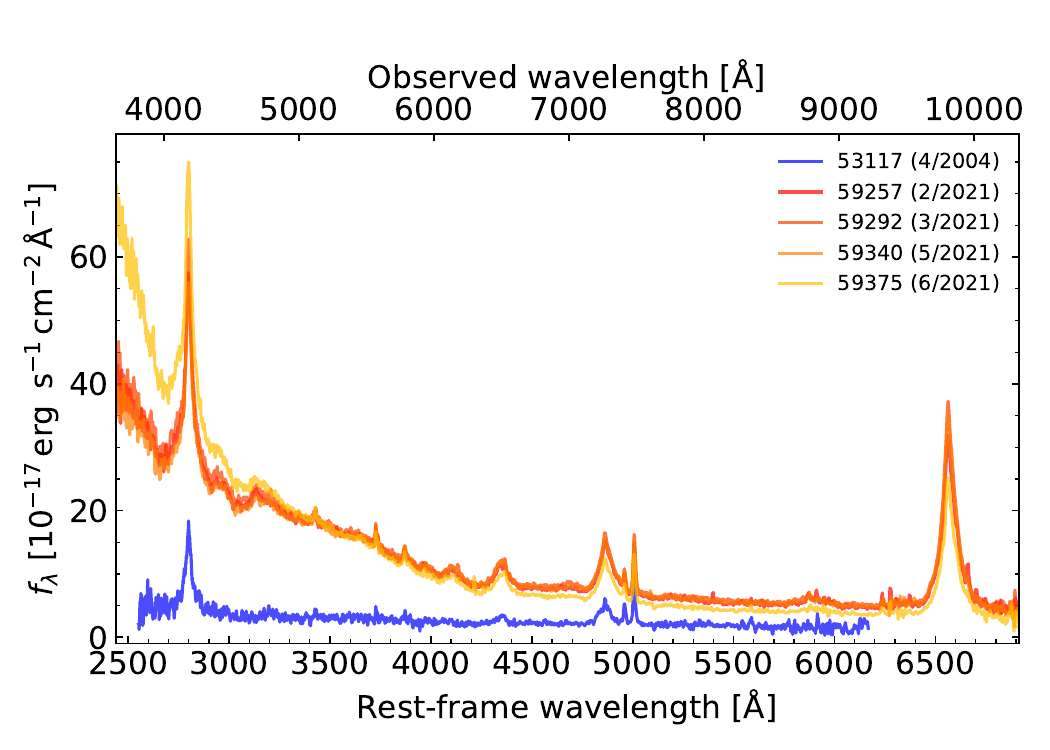}
\includegraphics[width=0.475\textwidth]{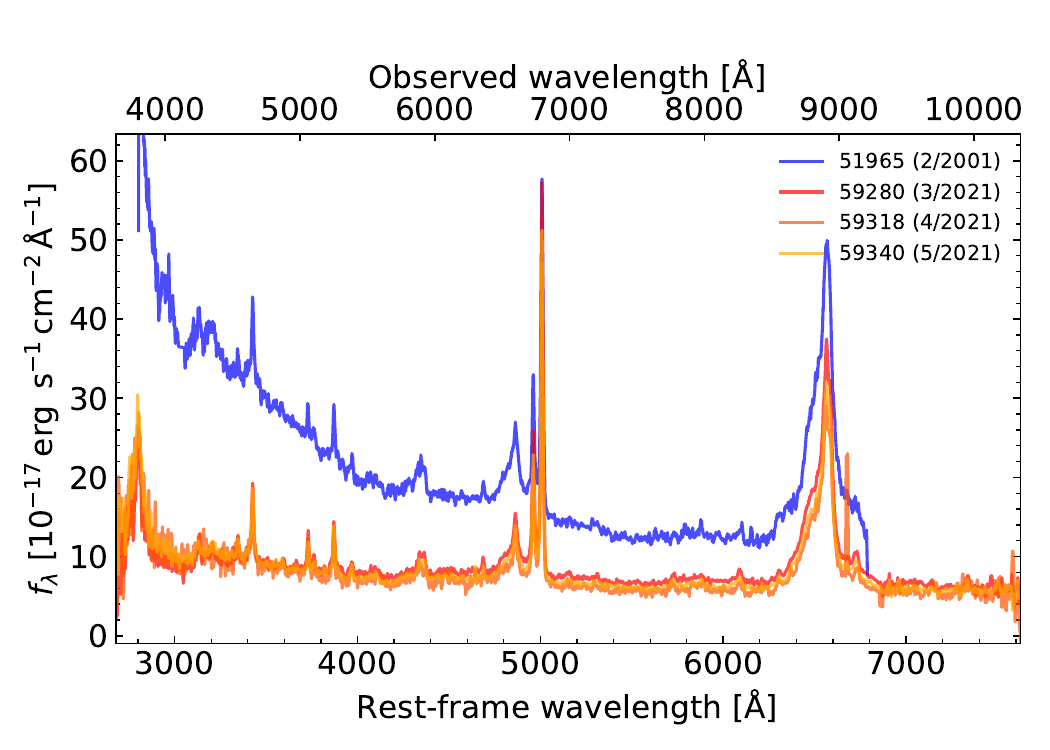}
\caption{Multi-epoch spectra demonstrating several steps and considerations in the process of CL-AGN selection. All the spectra were boxcar-smoothed over 7 pixels. All the sources shown here have passed our automated search criterion (Equation~\ref{eq:C_cut}), but did not enter our final core sample. See Appendix~\ref{app:selection} for details on each of the cases shown here.}
\label{fig:selection_examples}
\end{figure}

Figure~\ref{fig:selection_examples} shows the multi-epoch spectroscopy of several sources which demonstrate the importance of some of the considerations we took into account when constructing and analyzing our CL-AGN sample. Specifically, we show the following examples of a few objects that were flagged as CL-AGN candidates by our automated search, but that did not enter our final CL-AGN sample due to our visual inspection (VI), a calibration issue, or other reasons (as indicated below). 
In all panels, archival SDSS (DR16), SDSS-V, and our own follow-up spectra are shown in blue-cyan, red-orange-yellow, and black, respectively.

\paragraph{J000228.32$+$252744.7 -- low data quality (VI; $z=3.652071$; Figure~\ref{fig:selection_examples} - top-left)} This object was flagged as a CL-AGN candidate by our automated search procedure. However, upon a visual inspection of the spectra in hand, it became clear that there is no substantial spectral variability (i.e., changing-look behavior) in this object. Instead, this object was flagged following an incorrect estimation of the \ciii\ line flux change due to the significant noise in the relevant spectral region.

\paragraph{J022913.33$+$003430.8 -- calibration issue (VI; $z=0.151696$; Figure~\ref{fig:selection_examples} - top-right)} This object was flagged as a CL-AGN candidate by our automated search procedure. However, the visual inspection revealed that the \oiii\ narrow emission lines' strength varies substantially between the two epochs, pointing to a calibration issue in one of the spectra.

\paragraph{J160541.43$+$464554.1 -- calibration issue (follow-up; $z=1.558838$; Figure~\ref{fig:selection_examples} - middle-left)} In this case, the selection of this candidate by our automated search procedure can be attributed to a miscalibration of the SDSS-V spectrum. The follow-up LCOGT spectrum we obtained perfectly matches the earlier, archival SDSS spectrum. We note that we could not have identified this calibration issue through VI, given the lack of narrow emission lines in the spectral range covered by SDSS-V (as expected for a $z=1.558838$ object). 

\begin{figure*}[h]
\centering
\includegraphics[width=0.475\textwidth]{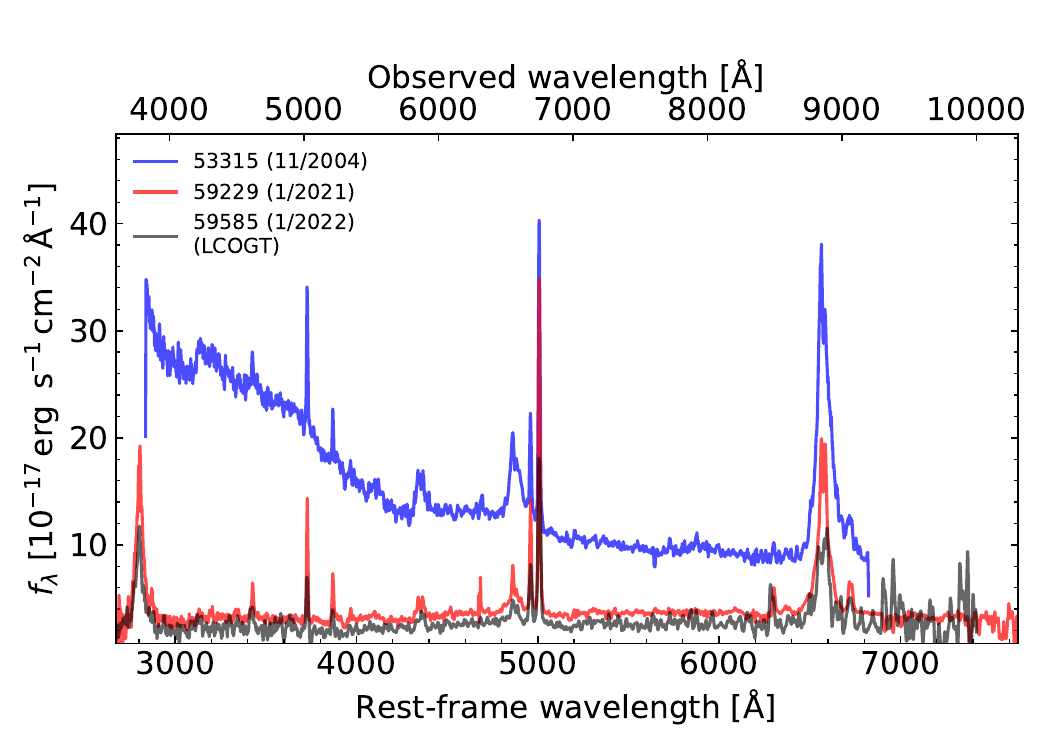}
\includegraphics[width=0.475\textwidth]{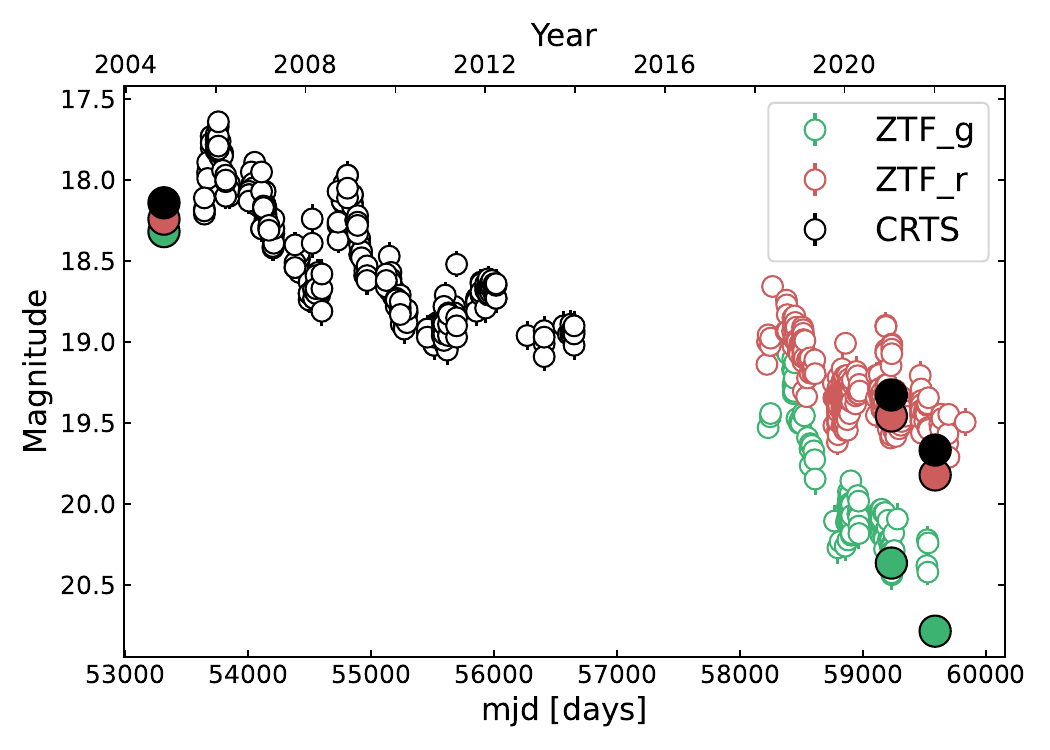}
\includegraphics[width=0.475\textwidth]{figs/4555074934_phot_spec.pdf}
\includegraphics[width=0.475\textwidth]{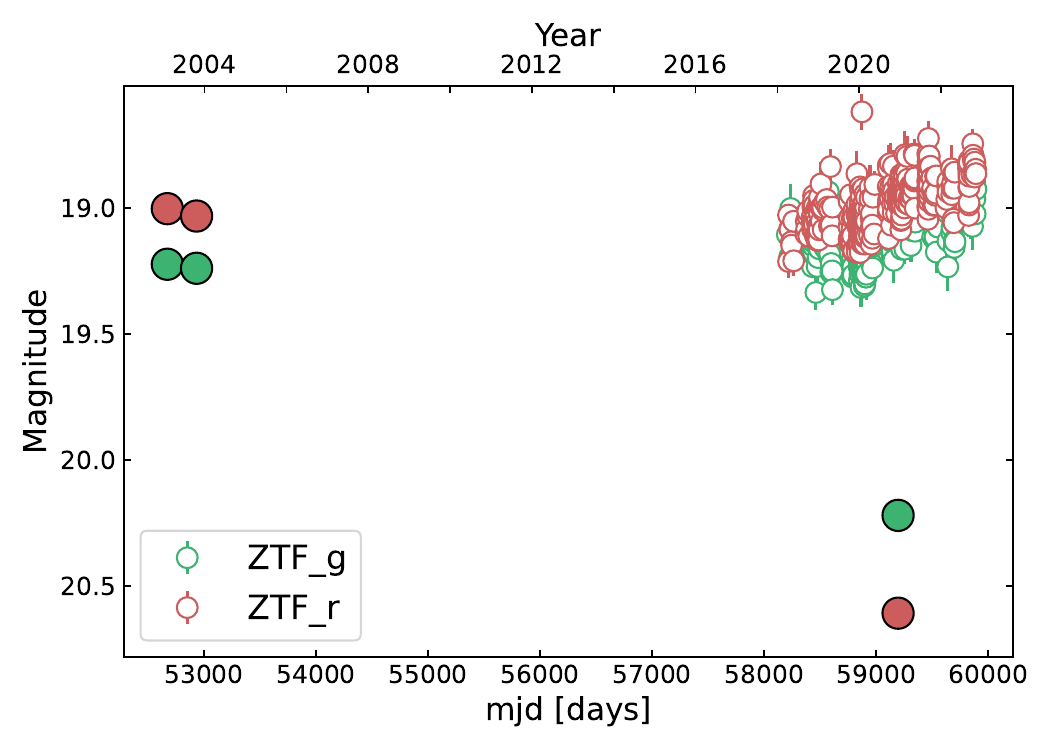}
\caption{SDSS spectra (left panels) and the corresponding optical light curves (right panels) demonstrating how these were used to either help confirm (top panels) or refute (bottom panels) CL-AGN candidates. 
All the spectra were boxcar-smoothed over 7 pixels.
In both panels, empty symbols represent photometric measurements obtained from various time-domain surveys (see legend), while filled symbols represent synthetic photometry obtained from SDSS spectra, using the respective filter curves.
See Appendix~\ref{app:selection} for details.}
\label{fig:selection_examples_phot}
\end{figure*}

\paragraph{J083050.94$+$295059.6 -- calibration issue (ZTF; $z=1.236$; Figure~\ref{fig:selection_examples} - middle-right)} Here, too, the selection of this candidate by our automated search procedure can be attributed to a miscalibration of the SDSS-V spectrum. 
The synthetic photometry obtained using the SDSS-V spectrum falls significantly below the actual ZTF light curve of this source (during the corresponding period; light curve shown in the bottom-right panel of  Figure~\ref{fig:selection_examples_phot}). 
Here, too, our VI could not have identified this calibration issue, given the lack of narrow emission lines (in this case, $z=1.236$). 

\paragraph{J121345.97$+$484419.4 -- EVQ ($z=0.495215$; Figure~\ref{fig:selection_examples} - bottom-left)} This object was flagged by our automated search for CL-AGNs and, based on a VI of narrow \oiii\ lines seems to indeed show an extreme brightening in both the broad-line and continuum emission. However, as all the (prominent) broad emission lines are clearly seen in the dim state of this object, we classify this object as an EVQ rather than a CL-AGN. 

\paragraph{J093149.46$+$014333.5 -- EVQ ($z=0.356283$; Figure~\ref{fig:selection_examples} - bottom-right)} Similar to the aforementioned case of J121345.97+484419.4, this object shows a genuine extreme spectral variability event, but in this case the line and continuum emission are dimming. Since all the broad lines are still clearly seen in the dimmest spectrum in hand, it is classified as an EVQ rather than a CL-AGN.

\clearpage

\clearpage

Figure~\ref{fig:selection_examples_phot} shows the spectra of two sources (left panels) along with the corresponding optical light curves (right panels), demonstrating how light curves were utilized to select or reject CL-AGN candidates.

\clearpage

\section{Examples of Changing-look Active Galactic Nuclei WISE light curves}
\label{app:wise}

\begin{figure*}[h]
\centering
\includegraphics[width=0.475\textwidth,clip,trim={0.4cm 0.5cm 0.4cm 1cm}]{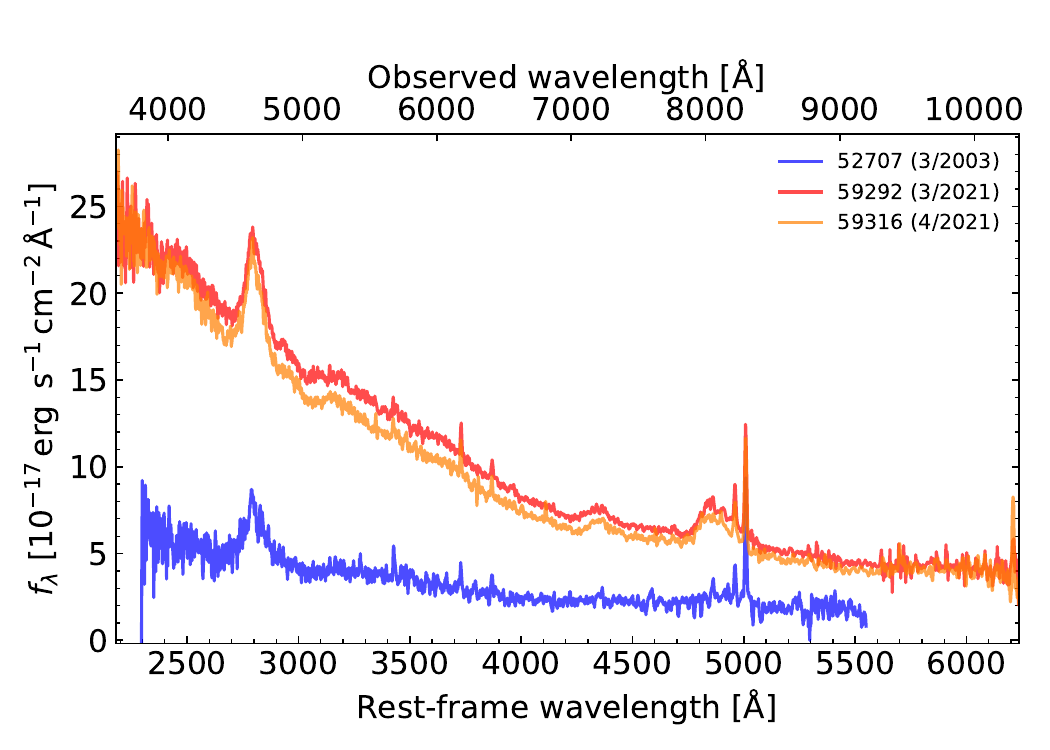}\hfill
\includegraphics[width=0.475\textwidth]{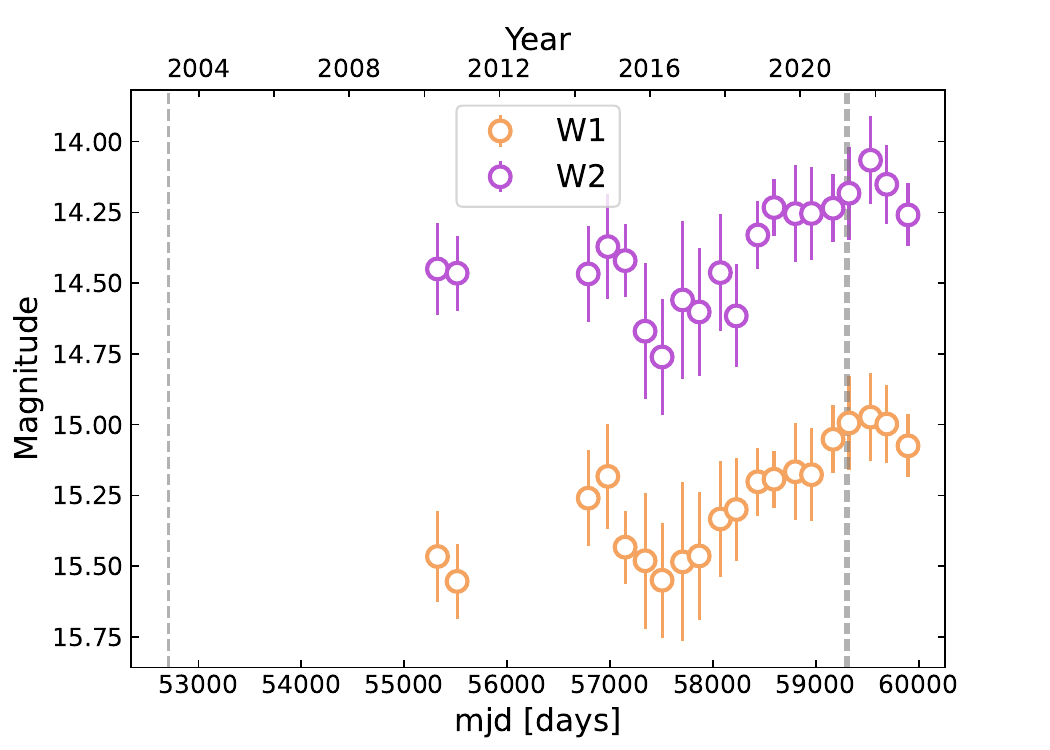}
\includegraphics[width=0.475\textwidth,clip,trim={0.4cm 0.5cm 0.4cm 1cm}]{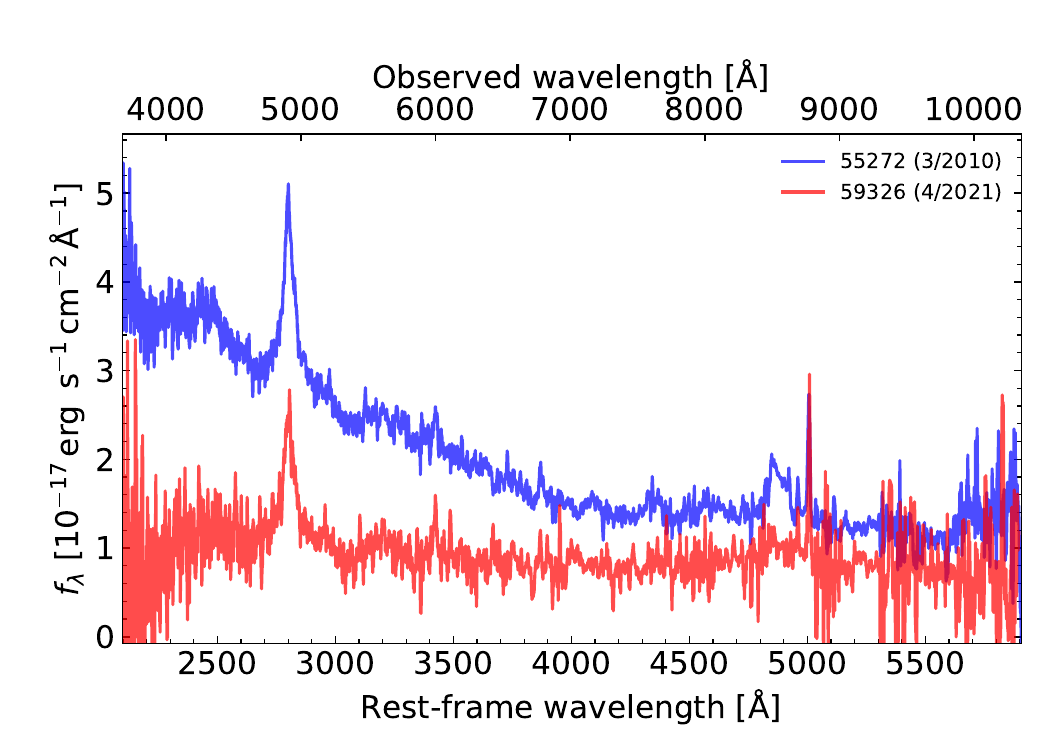}\hfill
\includegraphics[width=0.475\textwidth]{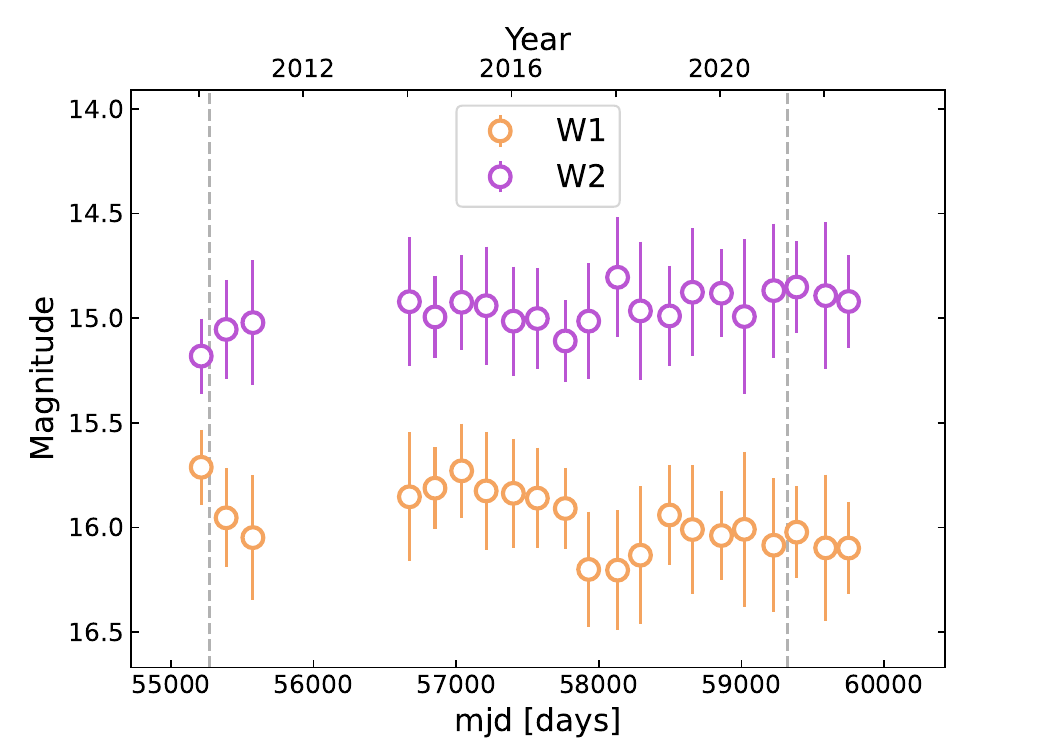}
\caption{
Examples of CL-AGN spectra (left panels) and their corresponding ALLWISE and NEOWISE light curves (right panels). 
The spectra panels show the archival DR16 spectrum of each CL-AGN (blue), and the more recent SDSS-V spectra (red-orange-yellow). All the spectra were boxcar-smoothed over 7 pixels.
The light-curve panels show all the available WISE data in the W1 and W2 bands, as well as the corresponding times at which the archival SDSS and the more recent SDSS-V spectra were taken (marked as vertical dashed gray lines).
}
\label{fig:wise_examples}
\end{figure*}

Figure~\ref{fig:wise_examples} shows the multi-epoch spectra of two CL-AGNs from our core sample (left panels) along with their corresponding WISE-based, IR light curves (right panels). 
The spectral panels include both the archival DR16 spectrum of each CL-AGN (blue), and the more recent SDSS-V spectra (red-orange-yellow). 
Each IR light-curve panel shows all the available WISE data in the W1 and W2 bands (orange and purple, respectively), as well as the epochs during which the optical spectra were obtained (vertical dashed lines).

\paragraph{J092313.53+043445.0
($z=0.65682$; Figure~\ref{fig:wise_examples} - top panels)} 
The spectra (top left) show that this CL-AGN brightened during the period 2003--2021. The WISE light curves (top-right panel) show a brightening in the IR flux from this CL-AGN during the period 2016--2021. 
These data suggest that this CL-AGN transition was driven by variations in the accretion flow.

\paragraph{J144024.60+345624.2
($z=0.752449$; Figure~\ref{fig:wise_examples} - bottom panels)} 
The spectra (bottom left) show that the continuum and \hb\ line emission in source dimmed during the period 2010--2021. However, the corresponding WISE light curves (bottom right) do not reflect this trend, showing an essentially constant IR flux over the same period.
While the lack of (extreme) variability in the WISE data may have been interpreted as supporting an obscuration-driven CL transition, we note that the bluer \mgii\ line has not dimmed considerably, as one may expect from (dusty) gas intervening our line of sight. 
A more detailed analysis is required in order to elucidate the nature of such CL-AGNs.

\clearpage

\section{Notes on individual sources}
\label{app:individual}

Here, we briefly discuss several noteworthy systems in our final sample. 

\begin{figure}
\centering

\includegraphics[width=0.475\textwidth]{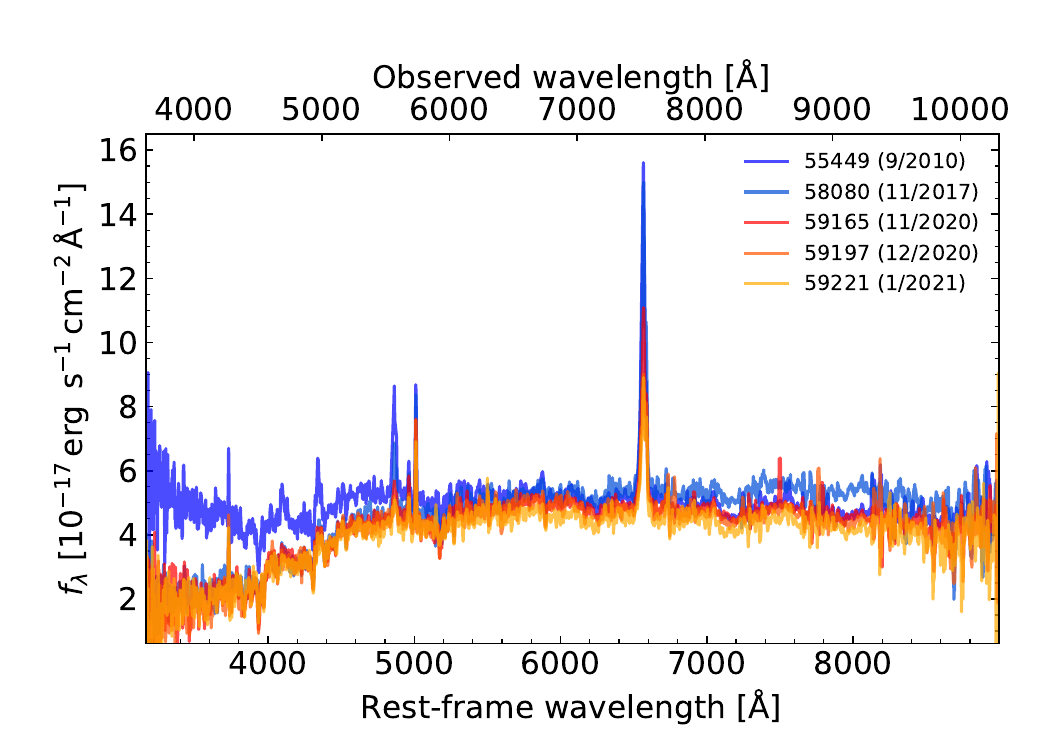}
\includegraphics[width=0.475\textwidth]{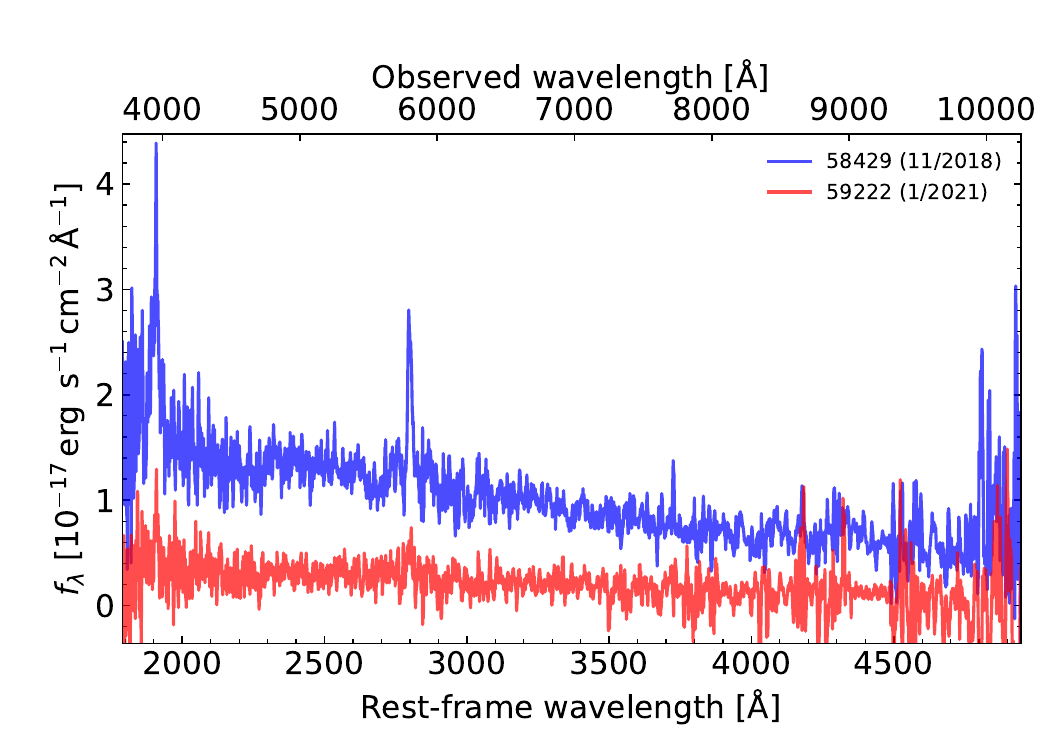}
\includegraphics[width=0.475\textwidth]{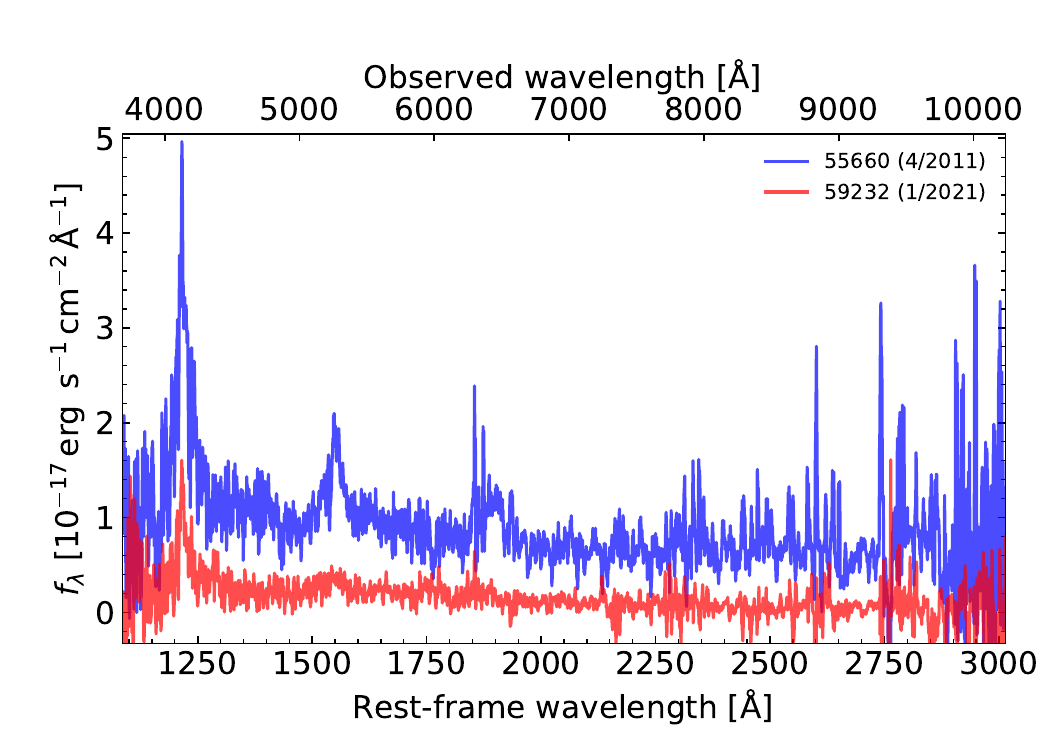}
\includegraphics[width=0.475\textwidth]{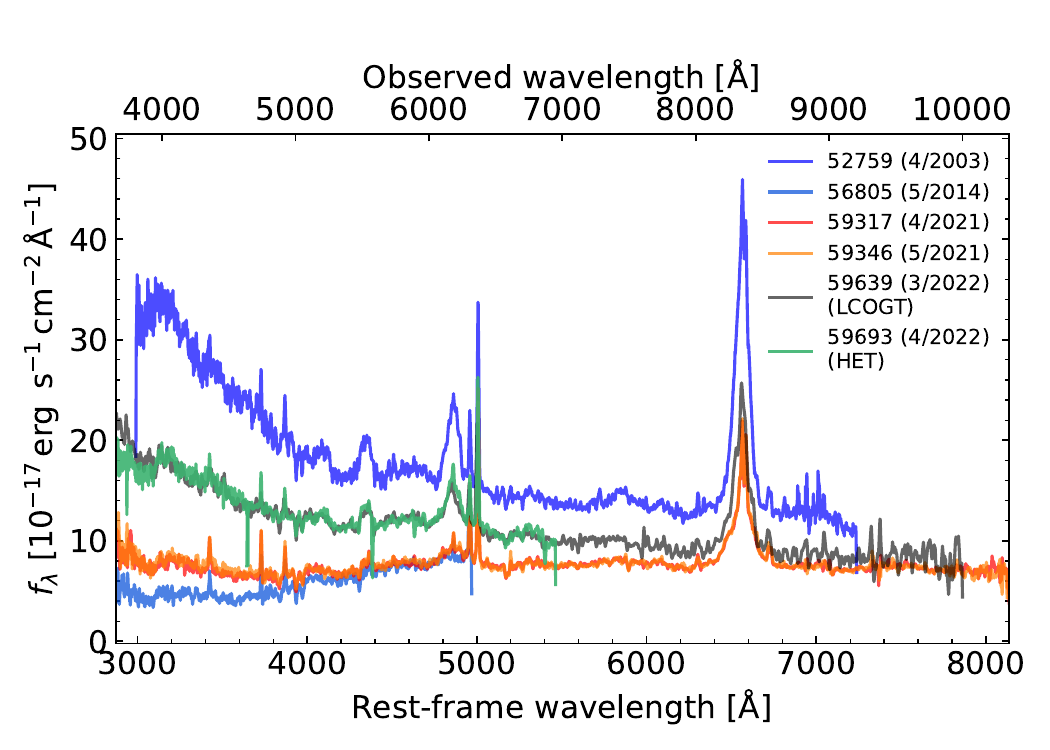}
\includegraphics[width=0.475\textwidth]{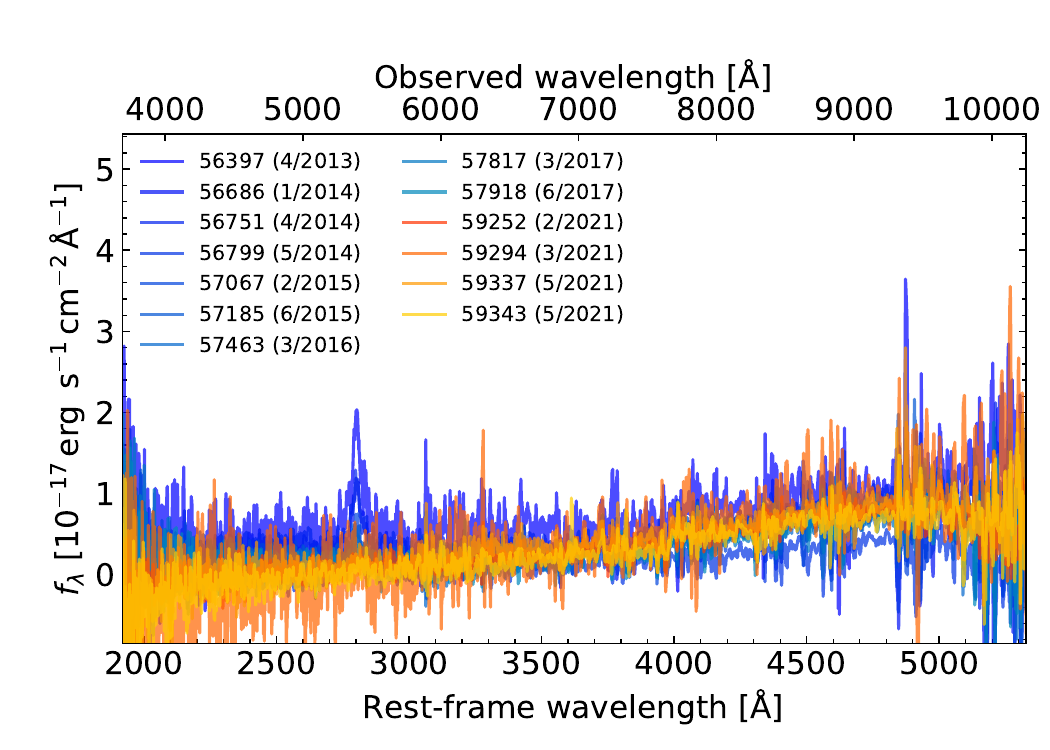}
\includegraphics[width=0.475\textwidth]{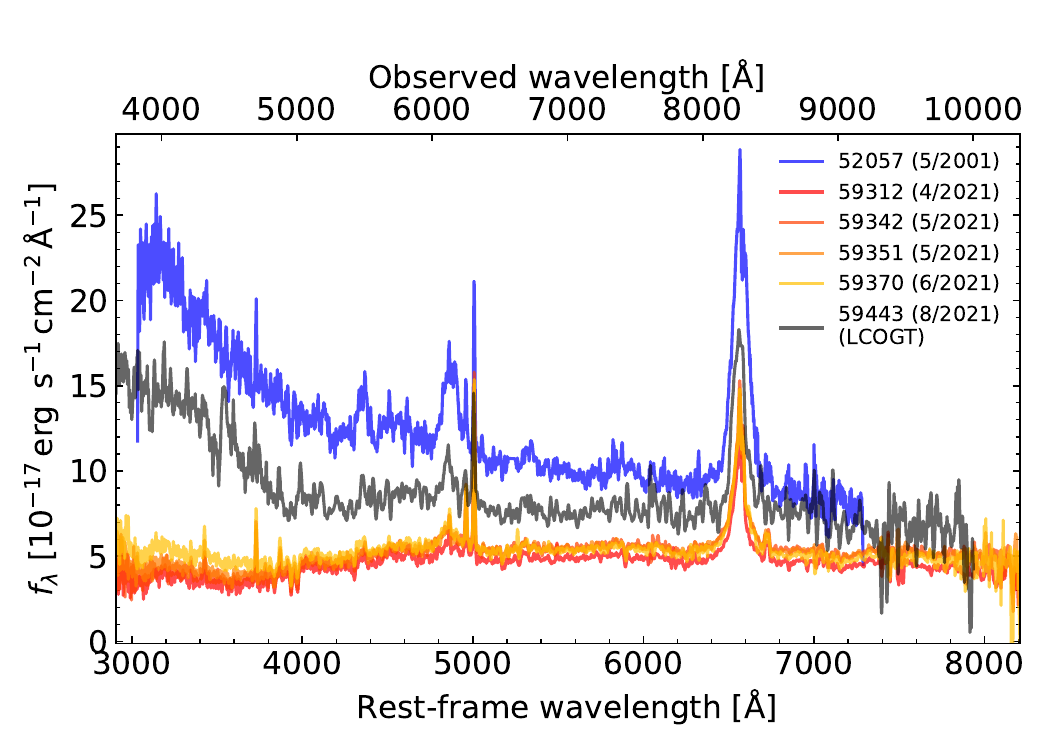}
\caption{Multi-epoch optical spectroscopy of all the CL-AGNs identified in this work. All the spectra were boxcar-smoothed over 7 pixels. Archival DR16 SDSS spectra are shown in blue-cyan, while SDSS-V spectra are shown in red-orange-yellow (early to late). Some panels include additional follow-up spectroscopy (see legend in each panel for details). The entire figure set is published online. A portion is shown here for guidance regarding its form and content.}
\label{fig:all_spec}
\end{figure}

Figure~\ref{fig:all_spec} shows the multi-epoch spectra of all the CL-AGNs identified in this work (full figure set is available online). In all panels, archival DR16 SDSS spectra are shown in blue-cyan, and SDSS-V spectra are shown in red-orange-yellow. Some panels include additional follow-up spectroscopy (see detailed description for each object below).

\paragraph{J015854.50+001307.3 ($z=0.145004$; Figure~\ref{fig:all_spec} - upper-left)}
This dimming CL-AGN exhibits a successive dimming of the various spectral components: in the period 2010--2017, the blue quasar-like continuum disappeared, the broad \hb\ line had somewhat dimmed, while the broad \ha\ did not change significantly. Then, in the period 2017--2020, \hb\ dimmed further, and \ha\ had started to dim, too. This delayed transition where the continuum dims first and is only later followed by the dimming of the broad emission lines is expected in the case of a drastic change in the accretion flow, albeit on somewhat long timescales \cite[see, e.g.,][for a similar example of a ``turn-on'' CL-AGN with such delayed spectral changes]{Trakhtenbrot19}.

\paragraph{J083414.18+270101.6 ($z=1.071886$; Figure~\ref{fig:all_spec} - upper-right)} 
This CL-AGN exhibits one of the fastest transitions we have observed. The dimming of this source occurred on a rest-frame timescale of $\Delta t_{\rm{rest}} \leq 383$ days.

\paragraph{J093041.65+011842.1 ($z=2.39625$; Figure~\ref{fig:all_spec} - middle-left)}
This is the highest-redshift system in our sample. It exhibits extreme spectral variations in both \civ\ and \ciii.

\paragraph{J132457.29+480241.3 ($z=0.271847$; Figure~\ref{fig:all_spec} - middle-right)} 
This CL-AGN had already transitioned, from a brighter quasar state to a dimmer state that has no visible broad \hb, during the period 2003--2014, and was already noted in \citet{MacLeod16, MacLeod19} and \citet{Hutsemekers19}. However, SDSS-V spectra taken in 2021 reveal the reappearance of a (weak) broad \hb\ and a weak quasar-like continuum. Subsequent LCOGT and HET spectra taken during 2022 show the source has kept rebrightening, reaching a type 1 quasar-like state.

\paragraph{J141144.12+531508.6 ($z=0.923187$; Figure~\ref{fig:all_spec} - bottom-left)} 
This is one of the three CL-AGNs we discovered in BHM-RM data. This source shows the gradual disappearance of the \mgii\ line over the period 2013--2021. For clarity, we show here only a subset of the 111 spectra of this object taken across this period.

\paragraph{J162829.17+432948.5 ($z = 0.2603$; Figure~\ref{fig:all_spec} - bottom-right)}
This source was presented and discussed in detail in \cite{Zeltyn22}. Its earliest SDSS-V spectra (2020--2021) showed it to be in a much dimmer state than archival SDSS-I spectroscopy (2001), with the continuum becoming host dominated and the broad \hb\ line significantly dimming. It then showed a remarkable ``recovery'' within $\sim$2 months, after which it regained its bright quasar-like appearance (as corroborated by the ZTF light curve). 
The detailed analysis presented in \cite{Zeltyn22} led to the claim that these dramatic events may be best described by variable obscuration.

\clearpage

\section{Alternative Distributions of Changing-look Active Galactic Nuclei Properties}
\label{app:alter_dists}

\begin{figure*}
  \centering
  \raisebox{70pt}{\parbox[b]{.1\textwidth}
  {\large $\Lbol\,\&\,z$ matched}}\hfill
  \includegraphics[width=.40\textwidth,trim={0 0 16.5cm 4cm}, clip]{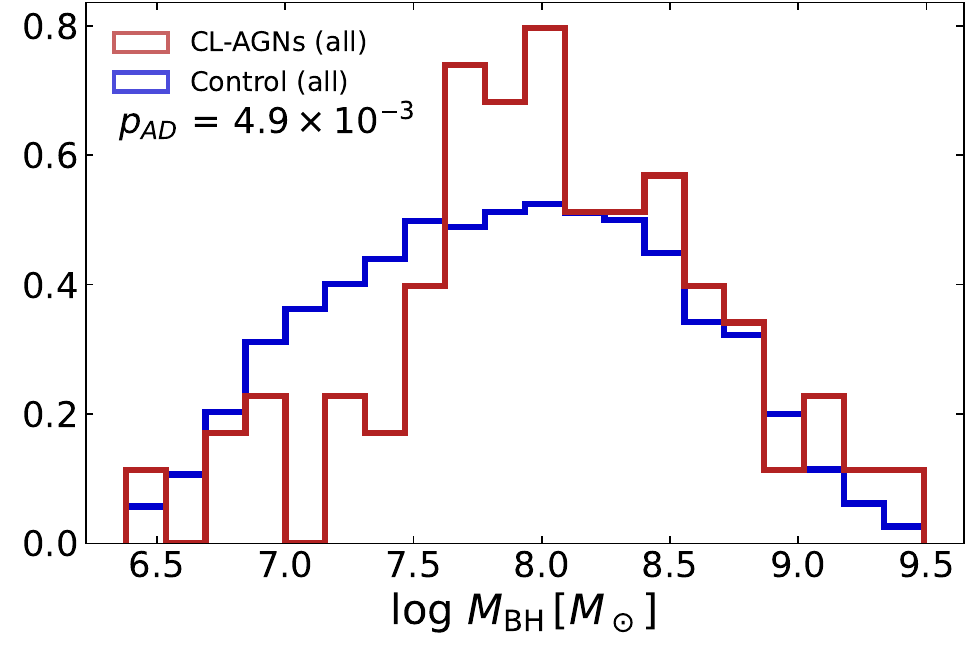}\hfill
  \includegraphics[width=.40\textwidth, trim={0 0 16.5cm 4cm}, clip]{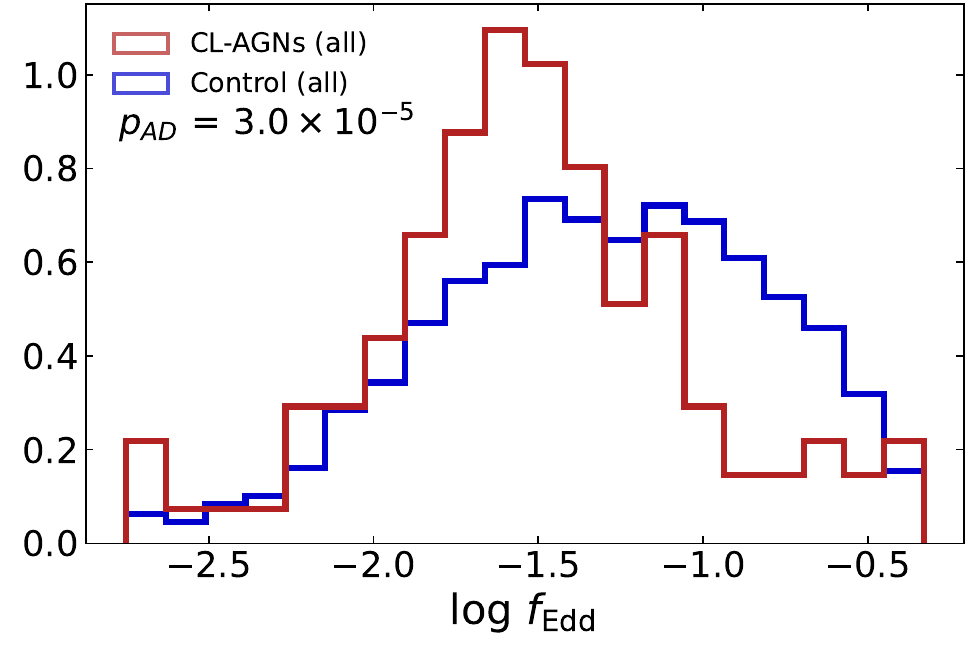}\\
  \raisebox{70pt}{\parbox[b]{.1\textwidth}
  {\large $\mbh\,\&\,z$ matched}}\hfill
  \includegraphics[width=.40\textwidth, trim={0 0 16.5cm 4cm}, clip]{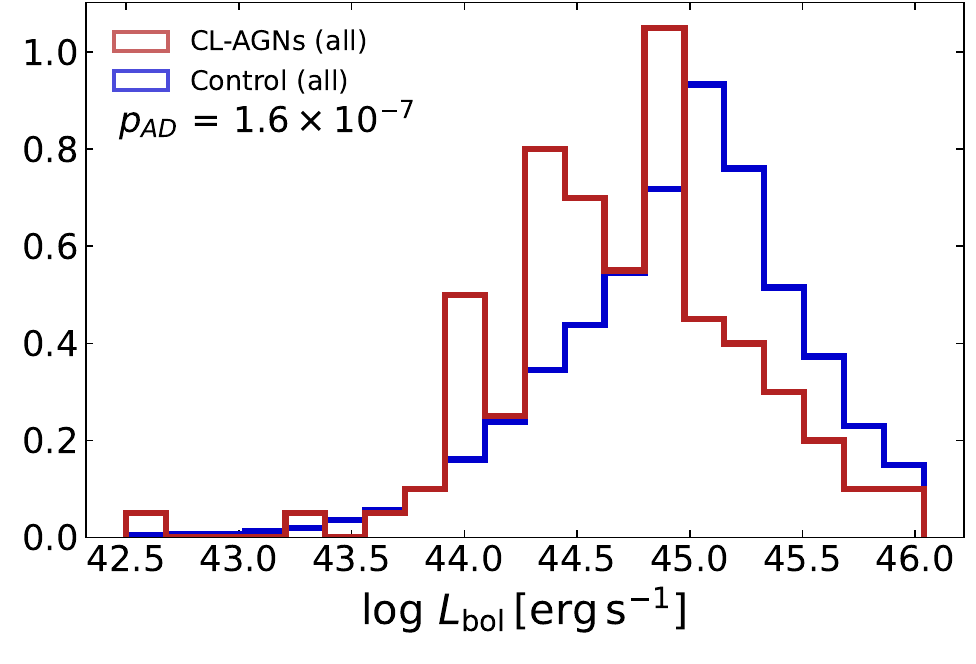}\hfill
  \includegraphics[width=.40\textwidth, trim={0 0 16.5cm 4cm}, clip]{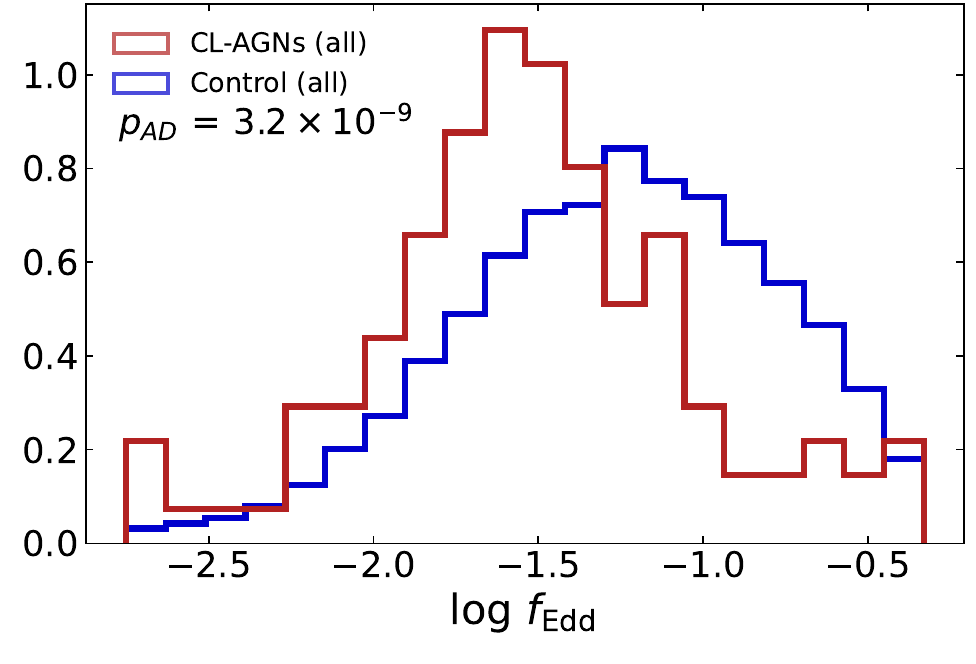}\\
  \raisebox{70pt}{\parbox[b]{.1\textwidth}{\large $\fedd\,\&\,z$ matched}}\hfill
  \includegraphics[width=.40\textwidth, trim={0 0 16.5cm 4cm}, clip]{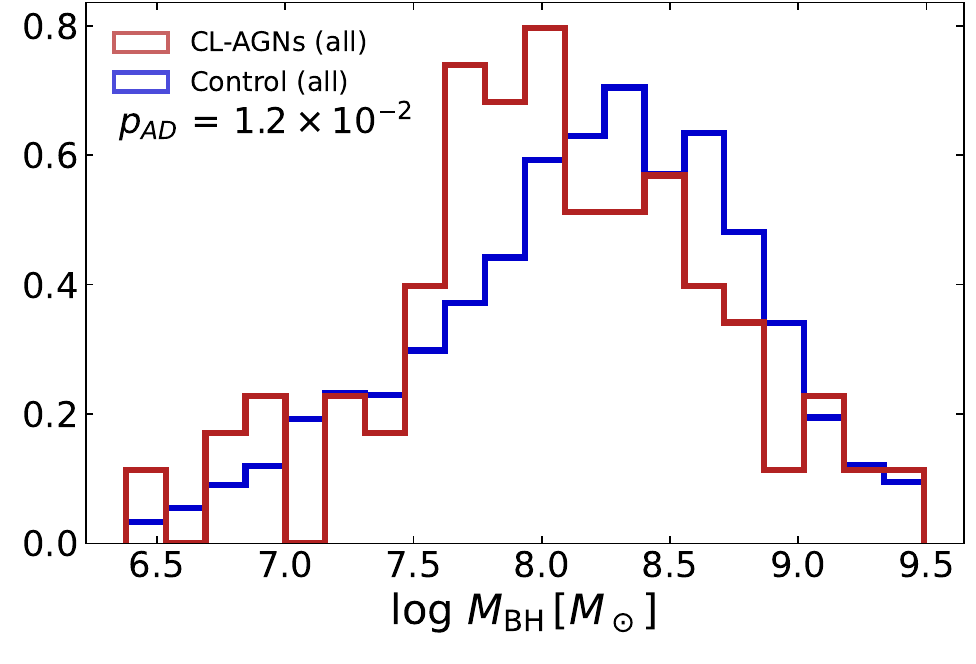}\hfill
  \includegraphics[width=.40\textwidth, trim={0 0 16.5cm 4cm}, clip]{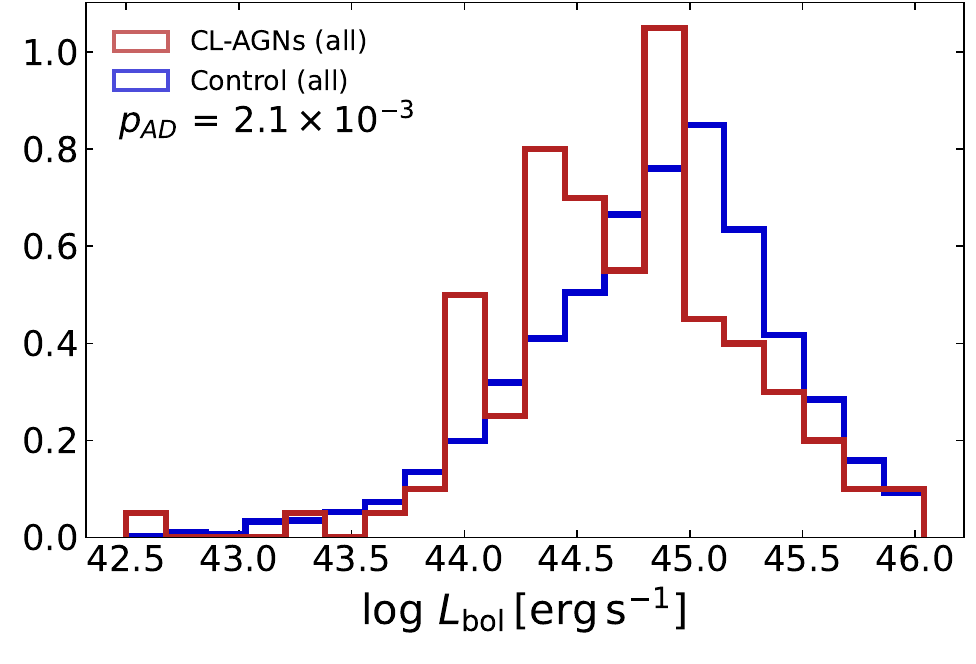}
\caption{Distributions of \Lbol, \mbh, and \fedd for our core sample of 113 CL-AGNs (red), and for matched control samples drawn from the WS22 catalog (blue). 
Unlike Figure~\ref{fig:histograms_ws22}, here the control samples are drawn to match the properties of our entire sample of 113 CL-AGNs.
Each row of panels shows control samples that are matched based on redshift and either \Lbol, \mbh, or \fedd\ (top, middle, and bottom panels, respectively).
Each panel includes the median $p$-value derived from AD tests between our core  sample of 113 CL-AGNs and 1000 realizations of the corresponding control sample.}
\label{fig:histograms_all}
\end{figure*}

Figure~\ref{fig:histograms_all} displays the distributions of \Lbol, \mbh, and \fedd for the CL-AGN sample (red) and the corresponding control sample (blue). 
Similar to Figure~\ref{fig:histograms_ws22}, the control samples are matched based on redshift and on either \Lbol, \mbh, or \fedd (top, middle, and bottom panels, respectively). 
However, unlike Figure~\ref{fig:histograms_ws22}, which utilizes the WS22 catalog values and corresponding derived quantities for the subset of 87 CL-AGNs from our sample that are present in that catalog, here we rely on own ``WS22-like'' spectral decomposition and corresponding derived quantities for the entire core CL-AGN sample (113 in total).

As can be seen in the top panels of Figure \ref{fig:histograms_all}, there is a statistically significant difference between the distributions of \fedd\ for the \Lbol-matched samples ($p_{\rm AD} = 3\times10^{-5}$), as well as between the distributions of \mbh, albeit with a somewhat lower statistical significance ($p_{\rm AD} = 5\times10^{-3}$). 
For the \mbh-matched samples (middle panels), there is a statistically significant difference between the distributions of both \Lbol\ and \fedd\ ($p_{\rm AD}=2\times10^{-7}$ and $3\times10^{-9}$, respectively).
For the \fedd-matched samples, shown in the bottom panels, there appears to be a statistically significant difference between the distribution of \Lbol\ ($p_{\rm AD} =2\times10^{-3}$) while the distributions of \mbh\ do not show a statistically significant difference ($p_{\rm AD}>0.01$).
This latter result for the \fedd-matched \Lbol\ distributions is different from what was found for the subset of 87 CL-AGNs that are part of the WS22 catalog.

\end{appendix}

\end{document}
